\documentclass[aps,pre,10pt,twocolumn,floatfix,amsmath]{revtex4-1}

\usepackage{epsfig}

\renewcommand{\Re}{\mathop{\mathrm{Re}}}
\renewcommand{\Im}{\mathop{\mathrm{Im}}}
\newcommand{\ep}{\epsilon}
\newcommand{\vep}{\varepsilon}
\newcommand{\sump}{\mathop{\sum\nolimits^\prime}\limits}

\begin{document}

\title{Kinetic theory of nonlinear diffusion
in a weakly disordered nonlinear Schr\"odinger chain
in the regime of homogeneous chaos}
\author{D. M. Basko}
\affiliation{Universit\'e Grenoble 1/CNRS, LPMMC UMR 5493,
25 rue des Martyrs, 38042 Grenoble, France}

\begin{abstract}
We study the discrete nonlinear Schr\"oinger equation with weak
disorder, focusing on the regime when the nonlinearity is, on the
one hand, weak enough for the eigenmodes of the linear problem to
remain well resolved, but on the other, strong enough for the
dynamics of the eigenmode amplitudes to be chaotic for almost all
modes.
We show that in this regime and in the limit of high temperature,
the macroscopic density $\rho$ satisfies the nonlinear diffusion
equation with a density-dependent diffusion coefficient,
$D(\rho)=D_0\rho^2$. An explicit expression for $D_0$ is obtained
in terms of the eigenfunctions and eigenvalues of the linear
problem, which is then evaluated numerically.
The role of the second conserved quantity (energy) in the transport
is also quantitatively discussed.
\end{abstract}

\maketitle

\section{Introduction}

Disordered classical nonlinear chains are convenient model
systems where the interplay between Anderson localization
and classical nonlinearity can be studied. Indeed, Anderson
localization is especially pronounced in one spatial
dimension, where even an arbitrarily weak disorder localizes
all normal modes of the linear system~%
\cite{Gertsenshtein1959,Mott1961}.
A nonlinearity couples the normal modes, which may lead to
chaotic dynamics~\cite{Zaslavsky,Lichtenberg} and destroy
Anderson localization
\cite{Shepelyansky1993,Molina1998,Kopidakis2008,Pikovsky2008},
although periodic solutions may be preserved with some
probability~\cite{Frohlich1986,Johansson2010}.
Viewed from the nonlinear dynamics side, disordered nonlinear
chains represent a class of dynamical systems with an infinite
number of  degrees of freedom, in which chaos may have a local
structure in the real space \cite{Oganesyan2009,Basko2011,%
Pikovsky2011,Mulansky2011a,Basko2012,Skokos2013},
and this structure can be controlled by the disorder strength.

Here we study the discrete
nonlinear Schr\"odinger equation with disorder (DNLSE) [see
Eq.~(\ref{DNLS=}) below]. It describes several physical
systems, such as light propagating in nonlinear photonic
lattices~\cite{Lahini2008} or cold bosonic atoms in optical
lattices in the mean-field approximation~\cite{Lucioni2012}.
One of the fundamental problems concerning DNLSE, as well as
disordered nonlinear chains in general,
is the evolution of an initially localized wave packet
at very long times (see Ref.~\cite{Fishman2012} for a review).
In a linear system with Anderson localization, the wave packet
remains exponentially localized at all times, due to the
superposition principle. In the presence of a nonlinearity,
the wave packet width was found to increase as a subdiffusive
power of time~$t$ in the direct numerical simulations of
DNLSE \cite{Shepelyansky1993,Molina1998,Kopidakis2008,%
Pikovsky2008,Flach2009,Skokos2010,Laptyeva2010,Bodyfelt2011}:
the second moment $m_2$ of the wave packet was growing as
$m_2\propto{t}^p$ with $p<1$ (for the usual diffusive
spreading, $p=1$). Specifically,
two regimes of such subdiffusive spreading have been
identified numerically: at longest observed times and at low
densities, $p\approx{1/3}$, but an intermediate range
of densities and times has also been found with
$p\approx{1/2}$. These regimes were called weak and strong
chaos, respectively~\cite{Laptyeva2010,Bodyfelt2011}.

At the same time, rigorous mathematical arguments suggest
that at long times the spreading, if any, should be slower
than any power of~$t$ \cite{Benettin1988,Wang2009}.
Analysis of perturbation theory in the nonlinearity suggests
that there is a front propagating as $\ln{t}$ beyond which
the wave packet is localized exponentially~\cite{Fishman2009}. 
These arguments can be reconciled with those of the direct
numerical simulations, if one assumes 
that the numerically
observed behavior $m_2\propto{t}^{1/3}$ is still an
intermediate asymptotics, which should break down at very
long times. An indication for slowing down of the power-law
subdiffusion has been observed in the scaling analysis of
numerical results~\cite{Mulansky2011,MulanskyNJP} and related
to the scaling of the probability of chaos with the
density~\cite{Mulansky2013}.
A possible mechanism for breakdown of subdiffusion at
long times has been suggested~\cite{Michaely2012}. However,
the progress in this direction is impeded by the absence of
a quantitative theory for the observed power-law
subdiffusion, which would (i)~elucidate the main mechanism
responsible for such subdiffusion, and (ii)~predict
quantitatively when this mechanism ceases to work.

The purpose of the present work is to construct such a theory
for the intermediate spreading regime 
where $m_2\propto{t}^{1/2}$, called ``strong chaos'' in
Refs.~\cite{Laptyeva2010,Bodyfelt2011}. However, the arguments
of the present work are very much analogous to those used in
the derivation of the kinetic equation in the theory of weak
wave turbulence~\cite{Zakharov}. Thus, to avoid the terms
``strong'' and ``weak'', which may be confusing, we use the
term ``homogeneous chaos'' to denote the studied regime. 
The main element of the physical picture of transport developed
here, is the large localization length of normal modes of the
linear problem, which occurs for weak enough disorder.
In this case, the nonlinearity couples each normal mode
to \textit{many} other modes, and if it is not too weak,
the dynamics of almost all modes is chaotic.
So, the chaos can be considered homogeneous both in the
real space and in the mode space. The same physical image
was proposed in Refs.~\cite{Laptyeva2010,Bodyfelt2011} to
identify the spreading regime with $m_2\propto{t}^{1/2}$.

It has already been argued on phenomenological grounds
that macroscopic transport of the conserved density~$\rho$
in DNLSE and other disordered nonlinear chains can be
described by a nonlinear diffusion equation, 
\begin{equation}\label{NLdif=}
\frac{\partial\rho}{\partial{t}}=
\frac\partial{\partial{x}}\left(D(\rho)\,
\frac{\partial\rho}{\partial{x}}\right),
\end{equation}
with a $\rho$-dependent diffusion coefficient $D(\rho)$~%
\cite{Flach2010,Mulansky2010}. Indeed, for a power-law
$D(\rho)=D_0\rho^a$ with some $a>0$ (as found in
Ref.~\cite{Flach2011}), Eq.~(\ref{NLdif=}) gives
$m_2\propto{t}^{2/(a+2)}$.
An expression interpolating between $D(\rho)\propto\rho^2$
at higher densities and $\rho^4$ at lower
densities has been proposed~\cite{Laptyeva2013}, 
giving a crossover from $m_2\propto{t}^{1/2}$ to
${t}^{1/3}$ at longer times. But, again, a
recent matematical work suggests that at lowest densities,
$D(\rho)$ should vanish faster
than any power of~$\rho$~\cite{Huveneers2013}. An explicit
expression satisfying this condition has been given for
the DNLSE in the limit of strong disorder and very low
density, based on the picture of chaos which very
inhomogeneous in space~\cite{Basko2011}. Detailed
quantitative understanding of the mechanisms leading to
Eq.~(\ref{NLdif=}) with a power-law $D(\rho)$ is still
lacking.

In the present paper, starting from the DNLSE with weak
disorder, the nonlinear diffusion equation,
Eq.~(\ref{NLdif=}) with $D(\rho)=D_0\rho^2$, is derived.
An explicit expression [Eq.~(\ref{D0=})] for $D_0$ is
given in terms of a certain average of eigenfunctions
and eigenvalues of the linear part of the DNLSE, which
is then evaluated numerically.
The resulting value of $D_0$ is in reasonable agreement
with that extracted from the direct numerical simulation
of Ref.~\cite{Laptyeva2010}. 
The conditions of validity of the present approach are
thoroughly discussed, and the corresponding interval of
densities is identified (Fig.~\ref{fig:rhoRange}).
Its lower boundary agrees with the value at which the
crossover between $m_2\propto{t}^{1/2}$ and $t^{1/3}$
behavior is observed in Ref.~\cite{Laptyeva2010}.

A specific property of the DNLSE is the presence of
\emph{two} conserved quantities: the total norm (action)
and the total energy. Its consequence for the
thermodynamics of the system is the existence of the
so-called non-thermal states which can be described in
the microcanonical ensemble, but not in the grand
canonical one~\cite{Rasmussen2000}. The consequence
for the transport is that Eq.~(\ref{NLdif=}) for the
norm density $\rho$ is not complete;
the complete macroscopic description should involve 
two coupled equations for the norm and energy.
Such coupled transport has received relatively
little attention so far: a numerical study for the
DNLSE without disorder is available~\cite{Iubini2012},
and in the limit of strong disorder, low density, and
high temperature, the off-diagonal transport coefficients
were estimated to be small~\cite{Basko2011}. Here, the
full $2\times{2}$ matrix of the transport coefficients
is calculated for 
weak disorder, intermediate
density, and high temperature. The analysis of the full
coupled equations shows that in the considered
regime the effect of energy transport on the norm
transport is small, so Eq.~(\ref{NLdif=}) for
$\rho$~only is consistent.

The paper is organized as follows. After introducing the
model in Sec.~\ref{sec:Model}, we derive a
Fokker-Planck-type master equation describing the
diffusive dynamics of the normal mode actions in
Sec.~\ref{sec:master}. Conditions for its validity and
its relation to chaos are discussed in
Sec.~\ref{sec:validity}.
Derivation of the macroscopic Eq.~(\ref{NLdif=}) from
the microscopic master equation, via a Boltzmann-type
kinetic equation for average actions, is done in
Sec.~\ref{sec:macroscopic}. In Sec.~\ref{sec:energy},
the role of the second conserved quantity (energy) is
discussed, and the coupled macroscopic equations are
analyzed. Finally, in Sec.~\ref{sec:numres}, the formal
expressions for the transport coefficients from
Secs.~\ref{sec:macroscopic},~\ref{sec:energy} are
evaluated for different disorder strength, and compared
to the numerics of Ref.~\cite{Laptyeva2010}.

\section{Model and assumptions}
\label{sec:Model}

The DNLSE reads as
\begin{equation}\label{DNLS=}
i\,\frac{d\psi_n}{dt}=-\Omega\left(\psi_{n+1}+\psi_{n-1}\right)
+\ep_n\psi_n+g|\psi_n|^2\psi_n,
\end{equation}
Here $n=1,\ldots,L$ labels sites of a one-dimensional
lattice (the limit $L\to\infty$ will be eventually taken).
To each site~$n$  a pair of complex conjugate variables
$\psi_n,\psi_n^*$ is associated.
The on-site frequencies $\ep_n$ are assumed to be random,
uncorrelated, and to have the flat box distribution
$\ep_n\in[-W/2,W/2]$ whose width~$W$ characterizes the
disorder strength. $\Omega$~and~$g$ measure the intersite
coupling and the strength of the nonlinearity. The sign of~$g$ is
not important, as Eq.~(\ref{DNLS=}) is invariant under the change
$g\to{-g}$, $\psi_n\to(-1)^n\psi_n^*$, $\ep_n\to-\ep_n$ (for the
latter it is important that the distribution of $\ep_n$ is
symmetric). 

Eq.~(\ref{DNLS=}) together with its complex conjugate
represent the Hamilton equations for the classical
Hamiltonian
\begin{equation}\begin{split}\label{Hpsi=}
H=\sum_n\left[\ep_n|\psi_n|^2
-\Omega\left(\psi_n^*\psi_{n+1}+\psi_{n+1}^*\psi_n\right)
+\frac{g}2|\psi_n|^4\right],
\end{split}\end{equation}
if $i\psi_n^*$ is treated as the canonical momentum
conjugate to the coordinate~$\psi_n$. 
One could measure time, frequency, action, and energy
in the units of $1/\Omega$, $\Omega$, $\Omega/g$, and
$\Omega^2/g$, respectively, thus setting $\Omega=1$, $g=1$
in Eq.~(\ref{DNLS=}), without loss of generality. However,
formal expressions are sometimes more transparent physically
when written in the dimensional form, so we assume $\Omega$
and $\psi_n$ to have the dimensionality of frequency
and $(\mathrm{action})^{1/2}$, respectively.

The linear part of the Hamiltonian can be diagonalized by an
orthogonal transformation
\begin{equation}\label{diagonalization=}
\psi_n(t)=\sum_{\alpha=1}^L{c}_\alpha(t)\,\phi_{\alpha{n}},
\end{equation}
where $\phi_{\alpha{n}}$ is the $\alpha$th eigenfunction
of the linear  problem,
\begin{equation}\label{anderson1D=}
\omega_\alpha\phi_{\alpha{n}}=\ep_n\phi_{\alpha{n}}
-\Omega\left(\phi_{\alpha,n+1}+\phi_{\alpha,n-1}\right),
\end{equation}
corresponding to the eigenvalue~$\omega_\alpha$, and
$c_\alpha$~is the complex amplitude of the $\alpha$th normal
mode.
The normal mode amplitudes satisfy the following equations
of motion:
\begin{equation}\label{EOMmodes=}
i\,\frac{dc_\alpha}{dt}=\omega_\alpha{c}_\alpha
+\sum_{\beta,\gamma,\delta}
V_{\alpha\beta\gamma\delta}c^*_\beta{c}_\gamma{c}_\delta.
\end{equation}
Here we introduced the overlap,
\begin{equation}\label{overlap=}
V_{\alpha\beta\gamma\delta}=g\sum_n
\phi_{\alpha{n}}\phi_{\beta{n}}\phi_{\gamma{n}}\phi_{\delta{n}},
\end{equation}
which is real and symmetric with respect to any permutations of
the mode indices. It is a random quantity, and its statistics
will play a crucial role in determining the macroscopic transport
properties of the system and their dependence on the disorder
strength.

The normal mode frequencies $\omega_\alpha$ lie in the interval
$|\omega_\alpha|<2\Omega+W/2$. Their distribution in this interval
is determined by the average spectral density per unit length,
$\nu_1(\omega)$.
All normal mode wave functions are localized, the localization
length $\xi(\omega)$ depending on the mode frequency.
The behavior of $\nu_1(\omega)$ and $\xi(\omega)$ for weak disorder
is discussed in detail in Appendix~\ref{app:DOS}. Here we only note 
that the localization length is the largest for modes close to the
center of the band, $\xi(\omega=0)\approx{100}(\Omega/W)^2$,
$\xi(\omega)$ is of the same order in the most of the band,
and becomes small near the band edges. 
The frequency spacing between the modes which are on the same
localization segment,
$\Delta_1(\omega)=[\nu_1(\omega)\,\xi(\omega)]^{-1}$ is the
smallest at $\omega=0$. 

For the transport mechanism discussed in the present work, it is
crucial that $V_{\alpha\beta\gamma\delta}$, defined in
Eq.~(\ref{overlap=}), couples many modes. This is only possible
when their localization lengths are much larger than the lattice
spacing, otherwise $V_{\alpha\beta\gamma\delta}$ is exponentially
suppressed. The representative estimate for the localization
length is $\xi(\omega=0)\equiv\xi$, since the few tightly localized
modes near the band edges contribute
little to the transport. Thus, the main assumption of the present
work which determines the applicability of the whole approach,
is $\xi\gg{1}$, corresponding to $W\ll{10}\,\Omega$.
Note that omitting the numerical factor from this condition would
result in an unnecessarily severe restriction; in fact, the wave
packet spreading with $m_2\propto{t}^{1/2}$ was observed in
Ref.~\cite{Laptyeva2010} for $W/\Omega=4$. At the same time,
it turns out that for several quantities determined by the
statistics of wave functions, the asymptotic behavior
corresponding to weakest disorder is reached at really small
$W\lesssim{0}.3\,\Omega$ corresponding to $\xi\gtrsim{1000}$
(see Sec.~\ref{ssec:high},  Appendix~\ref{app:shifts}, and Ref.~\cite{Kravtsov2011})

The change of variables~(\ref{diagonalization=}) is useful only
if the dynamics of the normal mode amplitudes, $c_\alpha$, due
to the last term in the right-hand side of
Eq.~(\ref{EOMmodes=}), is not too fast.
The relevant time scale at which the dynamics of the system
allows to resolve the individual modes, is determined by
the frequency spacing on one localization segment.
The representative value is
$\Delta_1(\omega=0)\equiv\Delta_1=2\pi\Omega/\xi$, so
the dynamics of $c_\alpha$ should be slow on the time scale
$1/\Delta_1$. Since the nonlinear dynamics is faster for
larger amplitudes, this condition imposes
an upper limit on the typical norm density,
$|\psi_n|^2\sim|c_\alpha|^2\sim\rho\ll\rho_\mathrm{max}$,
where the value of $\rho_\mathrm{max}$ depends on the disorder
strength, and is discussed in Sec.~\ref{ssec:high}. 

On the other hand, the approach developed below works if
the mode dynamics is sufficiently chaotic. For this, the
nonlinearity should be strong enough, so the density
should not be too low, $\rho\gg\rho_\mathrm{min}$.
This condition is discussed in Sec.~\ref{ssec:low}.
It will be seen that the two restrictions 
$\rho\ll\rho_\mathrm{max}$ and $\rho\gg\rho_\mathrm{min}$ are
not in conflict with each other when the condition $\xi\gg{1}$
is satisfied.

Finally, the temperature~$T$ is assumed to be sufficiently
high, $T\gg\Omega\rho$. This implies that there is no
correlation between the actions $|c_\alpha|^2$ and the
frequencies~$\omega_\alpha$ in the initial condition for
Eq.~(\ref{DNLS=}). It is this situation that was studied
in the numerical works
\cite{Shepelyansky1993,Molina1998,Kopidakis2008,%
Pikovsky2008,Flach2009,Skokos2010,Laptyeva2010,Bodyfelt2011}.
In the opposite case (low~$T$), one should
minimize Hamiltonian~(\ref{Hpsi=}) first, and then study the
dynamics of the low-energy excitations. The ground state of this
classical Hamiltonian at fixed extensive total norm is the Bose
condensate, which is spatially non-uniform due to the disorder.
The normal modes of Eq.~(\ref{DNLS=}), linearized around such
non-uniform condensate (the Bogolyubov modes), are strongly
different from the solutions of the eigenvalue problem
(\ref{anderson1D=})~\cite{Bilas2006,Gurarie2008}.
The assupmtion of high temperature enables one to disregard
the condensate and focus on the normal modes of the linear
problem~(\ref{anderson1D=}).

\section{Microscopic master equation}\label{sec:master}

\subsection{Nonlinear frequency shifts}
\label{ssec:shifts}

In the absence of the nonlinear coupling, the solution for the
normal mode amplitudes is
\begin{equation}\label{unperturb=}
c_\alpha=\sqrt{I_\alpha}\,e^{-i\theta_\alpha^0-i\omega_\alpha{t}},
\end{equation}
where $I_\alpha$ and $\theta_\alpha^0$ are the action and the
initial phase of the mode~$\alpha$, respectively.
With the nonlinearity included, among the $L^3$ terms contributing
to the sum on the right-hand side of  Eq.~(\ref{EOMmodes=}), there
are $2L-1$ terms oscillating at the frequency $\omega_\alpha$
(those for which either $\delta=\alpha$, $\gamma=\beta$ or
$\gamma=\alpha$, $\delta=\beta$). These, the so-called secular terms
are responsible for the nonlinear shift of the $\alpha$th frequency:
\begin{equation}\label{omegaHF=}
\tilde\omega_\alpha=\omega_\alpha
+2\sum_\beta{V}_{\alpha\beta\beta\alpha}I_\beta.
\end{equation}
This can be seen by splitting the full Hamiltonian, corresponding
to Eq.~(\ref{EOMmodes=}),
\begin{equation}\label{Hcccc=}
H=\sum_\alpha\omega_\alpha{c}_\alpha^*{c}_\alpha
+\frac{1}{2}\sum_{\alpha,\beta,\gamma,\delta}
V_{\alpha\beta\gamma\delta}c^*_\alpha{c}^*_\beta{c}_\gamma{c}_\delta,
\end{equation}
into
\begin{equation}\label{HHF=}
H_0=\sum_\alpha\omega_\alpha|c_\alpha|^2
+\sum_{\alpha\beta}V_{\alpha\beta\beta\alpha}|c_\alpha|^2|c_\beta|^2,
\end{equation}
which is integrable, and all the remaining terms, which represent
an integrability-breaking perturbation. 
The solution of the equations of motion
for the Hamiltonian $H_0$ is given by the same Eq.~(\ref{unperturb=}),
but with the modified frequencies, Eq.~(\ref{omegaHF=}), which are
nothing but $\tilde\omega_\alpha=\partial{H}_0/\partial(|c_\alpha|^2)$.
Since the sum over~$\beta$ in Eq.~(\ref{omegaHF=}) effectively contains
a large number of terms, of the order of $\xi\gg{1}$, $I_\beta$ can be
effectively replaced by its average,
$\langle{I}_\beta\rangle\equiv\rho$. Then, using the orthogonality
of the eigenmode wave functions $\phi_{\beta{n}}$, we obtain
\begin{equation}\label{omegaHFav=}
\tilde\omega_\alpha\approx\omega_\alpha+2g\rho.
\end{equation}
The precision of this approximation can be quantified by calculating
the average relative fluctuation~$\sigma^2$ of the nonlinear frequency
shift,
\begin{equation}\label{omegaHFfluc=}
\left\langle(\tilde\omega_\alpha-\omega_\alpha-2g\rho)^2\right\rangle
=4\rho^2\left\langle\sum_\beta{V}_{\alpha\beta\beta\alpha}^2
\right\rangle\equiv 4g^2\rho^2\sigma^2,
\end{equation}
where averaging of the quantities that depend on
${V}_{\alpha\beta\beta\alpha}$ is performed over the disorder
realizations, and of those which depend on $I_\beta$ is over the
thermal distribution $\prod_\beta\rho^{-1}{e}^{-I_\beta/\rho}$
(see Sec.~\ref{ssec:relaxation} below for details).
In particular, for this distribution,
$\langle{I}_\beta{I}_\gamma\rangle=(1+\delta_{\beta\gamma})\rho^2$,
which gives Eq.~(\ref{omegaHFfluc=}). 

Since the typical number of terms contributing to the sum
over~$\beta$ is $\sim\xi$ (at larger distances
$V_{\alpha\beta\beta\alpha}$ decays exponentially), it is natural
to expect the relative fluctuation $\sigma^2\sim{1}/\xi\ll{1}$.
The same result is obtained by considering the exact expression
\begin{equation}
\sum_\beta{V}_{\alpha\beta\beta\alpha}^2
=g^2\sum_\beta\sum_{n,n'}\phi_{\alpha{n}}^2\phi_{\alpha{n}'}^2
\phi_{\beta{n}}^2\phi_{\beta{n}'}^2,
\end{equation}
estimating $\phi_{\alpha{n}}^2\sim{1}/\xi$, and assuming each
summation to give a factor $\sim\xi$. The numerical test of this
argument is performed in Appendix~\ref{app:shifts}. 

\subsection{Diffusion in actions}
\label{ssec:delta}

The non-secular terms in the sum on the right-hand side of
Eq.~(\ref{EOMmodes=}) (i.~e., those, contained in $H-H_0$) lead to
exchange of energy between the normal modes, so that the actions
$I_\alpha$ change with time. Effectively, only those
$\beta,\gamma,\delta$ contribute to the sum, which correspond to the
modes separated from the mode $\alpha$ by a distance $\sim\xi$ (for
other modes, the overlap $V_{\alpha\beta\gamma\delta}$ is
exponentially small).
Among those, there are $O(\xi^3)$ terms for which all four indices
$\alpha,\beta,\gamma,\delta$ are different, and $O(\xi^2)$ terms
for which at least two indices coincide. For $\xi\gg{1}$, these
latter terms can be ignored, as their role is analogous to
those with $\alpha,\beta,\gamma,\delta$ are different, while their
number is parametrically smaller.

Let us now consider just one term with different indices,
that is, four oscillators described by the Hamiltonian
\begin{equation}
H(\{c_\alpha\})=\sum_{\alpha=1}^4\omega_\alpha|c_\alpha|^2
+2V_{1234}\left(c_1^*c_2^*c_3c_4+c_1c_2c_3^*c_4^*\right).
\end{equation}
In principle, there are two other terms coupling the same four
oscillators, $c_1^*c_3^*c_2c_4$ and $c_1^*c_4^*c_2c_3$ (with
their complex conjugates), but we do not consider them for the
moment. We have also neglected the nonlinear frequency shifts
for simplicity.
The first-order (in $V_{1234}$) correction to the unperturbed
solution~(\ref{unperturb=}) for $c_1$, is given by
\begin{equation}\begin{split}\label{Deltac1=}
&\Delta{c}_1=e^{-i\theta_1^0-i\omega_1t}\,2V_{1234}
\sqrt{I_2I_3I_4}\,e^{i\vartheta_{1234}}\,
\frac{e^{1-i\varpi_{1234}{t}}}{\varpi_{1234}},\\
&\varpi_{1234}\equiv\omega_1+\omega_2-\omega_3-\omega_4,\\
&\vartheta_{1234}\equiv\theta_1^0-\theta_2^0-\theta_3^0+\theta_4^0.
\end{split}\end{equation}
The solutions for the other three oscillators are obtained by
appropriate permutations. The change in the actions of the
oscillators,
$\Delta{I}_\alpha=c_\alpha^*\,\Delta{c}_\alpha
+c_\alpha\,\Delta{c}_\alpha^*$, is given by
\begin{subequations}\begin{eqnarray}
&&\Delta{I}_1=2V_{1234}\sqrt{I_1I_2I_3I_4}\times\nonumber\\
&&\qquad{}\times{}2\Re\left(e^{-i\vartheta_{1234}}\,
\frac{1-e^{i\varpi_{1234}{t}}}{\varpi_{1234}}\right),
\quad\\ \label{dI1dI2dI3dI4=}
&&\Delta{I}_2=-\Delta{I}_3=-\Delta{I}_4=\Delta{I}_1.
\end{eqnarray}\end{subequations}
The increment $\Delta{I}_1$ can be positive or negative,
depending on the phases $\theta_\alpha^0$. 

Let us now consider coupling of the mode $\alpha=1$ to
all other modes. The increment $\Delta{I}_\alpha$ will be
a sum of many terms with random signs, so the dynamics
of the action is diffusive. To quantify this dynamics,
let us calculate the phase-averaged $\Delta{I}_\alpha^2$:
\begin{equation}\begin{split}
\langle\Delta{I}_\alpha^2\rangle_\theta=
{}&{}\frac{1}2\sump_{\beta,\gamma,\delta}
4V_{\alpha\beta\gamma\delta}^2
I_\alpha{I}_\beta{I}_\gamma{I}_\delta\,
\frac{2\sin^2(\varpi_{\alpha\beta\gamma\delta}t/2)}%
{(\varpi_{\alpha\beta\gamma\delta}/2)^2}\approx\\
\approx{}&{}\frac{1}2\sump_{\beta,\gamma,\delta}
4V_{\alpha\beta\gamma\delta}^2
I_\alpha{I}_\beta{I}_\gamma{I}_\delta\,
4\pi{t}\,\delta(\varpi_{\alpha\beta\gamma\delta}).
\label{dIdelta=}
\end{split}\end{equation}
where the prefactor $1/2$ prevents from double counting which
originates from the symmetry $\gamma\leftrightarrow\delta$,
and the prime at the sum indicates that the secular terms
should be excluded. The key feature of Eq.~(\ref{dIdelta=})
is that $\langle\Delta{I}_\alpha^2\rangle\propto{t}$, which
is indeed a signature of the diffusive dynamics.

The $\delta$-function in Eq.~(\ref{dIdelta=}) deserves some
discussion. Formally, the presented derivation is fully
analogous to the derivation of the Fermi Golden Rule for
decay into a continuous spectrum, familiar from quantum
mechanics~\cite{LL3}. However, the sum in Eq.~(\ref{dIdelta=})
is discrete, so the $\delta$~function cannot be understood
in the strict sense.
For any finite $\varpi_{\alpha\beta\gamma\delta}$, each
individual term in the sum in the first line of
Eq.~(\ref{dIdelta=}) represents a growth $\propto{t}^2$ for
$t\ll1/|\varpi_{\alpha\beta\gamma\delta}|$,
and saturation (with oscillations) at
$t\gg{1}/|\varpi_{\alpha\beta\gamma\delta}|$,
so it never grows linearly in time. However, when considering
the whole sum, only the terms with
$|\varpi_{\alpha\beta\gamma\delta}|<1/t$ contribute to the
$t^2$ growth of $\Delta{I}_\alpha^2$. The number of such
terms decreases $\sim{1}/(\Delta\varpi{t})$, where $1/t$ is
the typical width of the peak in
$\varpi_{\alpha\beta\gamma\delta}$, and
$\Delta\varpi$ is the typical spacing between the values of
$\varpi_{\alpha\beta\gamma\delta}$ effectively contributing
to the sum for various $\beta,\gamma,\delta$. Thus,
Eq.~(\ref{dIdelta=}) only makes sense for not too long times,
$t\ll{1}/\Delta\varpi$, when the number of terms contributng
to the sum in the first line is large. This large number of
terms which enter with random phases also justifies the phase
averaging in Eq.~(\ref{dIdelta=}). This condition of large
number of terms is crucial for the validity of the whole
approach, and will be analyzed in detail in Sec.~\ref{ssec:low}.

\subsection{Master equation}

To describe the diffusion of actions, we introduce the joint
distribution function, $\mathcal{F}(\{I_\alpha\})$, which depends
on actions of all the normal modes. Eq.~(\ref{dIdelta=}) together
with the constraint
\begin{equation}
\Delta{I}_\alpha=\Delta{I}_\beta=-\Delta{I}_\gamma=-\Delta{I}_\delta,
\end{equation}
holding for each coupling term $V_{\alpha\beta\gamma\delta}$
[cf. Eq.~(\ref{dI1dI2dI3dI4=})], yield a diffusion equation of the
following form:
\begin{equation}\begin{split}\label{master=}
&\frac{\partial\mathcal{F}}{\partial{t}}=
\frac{1}8\sump_{\alpha,\beta,\gamma,\delta}
\left(\frac\partial{\partial{I}_\alpha}+\frac\partial{\partial{I}_\beta}
-\frac\partial{\partial{I}_\gamma}-\frac\partial{\partial{I}_\delta}\right)
\times\\ &\qquad{}\times{}
4V_{\alpha\beta\gamma\delta}^2
{I}_\alpha{I}_\beta{I}_\gamma{I}_\delta\,2\pi
\delta\!\left(\tilde\omega_\alpha+\tilde\omega_\beta
-\tilde\omega_\gamma-\tilde\omega_\delta\right)
\times\\ &\qquad{}\times{}
\left(\frac\partial{\partial{I}_\alpha}+\frac\partial{\partial{I}_\beta}
-\frac\partial{\partial{I}_\gamma}-\frac\partial{\partial{I}_\delta}\right)
\mathcal{F},
\end{split}\end{equation}
where the prefactor $1/8$ is introduced to prevent from double
counting which originates from the symmetry
$\alpha\leftrightarrow\beta$, $\gamma\leftrightarrow\delta$,
$(\alpha,\beta)\leftrightarrow(\gamma,\delta)$
(indeed, out of $24$ permutations of four different indices
$\alpha,\beta,\gamma,\delta$, only three produce distinct
diffusion operators).

In Eq.~(\ref{master=}) we have also included the nonlinear
frequency shifts in the $\delta$-function: each frequency
$\tilde\omega_\alpha$ depends on all actions $I_\beta$,
as given by Eq.~(\ref{omegaHF=}).
In fact, all reasoning of Sec.~\ref{ssec:delta} applies also
to the case of shifted frequencies. Indeed, since the
perturbative expressions of Sec.~\ref{ssec:delta} are valid
only as long as
$\sqrt{\langle\Delta{I}_\alpha^2\rangle}\ll{I}_\alpha$ 
anyway, in the derivation of Eq.~(\ref{dIdelta=}) the
frequencies $\tilde\omega_\alpha$ can be
considered fixed and determined by the instantaneous values
of the actions.

\subsection{Equilibrium and relaxation}
\label{ssec:relaxation}

The master equation~(\ref{master=}) conserves the total
action (norm) and total (unperturbed) energy,
\begin{subequations}\begin{eqnarray}
&&\frac{d\langle\mathcal{N}\rangle_\mathcal{F}}{dt}=0,\quad
\mathcal{N}=\sum_\alpha{I}_\alpha\,,\\
&&\frac{d\langle\mathcal{E}\rangle_\mathcal{F}}{dt}=0,\quad
\mathcal{E}=\sum_\alpha\omega_\alpha{I}_\alpha
+\sum_{\alpha,\beta}V_{\alpha\beta\beta\alpha}
I_\alpha{I}_\beta\,,\qquad
\end{eqnarray}\end{subequations}
the latter being ensured by the frequency $\delta$-function
with the frequencies given by Eq.~(\ref{omegaHF=}) (the
consequences of the finite width of the $\delta$~function are
discussed in Sec.~\ref{ssec:noshifts}, for the moment we
proceed formally).
Here we introduced the averaging over the distribution
function in the standard way:
\begin{equation}
\langle\ldots\rangle_\mathcal{F}=
\int\left(\ldots\right)\mathcal{F}(\{I_\alpha\})
\prod_\alpha{d}I_\alpha.
\end{equation}
Any distribution function $\mathcal{F}$ depending on
the actions through the combinations
$\mathcal{N},\mathcal{E}$, will be a stationary solution
of the master equation.
In particular, this is the case for the thermal distribution
function
\begin{equation}\label{Feq=}
\mathcal{F}_\mathrm{eq}\propto{e}^{-(\mathcal{E}-\mu\mathcal{N})/T},
\end{equation}
where $T$ is the temperature and $\mu$ is the chemical
potential. This distribution is the one that maximizes
the entropy $\langle\ln(e/\mathcal{F})\rangle_\mathcal{F}$,
[for which the $H$-theorem can be straightforwardly obtained
from Eq.~(\ref{master=})], under the constraint of fixed
average action $\langle\mathcal{N}\rangle_\mathcal{F}$
and energy $\langle\mathcal{E}\rangle_\mathcal{F}$ (provided
that the energy does not exceed a certain critical
value~\cite{Rasmussen2000}, see the discussion in
Sec.~\ref{ssec:coupled}).

We will be interested in the high-temperature limit
\begin{equation}\label{infiniteT=}
T\to\infty,\quad \mu\to-\infty,\quad
\frac\mu{T}=-\lambda=\mathrm{const}.
\end{equation}
Then the average action 
$\langle{I}_\alpha\rangle_{\mathcal{F}_\mathrm{eq}}$ or,
equivalently, the average density
$\langle|\psi_n|^2\rangle_{\mathcal{F}_\mathrm{eq},\theta}$
are simply given by $\rho=1/\lambda$.
(in the first case, the average is performed over the 
distribution function, while in the second one the
average over the phases is also implied)

Let us assume that all modes are in equilibrium, except one,
say, $\alpha=1$. This corresponds to the distribution function
of the form
\begin{equation}\label{F1alpha=}
\mathcal{F}(\{I_\alpha\})=f(I_1)\prod_{\alpha\neq{1}}
\frac{e^{-I_\alpha/\rho}}\rho,
\end{equation}
where $\rho$ is the equilibrium density, and we have taken
the limit~(\ref{infiniteT=}). Then, $f(I_1)$ satisfies the
following equation:
\begin{equation}\label{diff1=}
\frac{1}\Gamma_1\,\frac{\partial{f}}{\partial{t}}=
\frac\partial{\partial{I}_1}\,I_1
\left(\rho\,\frac\partial{\partial{I}_1}+1\right)f,
\end{equation}
where we denoted
\begin{equation}\label{GammaNL=}
\Gamma_\alpha=\frac{4\pi}\rho\sump_{\beta,\gamma,\delta}
V_{\alpha\beta\gamma\delta}^2\left\langle\delta\!
\left(\tilde\omega_\alpha+\tilde\omega_\beta
-\tilde\omega_\gamma-\tilde\omega_\delta\right)
I_\beta{I}_\gamma{I}_\delta\right\rangle_\mathcal{F}.
\end{equation}
Eq.~(\ref{diff1=}) coincides with the Fokker-Planck equation
for a damped oscillator subject to external noise (see
Appendix~\ref{app:oscnoise}), and has been studied in detail
in Sec.~8.3 of Ref.~\cite{Basko2011}. In particular,
$1/\Gamma_\alpha$ is the typical relaxation time
for the mode~$\alpha$.

When the nonlinear frequency shifts are self-averaging, as
given by Eq.~(\ref{omegaHFav=}), they disappear from the
$\delta$-function.
Also, since the sum in Eq.~(\ref{GammaNL=}) is contributed
by different $\beta,\gamma,\delta$, for the distribution
function~(\ref{F1alpha=}) we can split
$\langle{I}_\beta{I}_\gamma{I}_\delta\rangle_\mathcal{F}\to
\langle{I}_\beta\rangle_\mathcal{F}
\langle{I}_\gamma\rangle_\mathcal{F}
\langle{I}_\delta\rangle_\mathcal{F}$
(see also Sec.~\ref{ssec:Boltzmann} and Appendix~\ref{app:moments}).
Then Eq.~(\ref{GammaNL=}) reduces to
\begin{equation}\label{Gamma=}
\Gamma_\alpha=4\pi\rho^2\sump_{\beta,\gamma,\delta}
V_{\alpha\beta\gamma\delta}^2\,
\delta\!\left(\omega_\alpha+\omega_\beta
-\omega_\gamma-\omega_\delta\right).
\end{equation}

\subsection{Numerical evaluation of the relaxation rate}
\label{ssec:numGamma}

\begin{figure}
\includegraphics[width=8cm]{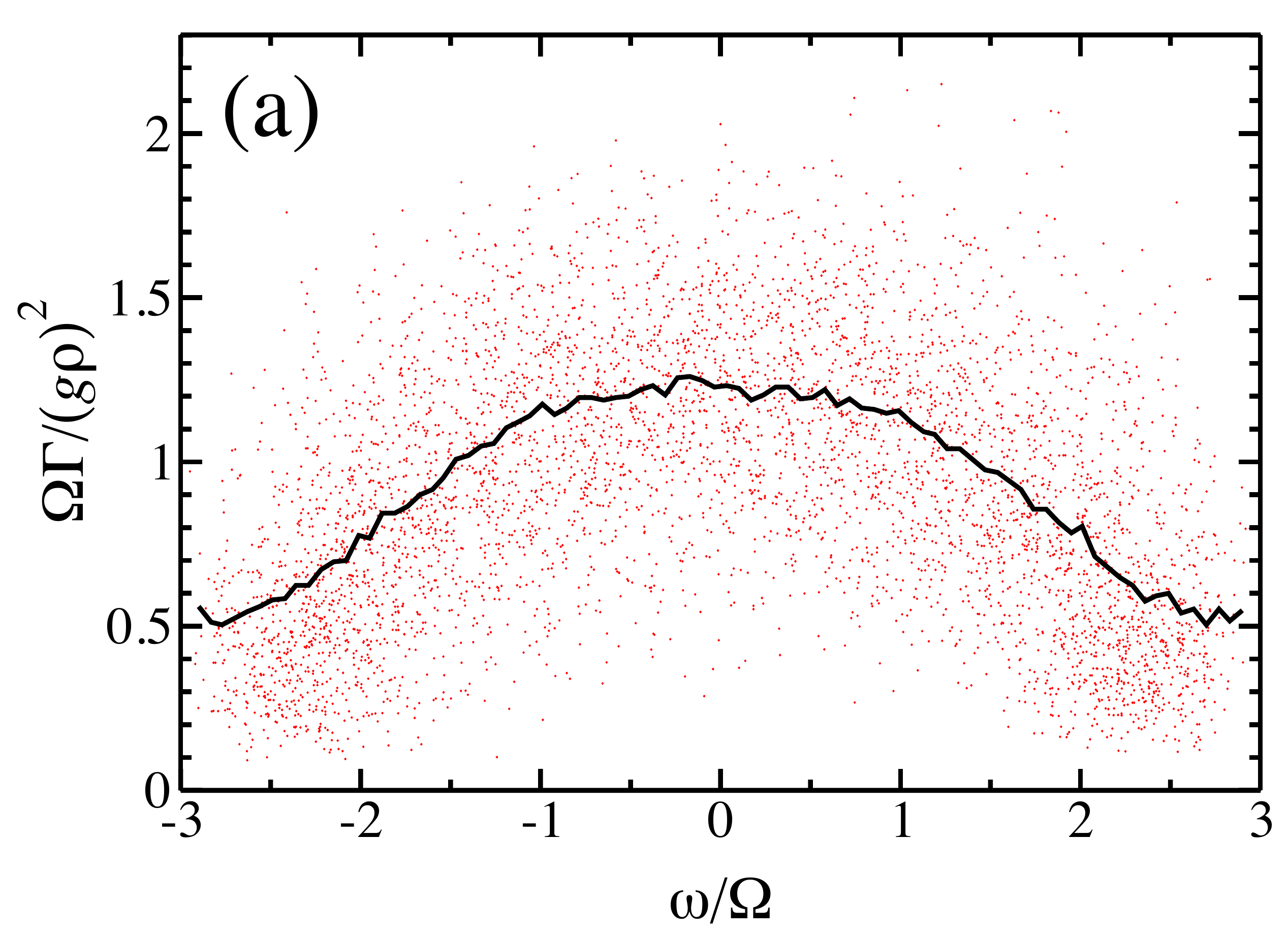}
\includegraphics[width=8cm]{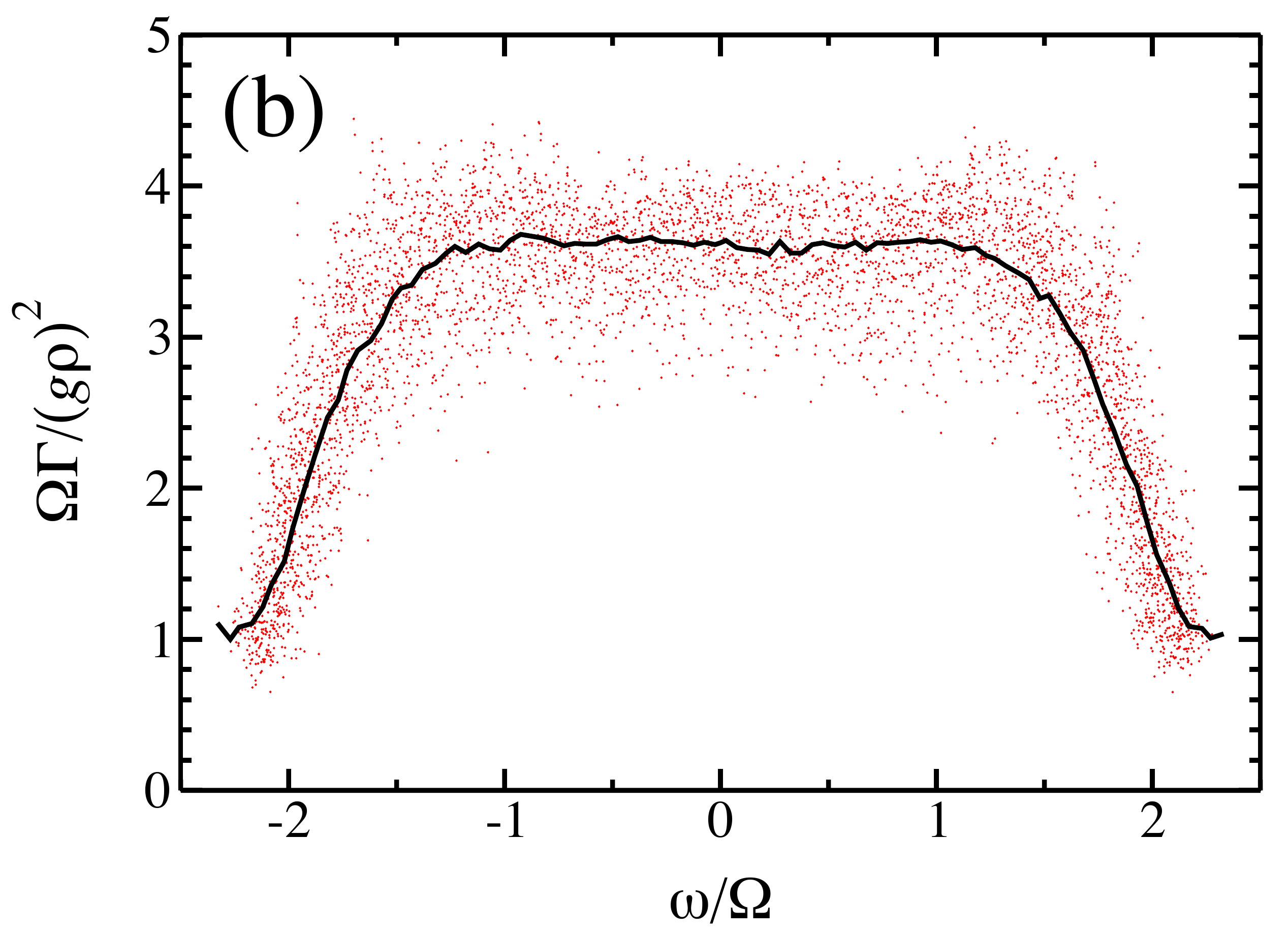}
\caption{\label{fig:GammaOm}
(a)~The relaxation rate in a chain of length $L=5000$ with
$W/\Omega=2.82$ obtained using Eq.~(\ref{deltaw=}) with
$w=0.1\,\Omega$ for individual eigenmodes from
Eq.~(\ref{Gamma=}) (dots),
and averaged over the modes at a given frequency,
Eq.~(\ref{Gammaw=}) with $\Delta=0.1\,\Omega$ (solid line).
(b)~The same for $W/\Omega=1$ and $w=0.01\,\Omega$.}
\end{figure}

The frequency $\delta$-function appearing in Eqs.~(\ref{GammaNL=})
and (\ref{Gamma=}) is implemented as a box of a finite width~$2w$,
\begin{equation}\label{deltaw=}
\delta(\varpi)\to\delta_w(\varpi)=\frac{\theta(w-|\varpi|)}{2w},
\end{equation}
where $\theta(x)$ is the unit step function. This implementation
is more efficient for numerical calculations than a smooth
function. Using this definition, one can evaluate $\Gamma_\alpha$
for each mode~$\alpha$ using Eq.~(\ref{Gamma=}). Two examples of
such calculation are shown in Fig.~\ref{fig:GammaOm}. 

Rather than looking at mode-specific quantities, we will be 
interested in some average characteristics. As the properties
of an eigenmode are determined by its frequency, it is convenient
to work with frequency-resolved averages; so we define
\begin{equation}\label{Gammaw=}
\overline{\Gamma}_\omega=\lim\limits_{L\to\infty}
\frac{\sum_{\alpha=1}^L\delta_\Delta(\omega_\alpha-\omega)\,
\Gamma_\alpha^{(w)}}%
{\sum_{\alpha=1}^L\delta_\Delta(\omega_\alpha-\omega)},
\end{equation}
where $\Delta$ is a sufficiently small interval, and averaging
over a long chain is assumed to be equivalent to averaging over
the disorder realizations. It is important that the limit
$L\to\infty$ should be taken prior to $\Delta\to{0}$.

It should be emphasized that the result of the calculation of
individual $\Gamma_\alpha$ depends on the width~$w$. Namely,
the same calculation with a larger~$w$ produces an analogous
picture, but the points in Fig.~\ref{fig:GammaOm} are more
squeezed towards the solid line, that is, $\Gamma_\alpha$
becomes more self-averaging. For smaller~$w$, the points
become more scattered. At the same time, the average
$\overline{\Gamma}_\omega$ is not sensitive to the value
of~$w$ as long as $w\ll\Omega$. However, at sufficiently
small~$w$ the values of $\Gamma_\alpha$ are so scattered that
the fluctuations become comparable to the average, and the
latter is no longer representative.
This dependence on~$w$ is crucial for the discussion in
Sec.~\ref{ssec:low} below.

\begin{figure}
\begin{center}
\includegraphics[width=8cm]{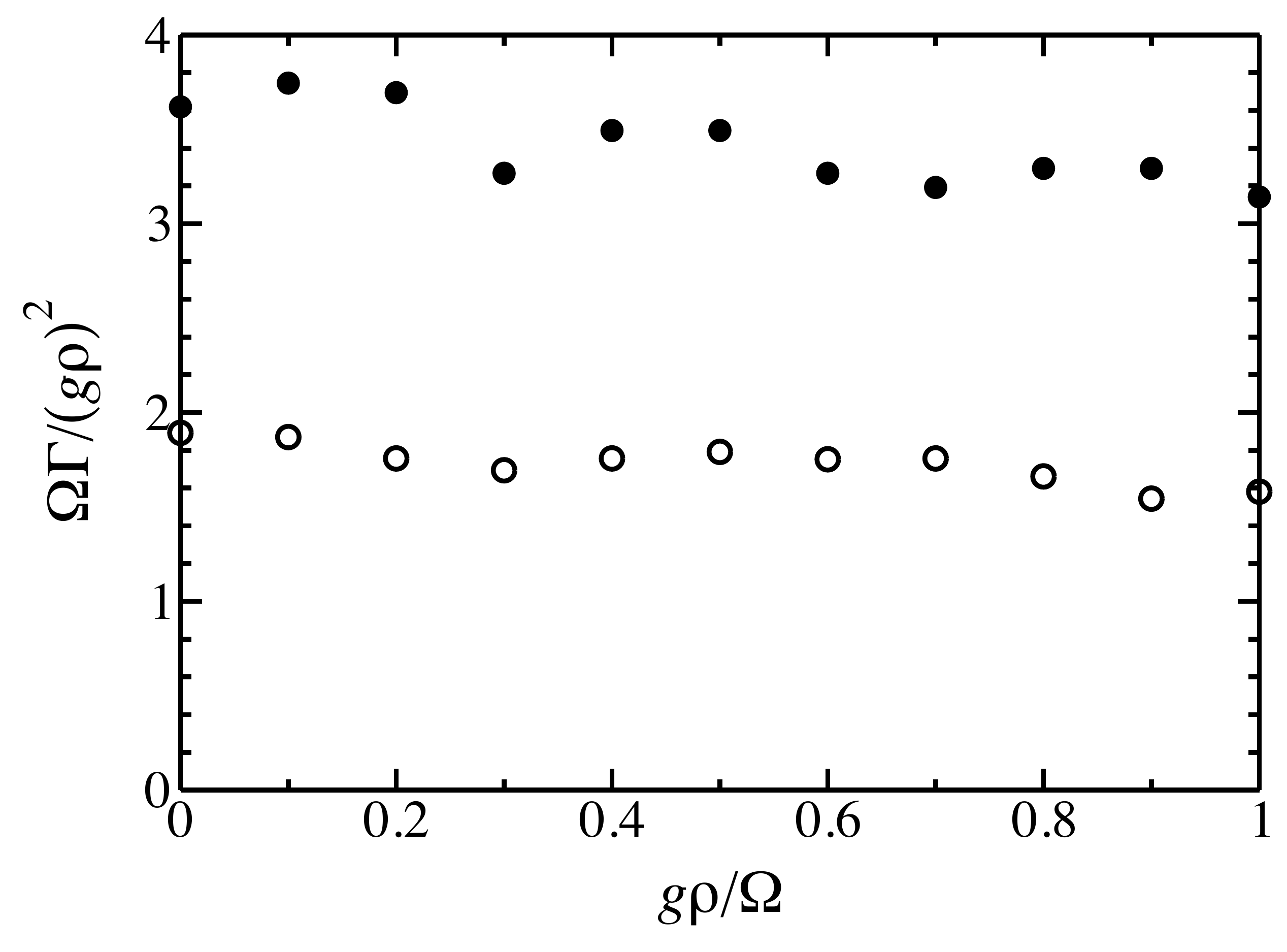}
\end{center}
\caption{\label{fig:GammaNL}
The relaxation rate
as a function of the dimensionless density $g\rho/\Omega$ for
$W/\Omega=1,2$ (filled and open circles, respectively), evaluated
from Eq.~(\ref{GammaNL=}) by numerically averaging over the
equilibrium distribution and over the modes around $\omega=0$
(the width $\Delta=0.1\,\Omega$), the $\delta$~function
given by Eq.~(\ref{deltaw=}) with $w=0.01\,\Omega$,
$\Delta=0.1\,\Omega$.
}
\end{figure}

We also check numerically the dependence $\Gamma_\alpha\propto\rho^2$ 
by evaluating Eq.~(\ref{GammaNL=}) [again, implementing the
$\delta$~function as in Eq.~(\ref{deltaw=}) and averaging over the 
modes at a given frequency as in Eq.~(\ref{Gammaw=})], where the
actions $I_\alpha$ are taken from the equilibrium distribution,
$\mathcal{F}\propto\prod_\alpha{e}^{-I_\alpha/\rho}$.  
As seen from Fig.~\ref{fig:GammaNL} for two values of the disorder
strength, $W/\Omega=1,2$, the average
$\langle\Gamma_\alpha\rangle/\rho^2$ is practically independent
of~$\rho$ up to the values of $\rho$ as large as~$\Omega/g$.

\section{Conditions for validity of the approach}
\label{sec:validity}

\subsection{Low densities}
\label{ssec:low}

The arguments of Sec.~\ref{ssec:delta} are based on the
perturbative expression (\ref{Deltac1=}), which is valid only
as long as the change in the actions is small compared to
actions themselves,
$\sqrt{\langle\Delta{I}_\alpha^2\rangle}\ll{I}_\alpha$.
This means that the time $t$ should be smaller than the typical
action relaxation time, $1/\Gamma_\alpha$, introduced in
Sec.~\ref{ssec:relaxation}. If at times
$t\lesssim{1}/\Gamma_\alpha$
the discreteness of the sum in Eq.~(\ref{dIdelta=}) is not
relevant, the arguments of Sec.~\ref{ssec:delta} are
self-consistent. In other words, if we take the effective
width of the $\delta$~function to be of the order of
$\Gamma_\alpha$ itself,
the sum in Eq.~(\ref{dIdelta=}) should be contributed by many
terms. Note that in this sense the sum in Eq.~(\ref{dIdelta=})
and that in Eq.~(\ref{Gamma=}) are fully analogous. In terms
of $\Delta\varpi$, the typical spacing between the values of
$\varpi_{\alpha\beta\gamma\delta}$ effectively contributing
to the sum for various $\beta,\gamma,\delta$, the condition
of consistency is thus simply $\Delta\varpi\ll\Gamma_\alpha$,
which should hold for most of the modes~$\alpha$.

The value of $\Delta\varpi$ is determined both
by the frequencies $\varpi_{\alpha\beta\gamma\delta}$
\emph{and} by the overlaps $V_{\alpha\beta\gamma\delta}$
[but \emph{not} by the actions~$I_\alpha$, as long as the
nonlinear shifts are self-averaging, Eq.~(\ref{omegaHFav=})].
In particular, only modes $\beta,\gamma,\delta$ which are
effectively within the same localization segment as
the mode $\alpha$ can give a significant contribution to the
sum, since at large distances the overlap
$V_{\alpha\beta\gamma\delta}$ is exponentially suppressed.
Thus, if there were no correlation between the frequencies
and the overlaps $V_{\alpha\beta\gamma\delta}$, we could
estimate $\Delta\varpi\sim\Delta_3$, where
$\Delta_3\approx{0}.14\,\Omega/\xi^3$ is the typical spacing
between different combinations of three frequencies,
$\omega_\delta+\omega_\gamma-\omega_\beta$
(see Appendix~\ref{app:DOS} for details), for modes
which are within the same localizaton length.

However, the correlation between the frequencies and the overlaps
turns out to be quite strong (see Appendix~\ref{app:correlations}
for details), so it is not obvious how to actually define
$\Delta\varpi$. This difficulty can be bypassed by noting that
when the sum in Eq.~(\ref{Gamma=}) is contributed by many terms,
$\Gamma_\alpha$~should be effectively self-averaging and its 
fluctuations should be small, and vice versa. This is directly
related to the dependence of the fluctuations of $\Gamma_\alpha$
(the vertical spread of the points in Fig.~\ref{fig:GammaOm})
on the effective $\delta$-function width~$w$, discussed in
Sec.~\ref{ssec:numGamma} above. Thus, the criterion
$\Delta\varpi\ll\Gamma_\alpha$, proposed above, can be
replaced by an equivalent one: the fluctuations of $\Gamma_\alpha$,
evaluated for modes with frequencies $\omega_\alpha\approx\omega$
at $w=\overline{\Gamma}_\omega$, should be small compared to the
average $\overline{\Gamma}_\omega$ itself.
Basically, it is the same criterion that was used in
Ref.~\cite{BAA2006} for the validity of the kinetic
equation on the metallic side of the quantum many-body
localization transition.

Specifically, we focus on $\omega=0$, assuming it to be
representative of the whole band. Indeed, the results shown in
Fig.~\ref{fig:GammaOm} suggest that the dependence on $\omega$
is weak as long as $\omega$ is away from the band edges (see,
hovewer, the discussion in Sec.~\ref{ssec:high}).
We then define the first and the second moments as
\begin{subequations}
\begin{eqnarray}
&&M_1(w)=\lim\limits_{L\to\infty}
\left[\sum_{\alpha=1}^L\delta_\Delta(\omega_\alpha)\right]^{-1}
\sum_{\alpha=1}^L\delta_\Delta(\omega_\alpha)
\times\nonumber\\ &&\qquad\qquad\times
\frac{2\Omega}{g^2}\sump_{\beta,\gamma,\delta}
V_{\alpha\beta\gamma\delta}^2\,
\delta_w(\varpi_{\alpha\beta\gamma\delta}),
\nonumber\\ &&\label{G1gamma=}\\
&&M_2(w)=\lim\limits_{L\to\infty}
\left[\sum_{\alpha=1}^L\delta_\Delta(\omega_\alpha)\right]^{-1}
\sum_{\alpha=1}^L\delta_\Delta(\omega_\alpha)
\nonumber\times\\
&&\qquad\qquad\times\left[\frac{2\Omega}{g^2}
\sump_{\beta,\gamma,\delta}V_{\alpha\beta\gamma\delta}^2\,
\delta_w(\varpi_{\alpha\beta\gamma\delta})\right]^2,\label{G2gamma=}
\end{eqnarray}
\end{subequations}
As mentioned in Sec.~\ref{ssec:numGamma} above, when the
limit $L\to\infty$ is taken, $M_1(w)$ does not depend on~$w$
at $w\ll\Omega$, so we can denote $M_1\equiv{M}_1(0)$, and
$\overline{\Gamma}_{\omega=0}=2\pi(g^2\rho^2/\Omega)M_1$. 
The fluctuations $M_2(w)-M_1^2$ still depend on~$w$, and we
denote by $w_\mathrm{min}$ the value of~$w$ at which the
fluctuations are equal to the average:
\begin{equation}
M_2(w_\mathrm{min})-M_1^2=M_1^2.
\end{equation}
The dimensionless quantities $M_1$ and
$w_\mathrm{min}/\Omega$ depend only on the dimensionless
disorder strength $W/\Omega$. We can now define the
minimal density $\rho_\mathrm{min}$ as the one for which
$2\pi(g^2\rho_\mathrm{min}^2/\Omega)M_1=w_\mathrm{min}$.
Thus, the low-density limit of the validity of the master
equation~(\ref{master=}) is
\begin{equation}\label{rhomin=}
\frac{g\rho}{\Omega}\gg
\sqrt{\frac{w_\mathrm{min}}{2\pi\Omega{M}_1}}
\equiv\frac{g\rho_\mathrm{min}}{\Omega}.
\end{equation}

\begin{figure}
\begin{center}
\vspace{0cm}
\includegraphics[width=8cm]{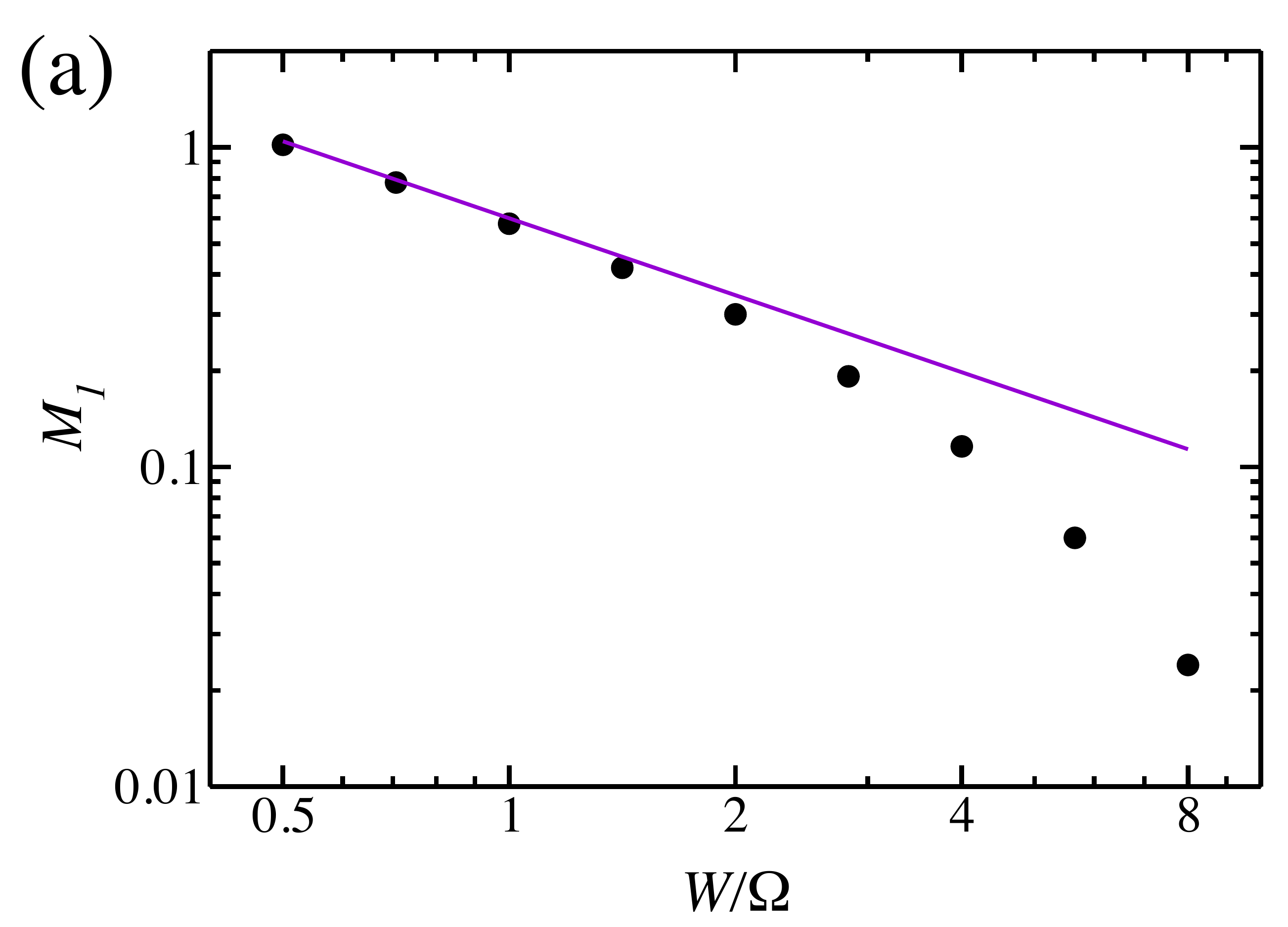}
\includegraphics[width=8cm]{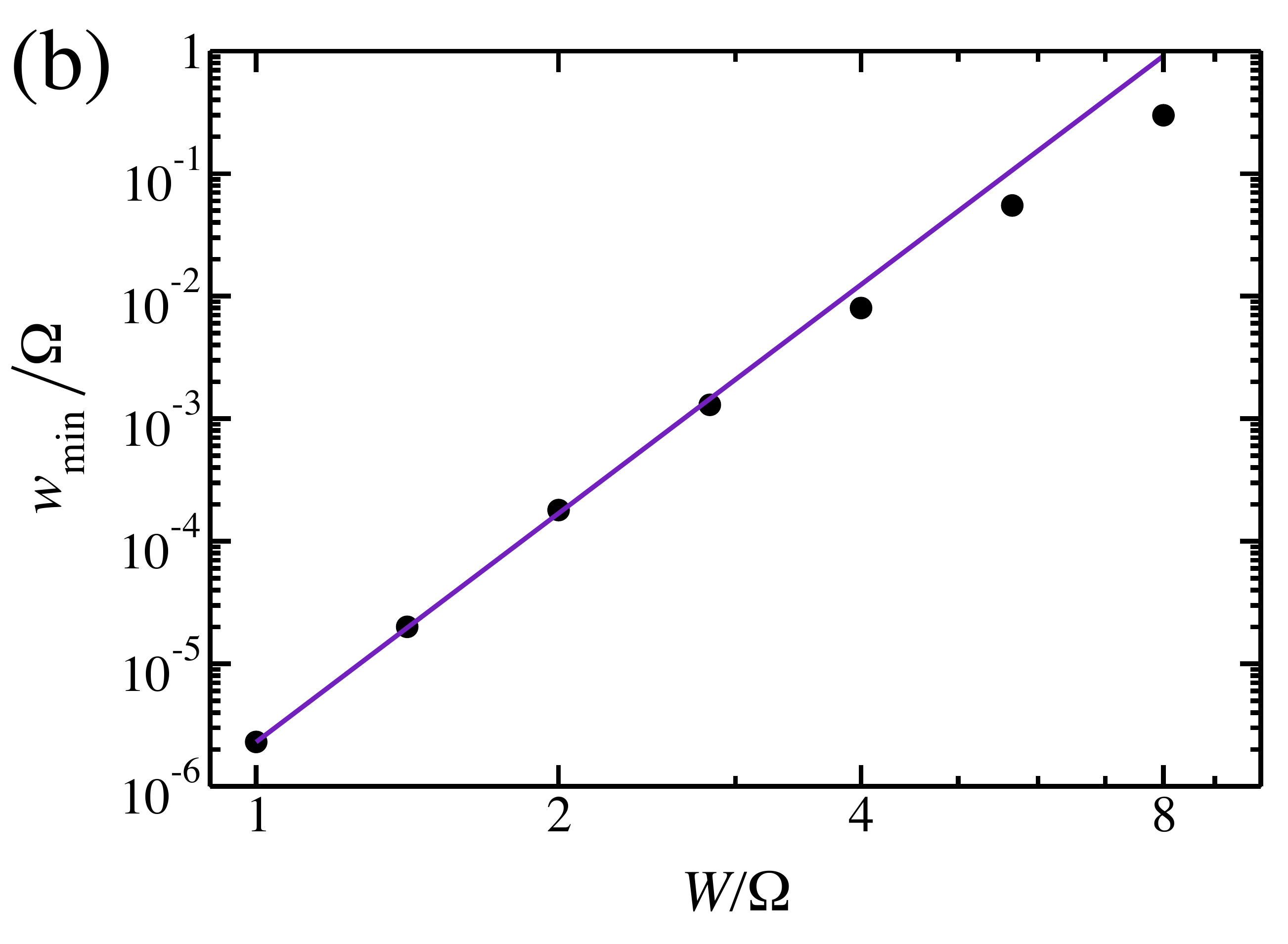}
\vspace{-0cm}
\end{center}
\caption{\label{fig:Gamma} (color online)
The dots show the values of (a)~$M_1$ and
(b)~$w_\mathrm{min}/\Omega$, calculated numerically.
The lines show the dependencies (a)~$M_1=0.6\,(\Omega/W)^{0.8}$
and (b)
$w_\mathrm{min}/\Omega=2.3\times{10}^{-6}\,(W/\Omega)^{6.2}$
which represent the best fits to the calculated values
at low~$W$.
}
\end{figure}

We have determined $M_1$ and $w_\mathrm{min}/\Omega$ numerically
for several values of the dimensionless disorder strength
$W/\Omega$ in the interval $0.5\leq{W}/\Omega\leq{8}$,
corresponding to $1.5\leq\xi\leq{400}$. The value
of $\Delta=0.1\,\Omega$ was checked to be sufficiently small,
and the chain length $L=16000$ was checked to be
sufficiently long to not affect the results (the value of
$w_\mathrm{min}$ could be reliably determined only for
$W/\Omega\geq{1}$, due to computational limitations). The
results are shown in Fig.~\ref{fig:Gamma}.
On the weak-disorder side, they can be fitted by
\begin{subequations}\begin{eqnarray}
&&{M}_1=\frac{0.6\pm{0.05}}{(W/\Omega)^{0.8\pm{0.1}}},
\label{M1num=}\\
&&\frac{w_\mathrm{min}}\Omega=(2.3\pm{0.3})\times{10}^{-6}
\left(\frac{W}\Omega\right)^{6.2\pm{0.2}}.\label{M2num=}
\end{eqnarray}\end{subequations}
The resulting dependence of $g\rho_\mathrm{min}/\Omega$
on $W/\Omega$
is shown in Fig.~\ref{fig:rhoRange} and can be fitted by
\begin{equation}
\frac{g\rho_\mathrm{min}}\Omega=(0.8\pm{0}.1)\times{10}^{-3}
\left(\frac{W}\Omega\right)^{3.5\pm{0}.2}.
\end{equation}

\begin{figure}
\begin{center}
\includegraphics[width=8cm]{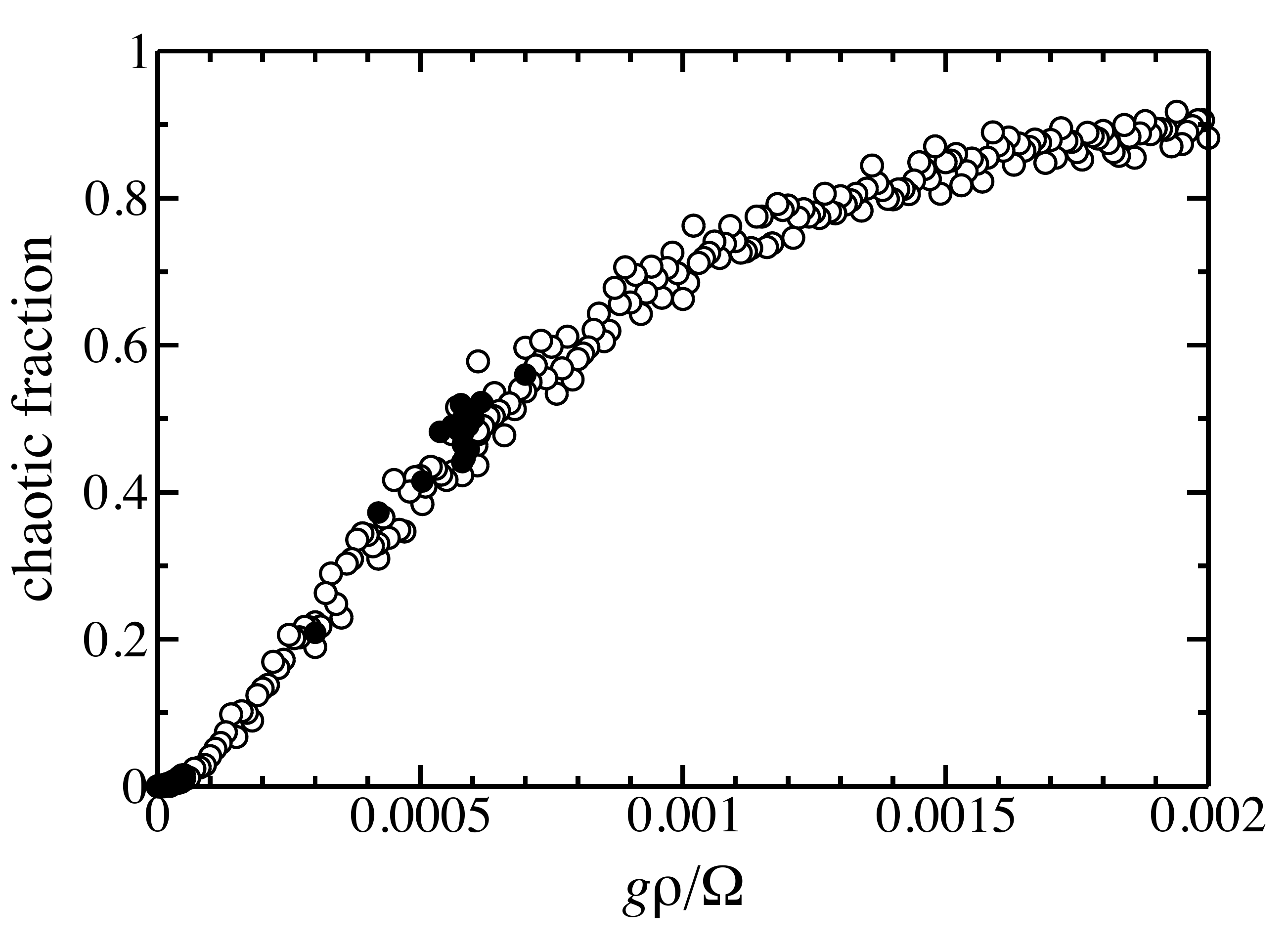}
\end{center}
\caption{\label{fig:fraction}
The fraction of chaotic modes for $W/\Omega=2$ as a function
of the dimensionless density $g\rho/\Omega$. The open and
filled circles correspond to $L=2000$ and $L=4000$, respectively.}
\end{figure}

\subsection{Relation to chaos}
\label{ssec:chaos}

Along with $\rho_\mathrm{min}$, defined above by
Eq.~(\ref{rhomin=}), we also plot the value $\rho_{1/2}$,
defined as the density at which half of the modes of the chain
are chaotic and half are not. Even though only the motion of
the whole coupled chain can be, strictly speaking, characterized
as chaotic or not (in the sense of positive Lyapunov exponent),
for given initial conditions one can focus on the dynamics of a
single mode assuming the actions and phases of all other modes
to be frozen, so that they represent an external (quasi-periodic)
force acting on the chosen mode.
For the problem of a single nonlinear oscillator under the
action of an external force, one can clearly define chaotic
and regular motion.
Then, for the whole chain with given initial conditions, one
can calculate the fraction of modes whose motion is chaotic
in the above sense (see Appendix~\ref{app:fraction} for the
details of numerical implementation). This fraction depends
on the density: it vanishes when $\rho\to{0}$ (since for a
linear system no modes are chaotic) as $\propto\rho^2$~%
\cite{Basko2011,Pikovsky2011,Basko2012}, and approaches unity
for sufficiently large density, as shown in
Fig.~\ref{fig:fraction} for $W/\Omega=2$.

Let $\rho_{1/2}$ be the value of~$\rho$ when the fraction is
$1/2$. In Fig.~\ref{fig:rhoRange}, we plot the dimensionless
quantity $g\rho_{1/2}/\Omega$ versus disorder strength. The
extracted dependence,
\begin{equation}\label{grho12=}
\frac{g\rho_{1/2}}\Omega=(4\pm{0}.5)\times{10}^{-5}
\left(\frac{W}\Omega\right)^{3.5\pm{0}.2},
\end{equation}
has the same power law as that for $\rho_\mathrm{min}$, and
differs only in the prefactor. This strongly suggests that
the two seemingly unrelated criteria (namely,
$\rho\gg\rho_\mathrm{min}$,
responsible for the self-consistency of the Fermi Golden Rule
where chaos simply did not enter the discussion at all, and
$\rho\gg\rho_{1/2}$, ensuring that most modes are chaotic)
are, in fact, closely connected. Thus, the situation
analyzed in the present work corresponds to the regime when
most of the modes are chaotic. This condition was used in
Ref.~\cite{Laptyeva2010} to define the regime of strong chaos,
where the $\rho^2$-dependence of the diffusion coefficient was
observed.

\subsection{High densities}
\label{ssec:high}

\begin{figure}
\begin{center}
\vspace{0cm}
\includegraphics[width=8cm]{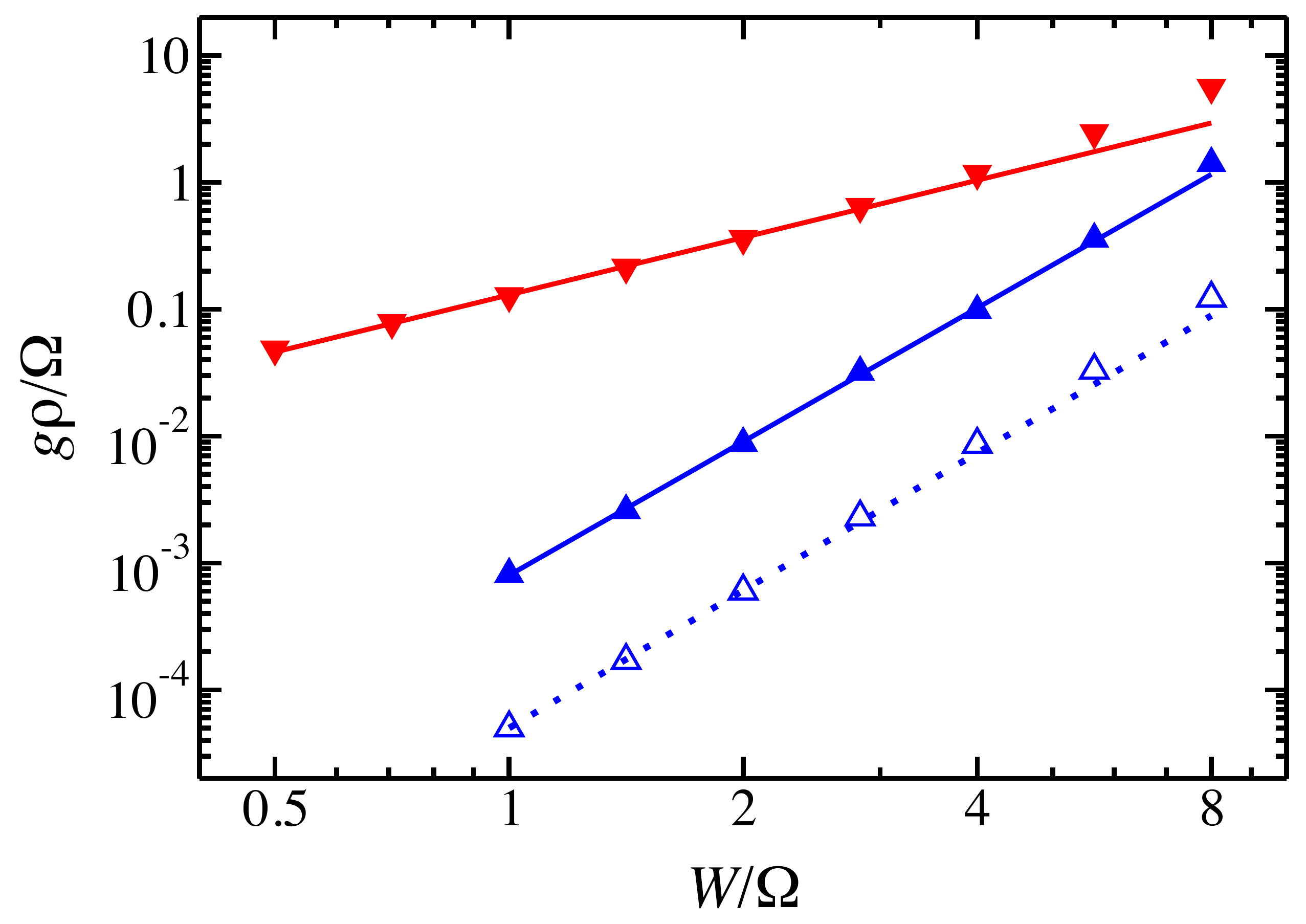}
\vspace{-0cm}
\end{center}
\caption{\label{fig:rhoRange}
The limits of validity of the master equation approach.
The filled upward triangles and the solid line passing
through them, which corresponds to
$g\rho_\mathrm{min}/\Omega=0.8\times{10}^{-3}\,(W/\Omega)^{3.5}$,
represent the low-density limitation expressed by
Eq.~(\ref{rhomin=}).
The open upward triangles and the dotted line passing
through them, which corresponds to
$g\rho_{1/2}/\Omega=4\times{10}^5\,(W/\Omega)^{3.5}$,
represent the density at which half of the modes of
the chain are chaotic.
The filled downward triangles and the solid line passing
through them, which corresponds to
$g\rho_\mathrm{max}/\Omega=0.13\,(W/\Omega)^{1.5}$,
represent the high-density limitation expressed by
Eq.~(\ref{rhomax=}).
}
\end{figure}

The condition for the validity of the master
equation~(\ref{master=}) from the high-density side is that
the localized normal modes of the linear problem should be
well defined on the relaxation time scale $1/\Gamma$, i.~e.,
this time should be longer than the inverse spacing between
mode frequencies on the same localization segment,
$1/\Delta_1=\xi/(2\pi\Omega)$. In the opposite case, the
discrete localized normal modes are not well resolved, and do
not represent a good starting basis.
Using $\Gamma=2\pi(g^2\rho^2/\Omega)M_1$, we can write the
condition $\Gamma\ll\Delta_1$ as
\begin{equation}\label{rhomax=}
g\rho\ll\sqrt{\frac{\Omega\Delta_1}{2\pi M_1}}
\equiv{g\rho_\mathrm{max}}.
\end{equation}
The dependence of $g\rho_\mathrm{max}/\Omega$ on the disorder
strength, obtained from the numerical results for $M_1$ and
$\Delta_1$, is shown in Fig.~\ref{fig:rhoRange}.
It can be fitted by the expression
\begin{equation}\label{grhomax=}
\frac{g\rho_\mathrm{max}}\Omega=(0.13\pm{0}.02)
\left(\frac{W}\Omega\right)^{1.5\pm{0}.1}.
\end{equation}

It is instructive to approach the same condition from the clean
side, treating the disorder as a perturbation. Without disorder,
the normal mode wave functions and frequencies are given by
\begin{equation}\label{cleanwf=}
\phi_{\alpha{n}}=\sqrt{\frac{2}{L+1}}\,\sin\frac{\pi\alpha{n}}{L+1},\quad
\omega_\alpha=-2\Omega\cos\frac{\pi\alpha}{L+1},
\end{equation}
and for $L\gg{1}$ it is convenient to introduce the wave vector
$k=\pi\alpha/(L+1)$ and velocity $v_k=2\Omega\sin{k}$. It is
well known that perturbative treatment of the disorder gives
the elastic backscattering rate
$\Gamma^{(\mathrm{bs})}_k=2v_k/\xi(\omega_k)$.
In other words, the backscattering mean free path
$v_k/\Gamma^{(\mathrm{bs})}_k$ is twice shorter than the
localization length $\xi(\omega_k)$ at the same frequency
$\omega_k=-2\Omega\cos{k}$~\cite{Thouless1979}. At the same
time, the wave functions (\ref{cleanwf=}) of the clean chain
can be used to evaluate the mode relaxation rate
$\Gamma_k^{(\mathrm{nl})}$ due to the nonlinearity from
Eq.~(\ref{Gamma=}), whose derivation did not assume any
specific form of the wave functions. If
$\Gamma_k^{(\mathrm{nl})}\gg\Gamma_k^{(\mathrm{bs})}$, the
elastic scattering on the disorder and Anderson localization
are not important. Noting that the mean mode spacing on the
localization length $\Delta_1=(\pi/2)\Gamma^{(\mathrm{bs})}_k$,
we conclude that the conditions $\Gamma\ll\Delta_1$ mentioned
above, and $\Gamma_k^{(\mathrm{nl})}\ll\Gamma_k^{(\mathrm{bs})}$
are equivalent, provided that the relaxation rates~(\ref{Gamma=}),
evaluated on clean and localized wave functions, match.

\begin{figure}
\begin{center}
\vspace{0cm}
\includegraphics[width=8cm]{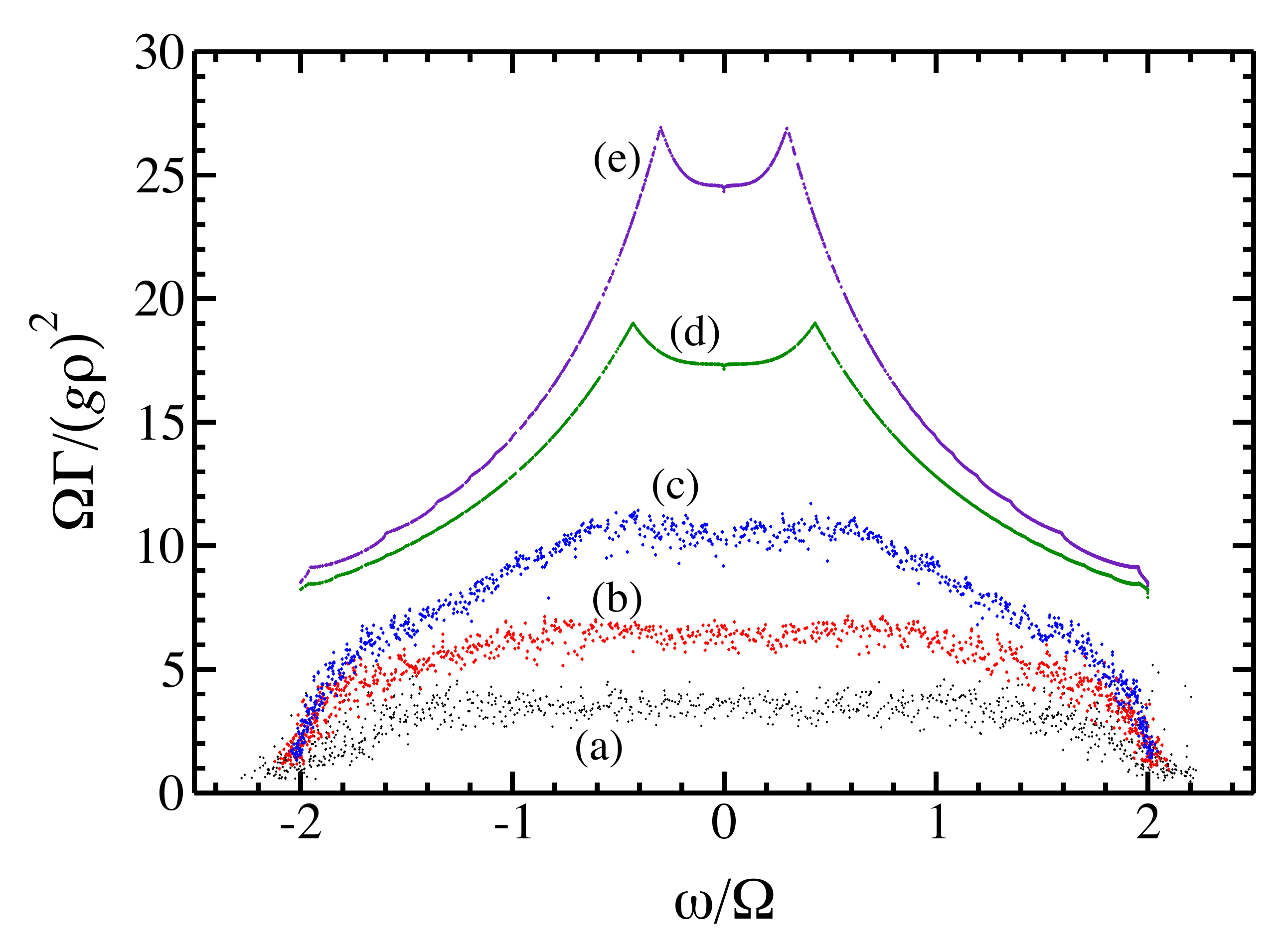}
\vspace{-0cm}
\end{center}
\caption{\label{fig:GammaClean}
The relaxation rates of individual normal modes in a chain
of $L=5000$ sites. Different sets of points correspond to 
different disorder strengths: $W/\Omega=1$~(a), 
$W/\Omega=0.5$~(b), $W/\Omega=0.25$~(c), $W=0$~(d,e). 
The spikes on the sets (d) and~(e) result from
the $k=\pi/2$ singularity being cut off by the finite width
of the $\delta$~function. This width is $w=10^{-3}$ for sets
(a), (b), (c), (e), and $w=3.16\cdot{10}^{-3}$ for
set~(d). 
}
\end{figure}

Eq.~(\ref{Gamma=}) with the wave functions (\ref{cleanwf=}) gives
\begin{equation}\label{Gammaclean=}
\Gamma_k^{(\mathrm{nl})}=\frac{(g\rho)^2}{2\pi\Omega}
\int\limits_{-\pi}^{\pi}\frac{dk'}{|\sin{k}'-\sin{k}|}
\approx\frac{2(g\rho)^2}{\pi\Omega|\cos{k}|}
\ln\frac{|\cos{k}|}{\Delta{k}},
\end{equation}
where $\Delta{k}$ is a small uncertainty in the wave vector,
needed to cut off the divergence at $k'\to{k}$. For consistency,
it should be taken of the order of the inverse mean free path,
$\Delta{k}\sim\Gamma_k^{(\mathrm{nl})}/v_k$. Then, for most of
the band one can write with logarithmic precision
\begin{equation}
\Gamma_k^{(\mathrm{nl})}\approx\frac{4(g\rho)^2}{\pi\Omega|\cos{k}|}
\ln\frac\Omega{g\rho}.
\end{equation}
However, at $k\to\pi/2$, expression~(\ref{Gammaclean=}) is
divergent.
This divergence should be smeared on the scale
$|k-\pi/2|\sim\Delta{k}$, so one can estimate (up to a numerical
coefficient)
\begin{equation}
\Gamma_k^{(\mathrm{nl})}\sim{g}\rho,\quad
|k-\pi/2|\lesssim\frac{g\rho}\Omega.
\end{equation}
These results are confirmed by the direct numerical evaluation
of Eq.~(\ref{Gamma=}) for different values of disorder including
the disorder-free case, as shown in Fig.~\ref{fig:GammaClean}.
For the disorder-free case the cutoff scale $\Delta{k}$ in the
numerical calculation is effectively provided by the
$\delta$~function width, $\Delta{k}\sim{w}/v_k$, as seen by
comparing curves (d) and~(e) which differ by the value of~$w$
used in the calculation.

Thus, upon increasing the disorder at fixed~$\rho\ll\Omega/g$,
when $1/\xi\gtrsim(g\rho/\Omega)^2$ (up to numerical and
logarithmic  factors), the localization condition
$\Gamma_k^{(\mathrm{nl})}\sim\Gamma_k^{(\mathrm{bs})}$ becomes
fulfilled for most of the normal modes, except for those in a
relatively narrow frequency range
$|\omega|\lesssim(g\rho)^2/(\Omega\xi)$; this range shrinks
completely at a stronger disorder such that
$1/\xi\sim{g}\rho/\Omega$. These two conditions correspond to
$g\rho_\mathrm{max}/\Omega\sim\xi^{-1/2}$ and 
$g\rho_\mathrm{max}/\Omega\sim\xi^{-1}$, respectively, which
should be contrasted to
$g\rho_\mathrm{max}/\Omega\sim\xi^{-0.75}$ following from
Eq.~(\ref{grhomax=}).
This strongly suggests that Eqs.~(\ref{M1num=}), (\ref{M2num=}),
(\ref{grho12=}), (\ref{grhomax=}) [as well as Eq.~(\ref{D0K0K1=})
below] do not represent the scaling at lowest disorder strengths,
but rather intermediate asymptotics. 
The results shown in Fig.~\ref{fig:GammaClean} suggest that the
behaviour of relaxation rates in the disordered system starts to
resemble that of the clean one at $W/\Omega>0.25$, corresponding
to extremely large localization lengths $\xi>1000$. The detailed
investigation of this issue requires significant computational
effort and is beyond the scope of the present work. We only note
that similarly large localization lengths were found to be
necessary to reach the weak-deisorder asymptotics in the
statistics of a single normal mode wave
function~\cite{Kravtsov2011}.

\section{Macroscopic diffusion equation}
\label{sec:macroscopic}

The main task of the present section is to pass from the
master equation~(\ref{master=}), which is still microscopic
as it includes the dynamics of each individual normal mode,
to the macroscopic description in terms of the density.
In this section, we neglect the nonlinear frequency shifts in
Eq.~(\ref{omegaHF=}). As discussed in Secs.~\ref{ssec:shifts}
and~\ref{ssec:numGamma}, they produce just small corrections,
$\sim{g}\rho/\Omega$, to the main result.

\subsection{Classical Boltzmann equation}
\label{ssec:Boltzmann}

Multiplying Eq.~(\ref{master=}) by $I_\alpha$ and integrating
over all actions, we obtain an equation for the average:
\begin{subequations}\begin{eqnarray}
&&\frac{\partial\langle{I}_\alpha\rangle_\mathcal{F}}{\partial{t}}
=\sump_{\beta,\gamma,\delta}
R_{\alpha\beta\gamma\delta}{}\times
\nonumber\\ &&\qquad\qquad{}\times{}\left\langle
{I}_\beta{I}_\gamma{I}_\delta+{I}_\alpha{I}_\gamma{I}_\delta
-{I}_\alpha{I}_\beta{I}_\delta-{I}_\alpha{I}_\beta{I}_\gamma
\right\rangle_\mathcal{F},\nonumber\\ && \\
&&R_{\alpha\beta\gamma\delta}=
4\pi{V}_{\alpha\beta\gamma\delta}^2\,\delta\!
\left(\omega_\alpha+\omega_\beta-\omega_\gamma-\omega_\delta\right).
\label{Rabcd=}
\end{eqnarray}\end{subequations}
Next, we note that the action of each mode is changed
due to random resonant interactions with many other
modes, so any two modes effectively see mostly different
resonances. Thus, one can neglect the correlations between
different actions and decouple
$\langle{I}_\beta{I}_\gamma{I}_\delta\rangle_\mathcal{F}\to
\langle{I}_\beta\rangle_\mathcal{F}
\langle{I}_\gamma\rangle_\mathcal{F}
\langle{I}_\delta\rangle_\mathcal{F}$, \emph{etc}.
This can be done when the number of terms contributing to
the sum over modes is large (see Appendix~\ref{app:moments}).
The latter condition was discussed in detail in
Sec.~\ref{ssec:low}.

As a result, we obtain a closed kinetic equation for the averages
$\langle{I}_\beta\rangle_\mathcal{F}\equiv\bar{I}_\alpha$
(the new notation is introduced for compactness):
\begin{equation}\label{kinetic=}
\frac{d\bar{I}_\alpha}{dt}=
\sump_{\beta,\gamma,\delta}R_{\alpha\beta\gamma\delta}
\left[\left(\bar{I}_\alpha+\bar{I}_\beta\right)
\bar{I}_\gamma\bar{I}_\delta
-\bar{I}_\alpha\bar{I}_\beta
\left(\bar{I}_\gamma+\bar{I}_\delta\right)\right].
\end{equation}
This kinetic equation conserves the total action
$\sum_\alpha\bar{I}_\alpha$ and the total energy
$\sum_\alpha\omega_\alpha\bar{I}_\alpha$ (due to
the frequency $\delta$~function in
$R_{\alpha\beta\gamma\delta}$).
The equilibrium state, which nullifies the collision
integral in the right-hand side, is given by the
Rayleigh-Jeans distribution,
\begin{equation}\label{barIeq=}
\bar{I}_\alpha^\mathrm{eq}=\frac{T}{\omega_\alpha-\mu},
\end{equation}
or simply $\bar{I}_\alpha^\mathrm{eq}=1/\lambda$ in
the limit~(\ref{infiniteT=}). Eq.~(\ref{barIeq=})
correctly reproduces the thermodynamics of the
chain with the required precision, as discussed
in Appendix~\ref{app:thermodynamics}.

It is worth noting that
Eq.~(\ref{barIeq=}) represents the classical limit
$\hbar\to{0}$ of the Bose-Einstein distribution
$N_\alpha=1/[e^{-\hbar(\omega_\alpha-\mu)/T}-1]$, 
and Eq.~(\ref{kinetic=}) has the same form as the
quantum Boltzmann equation for bosonic occupation
numbers $N_\alpha=\bar{I}_\alpha/\hbar$ changing due
to pair collisions in the limit $N_\alpha\gg{1}$.
Indeed, in this limit, the combination
$(N_\alpha+1)(N_\beta+1)N_\gamma{N}_\delta
-N_\alpha{N}_\beta(N_\gamma+1)(N_\delta+1)$ reduces
to the one in Eq.~(\ref{kinetic=}), while
$R_{\alpha\beta\gamma\delta}$ is the rate of a pair
collision, as obtained from the Fermi Golden Rule
for quantum Hamiltonian~(\ref{Hpsi=}) with
$\psi_n,\psi_n^*$ treated as bosonic field operators.
Such quantum Boltzmann equation can be applied to
describe the dynamics of disordered bosons on the
metallic side of the many-body localization
transition~\cite{AAS2010} in the regime, analogously
to the power-law hopping regime considered earlier
for fermions~\cite{GMR,Basko2003,BAA2006}.

\subsection{Macroscopic current}
\label{ssec:current}

Generally, the diffusion equation (\ref{NLdif=}) is
obtained from the continuity equation,
\begin{subequations}
\begin{equation}\label{continuity=}
\frac{\partial\rho}{\partial{t}}=
-\frac{\partial{J}}{\partial{x}},
\end{equation}
supplemented by the Fick's law,
\begin{equation}\label{Fick=}
J=-D(\rho)\,\frac{\partial\rho}{\partial{x}}.
\end{equation}
\end{subequations}
Eq.~(\ref{continuity=}) expresses the conservation of
the total norm $\mathcal{N}=\int\rho\,dx$, and
Eq.~(\ref{Fick=}) represents the first term of the
gradient expansion of the current in local equilibrium,
characterized by a spatially dependent density
(in the global equilibrium, where the density is constant
along the chain, the current must vanish).

The continuous functions
$\rho(x,t)$ and $J(x,t)$ entering Eqs.~(\ref{continuity=}),
(\ref{Fick=}) are the macroscopic density and current,
whose dependence on the coordinate~$x$ and time~$t$ is
smooth enough compared to some microscopic scales. As we
are studying the dynamics of energy and action exchange
between different normal modes, the corresponding length
scale is the mode localization length~$\xi$. 
We are interested in the case when local equilibrium is
reached, so the corresponding time scale is the inverse
of the relaxation rate~$\Gamma$, introduced in
Sec.~\ref{ssec:relaxation} (assumed to be longer that the
frequency spacing between the normal modes, $1/\Delta_1$,
as discussed in Sec.~\ref{ssec:high}).
Formally, we can define
\begin{equation}\begin{split}
&\rho(x,t)={}{}\int{d}t'\,\mathcal{T}(t-t')
\sum_n\mathcal{S}(x-n)
\left\langle|\psi_n(t')|^2\right\rangle_\mathcal{F}=\\
&\qquad{}={}{}\int{d}t'\,\mathcal{T}(t-t')
\sum_\alpha\mathcal{S}(x-X_\alpha)\,
\bar{I}_\alpha(t')+O(\xi^2/\ell^2),
\label{rhoxt=}
\end{split}\end{equation}
where the ``center of mass'' of the mode~$\alpha$ is
defined as
\begin{equation}
X_\alpha=\sum_nn\phi_{\alpha{n}}^2,
\end{equation}
and the spatial and temporal smoothing functions
$\mathcal{S}(x)$ and
$\mathcal{T}(t)$ can be taken, e.~g., Gaussian:
\begin{equation}
\mathcal{S}(x)=\frac{e^{-x^2/(2\ell^2)}}{\sqrt{2\pi\ell^2}},
\quad
\mathcal{T}(t)=\frac{e^{-t^2/(2\tau^2)}}{\sqrt{2\pi\tau^2}}.
\end{equation}
Here the smoothing length and time scales $\ell\gg\xi$
and $\tau\gg{1}/\Gamma$, as discussed above;
the diffusion equation is then valid at length scales
$x\gtrsim\ell$ and $t\gtrsim\tau$.
The relation between the first and the second expression in
Eq.~(\ref{rhoxt=}) is discussed in detail in
Appendix~\ref{app:smooth}.


The macroscopic current is defined in order to identically
satisfy the continuity equation,
\begin{equation}\begin{split}
J(x,t)={}&{}\int{d}t'\,\mathcal{T}(t-t')
\sum_\alpha\tilde{\mathcal{S}}(x-X_\alpha)\,
\frac{d\bar{I}_\alpha(t')}{dt'}+{}\\&{}+O(\xi^2/\ell^2),
\end{split}\end{equation}
where
\begin{equation}
\tilde{\mathcal{S}}(x)\equiv-\int\limits_0^x
\mathcal{S}(x')\,dx'.
\end{equation}
Substituting $d\bar{I}_\alpha/dt$ from Eq.~(\ref{kinetic=})
and symmetrizing with respect to $\alpha\leftrightarrow\beta$,
$\alpha\beta\leftrightarrow\gamma\delta$, one obtains
\begin{equation}\begin{split}\label{Jxt=}
J(x,t)={}&{}\int{d}t'\,\mathcal{T}(t-t')\,
\sump_{\alpha,\beta,\gamma,\delta}
\frac{R_{\alpha\beta\gamma\delta}}4\times{}\\
&{}\times\left[\left(\bar{I}_\alpha+\bar{I}_\beta\right)
\bar{I}_\gamma\bar{I}_\delta-\bar{I}_\alpha\bar{I}_\beta
\left(\bar{I}_\gamma+\bar{I}_\delta\right)\right]\times{}\\
&{}\times\left[\tilde{\mathcal{S}}(x-X_\alpha)
+\tilde{\mathcal{S}}(x-X_\beta)\right.-\\
&\quad{}-\left.\tilde{\mathcal{S}}(x-X_\gamma)
-\tilde{\mathcal{S}}(x-X_\delta)\right].
\end{split}\end{equation}
For each term in the sum, it is convenient to introduce
the short-hand notations for the ``center-of-mass''
coordinate and the ``displacement'':
\begin{subequations}\begin{eqnarray}
&&X_{\alpha\beta\gamma\delta}\equiv
\frac{X_\alpha+X_\beta+X_\gamma+X_\delta}4,\\
&&{d}_{\alpha\beta\gamma\delta}
=X_\alpha+X_\beta-X_\gamma-X_\delta.\label{tildex=}
\end{eqnarray}\end{subequations}
As a final step, we expand each
$\tilde{\mathcal{S}}(x-X_\alpha)$ to the first order around
$x-X_{\alpha\beta\gamma\delta}$, which gives
\begin{equation}\begin{split}
J(x,t)={}&{}\int{d}t'\,\mathcal{T}(t-t')\,
\sump_{\alpha,\beta,\gamma,\delta}
\mathcal{S}(x-X_{\alpha\beta\gamma\delta})\,
\frac{R_{\alpha\beta\gamma\delta}}4\times{}\\
&{}\times{d}_{\alpha\beta\gamma\delta}
\left[\left(\bar{I}_\alpha+\bar{I}_\beta\right)
\bar{I}_\gamma\bar{I}_\delta-\bar{I}_\alpha\bar{I}_\beta
\left(\bar{I}_\gamma+\bar{I}_\delta\right)\right].
\end{split}\end{equation}
In this expression, the time argument~$t'$ is implied
for all actions;
it has been omitted for the sake of compactness.

To illustrate how the formalism developed above works
for a very simple toy model, in Appendix~\ref{app:circuit}
it is used to calculate the conductivity of a disordered
electric $RC$-circuit.

\subsection{Diffusion coefficient}
\label{ssec:diffcoeff}

The diffusion coefficient $D(\rho)$ should be found by
calculating the linear response of the current~$J$ to
an infinitesimal gradient of the density,
$\partial\rho/\partial{x}=-\kappa$. For this, one should
look for a stationary solution of Eq.~(\ref{kinetic=})
in the form
\begin{equation}\label{densityprofile=}
\bar{I}_\alpha=\rho-\kappa{X}_\alpha+r_\alpha,
\end{equation}
to the linear order order in $\kappa$, where $r_\alpha$ is
such that
\begin{equation}\label{sumralpha=}
\sum_\alpha \mathcal{S}(x-X_\alpha)\,r_\alpha
=O(\kappa\xi^2/\ell^2)
\end{equation}
and does not grow with~$x$.
Indeed,
\begin{subequations}\begin{eqnarray}
&&\sum_\alpha\frac{d\mathcal{S}(x-X_\alpha)}{dx}=O(\xi^2/\ell^2),
\label{dSdx1=}\\
&&\sum_\alpha\frac{d\mathcal{S}(x-X_\alpha)}{dx}\,X_\alpha=
1+O(\xi^2/\ell^2),\label{dSdx2=}
\end{eqnarray}\end{subequations}
as discussed in Appendix~\ref{app:smooth}.


With the substitution (\ref{densityprofile=}), the linearized 
stationary Eq.~(\ref{kinetic=}) becomes a system of linear
equations for $r_\alpha$, which can be written as
\begin{equation}
\sump_{\beta,\gamma,\delta}R_{\alpha\beta\gamma\delta}
\left(r_\alpha+r_\beta-r_\gamma-r_\delta\right)=
\kappa\sump_{\beta,\gamma,\delta}
R_{\alpha\beta\gamma\delta}{d}_{\alpha\beta\gamma\delta}.
\end{equation}
The typical value of ${d}_{\alpha\beta\gamma\delta}$ is 
${d}_{\alpha\beta\gamma\delta}\sim\xi$, since otherwise
the overlap $V_{\alpha\beta\gamma\delta}$ is exponentially
suppressed.
Moreover, because of the large number of terms contributing
to the sum on the right-hand side, the effective self-averaging
occurs, so the typical value of sum should be close to its
statistical average over the disorder realizations. But the
latter is zero because on the average, the chain is symmetric
with respect to translations and spatial inversion; in other
words, there are as many terms with
${d}_{\alpha\beta\gamma\delta}>0$, as with
${d}_{\alpha\beta\gamma\delta}<0$. As a result, the
typical value of $r_\alpha+r_\beta-r_\gamma-r_\delta$ is
much smaller than $\kappa{d}_{\alpha\beta\gamma\delta}$.

For the solution given by Eq.~(\ref{densityprofile=}), the
current becomes
\begin{equation}\begin{split}
&J(x)=\rho^2\sump_{\alpha,\beta,\gamma,\delta}
\mathcal{S}(x-X_{\alpha\beta\gamma\delta})\,
\frac{R_{\alpha\beta\gamma\delta}}4\times{}\\
&\qquad{}\times{d}_{\alpha\beta\gamma\delta}
\left[\kappa{d}_{\alpha\beta\gamma\delta}
-(r_\alpha+r_\beta-r_\gamma-r_\delta)\right].
\end{split}\end{equation}
As discussed above, $r_\alpha+r_\beta-r_\gamma-r_\delta$
can be neglected with respect to
$\kappa{d}_{\alpha\beta\gamma\delta}$. Finally, as
$\mathcal{S}(x)$ is a slowly varying function with unit
integral, the convolution is equivalent to spatial averaging,
so the current is $J=\kappa{D}_0\rho^2$, with $D_0$ given by
\begin{equation}\label{D0=}
D_0=\frac\pi{L}\sump_{\alpha,\beta,\gamma,\delta}
d_{\alpha\beta\gamma\delta}^2
V_{\alpha\beta\gamma\delta}^2\,\delta\!
\left(\omega_\alpha+\omega_\beta-\omega_\gamma-\omega_\delta\right).
\end{equation}
In combination with the continuity equation,
Eq.~(\ref{continuity=}), this immediately gives Eq.~(\ref{NLdif=}).

\section{Energy transport}\label{sec:energy}

\subsection{General remarks}

In Sec.~\ref{sec:macroscopic}, the macroscopic nonlinear
diffusion equation for the norm density was derived.
However, the original nonlinear Schr\"odinger equation~%
(\ref{DNLS=}),
as well as the master equation~(\ref{master=}) and the
Boltzmann equation~(\ref{kinetic=}) have two conserved
quantities: norm (action) and energy. The transport of
energy was ignored in Sec.~\ref{sec:macroscopic} for
simplicity.
The purpose of the present section is to derive and study
the full system of the macroscopic transport equations
for the two conserved quantities.

Very generally, the macroscopic norm and energy currents,
$J$~and~$Q$, vanish in the state of global thermal
equilibrium, characterized by the values of the
temperature~$T$ and the chemical potential~$\mu$, which
are constant along the chain. 
Under the assumption of \emph{local} equilibrium, when
$\mu$~and~$T$ are slowly changing with the position, the
currents can be evaluated by performing the expansion in
the spatial gradients of $\mu$~and~$T$. Restricting the
expansion to the first term, one obtains
\begin{equation}
\left(\begin{array}{c} J \\ Q \end{array}\right)=
\hat{\mathcal{L}}(\mu,T)\,
\frac\partial{\partial{x}}
\left(\begin{array}{c} -\mu/T \\ 1/T \end{array}\right)
\end{equation}
where $\hat{\mathcal{L}}$ is a $2\times{2}$ matrix which
depends on the local values of $\mu$ and $T$. It is
symmetric by virtue of the Onsager relations.

Upon substitution of these currents to the corresponding
continuity equations, the equations for the macroscopic
norm and energy densities, $\rho(x,t)$ and $\vep(x,t)$,
are obtained:
\begin{equation}\label{dt=Ldx}
\frac\partial{\partial{t}}
\left(\begin{array}{c} \rho \\ \vep \end{array}\right)=
\frac\partial{\partial{x}}\,
\hat{\mathcal{L}}(\mu,T)\,
\frac\partial{\partial{x}}
\left(\begin{array}{c} -\mu/T \\ 1/T \end{array}\right).
\end{equation}
To close the equations, the local equilibrium relations
between $\rho,\vep$ and $\mu,T$ should be supplied. For
weak disorder they are approximately the same as for the
clean case, the latter analyzed in
Ref.~\cite{Rasmussen2000}. The explicit expressions in
the high-temperature limit are given in
Appendix~\ref{app:thermodynamics}.

\subsection{Neglecting the nonlinear frequency shifts}
\label{ssec:noshifts}

Let us first neglect the nonlinear frequency shifts and
study Eq.~(\ref{kinetic=}) which conserves the total
energy
\begin{equation}
\mathcal{E}=\sum_\alpha\omega_\alpha\bar{I}_\alpha
=\sum_n\langle{h}_n\rangle_\mathcal{F},
\end{equation}
where the on-site energy is defined as
\begin{equation}\begin{split}
h_n={}&{}\ep_n|\psi_n|^2-\frac\Omega{2}
\left(\psi^*_n\psi_{n-1}+\psi^*_{n-1}\psi_n\right)-{}\\
&{}-\frac\Omega{2}
\left(\psi^*_n\psi_{n+1}+\psi^*_{n+1}\psi_n\right).
\end{split}\end{equation}
The expression for the macroscopic energy density is obtained
analogously to Eq.~(\ref{rhoxt=}):
\begin{equation}\begin{split}\label{vepxt=}
\vep(x,t)={}&{}\int{d}t'\,\mathcal{T}(t-t')
\sum_n\mathcal{S}(x-n)
\left\langle{h}_n(t')\right\rangle_\mathcal{F}=\\
={}&{}\int{d}t'\,\mathcal{T}(t-t')
\sum_\alpha\mathcal{S}(x-X_\alpha)\,
\omega_\alpha\bar{I}_\alpha(t')+{}\\&{}+O(\xi^2/\ell^2),
\end{split}\end{equation}
where we have used the fact that
\begin{equation}
\sum_nn\phi_{\alpha{n}}\left[\ep_n\phi_{\alpha{n}}-\Omega
\left(\phi_{\alpha,n-1}+\phi_{\alpha,n+1}\right)\right]
=\omega_\alpha{X}_\alpha.
\end{equation}
The energy current is defined analogously to Eq.~(\ref{Jxt=}):
\begin{equation}\begin{split}\label{Qxt=}
Q(x,t)={}&{}\int{d}t'\,\mathcal{T}(t-t')\,
\sump_{\alpha,\beta,\gamma,\delta}
\mathcal{S}(x-X_{\alpha\beta\gamma\delta})\,
\frac{R_{\alpha\beta\gamma\delta}}4\times{}\\
&{}\times{v}_{\alpha\beta\gamma\delta}
\left[\left(\bar{I}_\alpha+\bar{I}_\beta\right)
\bar{I}_\gamma\bar{I}_\delta-\bar{I}_\alpha\bar{I}_\beta
\left(\bar{I}_\gamma+\bar{I}_\delta\right)\right],
\end{split}\end{equation}
where the time argument $t'$ of the actions has been
omitted for compactness, and we have denoted
\begin{equation}\label{tildev=}
{v}_{\alpha\beta\gamma\delta}=
\omega_\alpha{X}_\alpha+\omega_\beta{X}_\beta
-\omega_\gamma{X}_\gamma-\omega_\delta{X}_\delta.
\end{equation}
The matrix elements of $\hat{\mathcal{L}}$ are obtained
by calculating the response of the two currents to the
gradients $\chi_1=\partial(-\mu/T)/\partial{x}$ and
$\chi_2=\partial(1/T)/\partial{x}$. If one seeks the solution
of the stationary Boltzmann equation in the form
\begin{equation}
\bar{I}_\alpha=\frac{T}{\omega_\alpha-\mu}
-\left(\frac{T}{\omega_\alpha-\mu}\right)^2
\left(\chi_1X_\alpha+\chi_2\omega_\alpha{X}_\alpha+r_\alpha\right),
\end{equation}
where $r_\alpha$ again satisfies condition~(\ref{sumralpha=})
and does not grow with $X_\alpha$, the linearized equation
becomes
\begin{equation}\begin{split}
&\sump_{\beta,\gamma,\delta}
\frac{T^4R_{\alpha\beta\gamma\delta}(r_\alpha+r_\beta-r_\gamma-r_\delta)}%
{(\omega_\alpha-\mu)(\omega_\beta-\mu)
(\omega_\gamma-\mu)(\omega_\delta-\mu)}=\\
&{}=-\sump_{\beta,\gamma,\delta}\frac{T^4R_{\alpha\beta\gamma\delta}
(\chi_1{d}_{\alpha\beta\gamma\delta}
+\chi_2{v}_{\alpha\beta\gamma\delta})}%
{(\omega_\alpha-\mu)(\omega_\beta-\mu)
(\omega_\gamma-\mu)(\omega_\delta-\mu)},\label{rh1h2=}
\end{split}\end{equation}
where ${d}_{\alpha\beta\gamma\delta}$ is defined in
Eq.~(\ref{tildex=}).

In full analogy with Sec.~\ref{ssec:diffcoeff}, we neglect
the contribution of $r_\alpha$ to the currents. Also, we take
the limit $T\to\infty$, $\mu/T=-1/\rho$. This limit for
$\hat{\mathcal{L}}$ is regular, and is given by
\begin{equation}\label{Lnoshifts=}
\hat{\mathcal{L}}=\frac{\rho^4}{L}
\sump_{\alpha,\beta,\gamma,\delta}\frac{R_{\alpha\beta\gamma\delta}}4
\left(\begin{array}{cc} {d}_{\alpha\beta\gamma\delta}^2
& {d}_{\alpha\beta\gamma\delta}
{v}_{\alpha\beta\gamma\delta} \\
{v}_{\alpha\beta\gamma\delta}
{d}_{\alpha\beta\gamma\delta} &
{v}_{\alpha\beta\gamma\delta}^2 \end{array}\right).
\end{equation}
Corrections to the $T\to\infty$ limit can be obtained by
expanding $T/(\omega_\alpha-\mu)$ in $\omega_\alpha/\mu$,
and the first correction has a relative smallness
$\sim\Omega\rho/T$.

The formal construction, presented above, has a caveat.
Suppose we choose a different origin for counting the
chain sites, that is, $n\to{n}-n_0$. Then,
$X_\alpha\to{X}_\alpha-n_0$ and
${v}_{\alpha\beta\gamma\delta}\to{v}_{\alpha\beta\gamma\delta}
-n_0(\omega_\alpha+\omega_\beta-\omega_\gamma-\omega_\delta)$.
The unphysical dependence on the choice of the origin
disappears only when 
$\varpi_{\alpha\beta\gamma\delta}=
\omega_\alpha+\omega_\beta-\omega_\gamma+\omega_\delta$
vanishes exactly. If the frequency $\delta$~function entering
$R_{\alpha\beta\gamma\delta}$ has a small but finite width,
${v}_{\alpha\beta\gamma\delta}$ has a component which
depends on the origin, and makes the limit $L\to\infty$ in
Eq.~(\ref{Lnoshifts=}) ill-defined.

Formally, the problem arises because the original dynamical
system, Eq.~(\ref{DNLS=}), does not conserve the unperturbed
energy $\sum_\alpha\omega_\alpha{I}_\alpha$, but the total
one, Eq.~(\ref{Hcccc=}). However, on physical grounds, we
expect the energy contained in the perturbation terms to be
less important than that in the unperturbed part, as long as
the nonlinearity is small, $g\rho\ll\Omega$, and then the
macroscopic description which neglects the difference between
the total and the unperturbed energy, should still be
meaningful.
To obtain such description, one has to redefine
${v}_{\alpha\beta\gamma\delta}$, introducing a term which
would vanish when $\varpi_{\alpha\beta\gamma\delta}=0$,
but would eliminate the unphysical dependence on the choice
of the origin at small but finite
$\varpi_{\alpha\beta\gamma\delta}$. We choose
\begin{equation}\label{tildevX=}
{v}_{\alpha\beta\gamma\delta}=
\omega_\alpha{X}_\alpha+\omega_\beta{X}_\beta
-\omega_\gamma{X}_\gamma-\omega_\delta{X}_\delta
-\varpi_{\alpha\beta\gamma\delta}X_{\alpha\beta\gamma\delta}.
\end{equation}
As  $X_{\alpha\beta\gamma\delta}\to
{X}_{\alpha\beta\gamma\delta}-n_0$ upon $n\to{n}-n_0$,
Eq.~(\ref{tildevX=}) remains invariant. To estimate the
the error introduced by the last term, we note that since
$\hat{\mathcal{L}}$ does not depend on the choice of the
origin, one can focus on the region near $n\sim\xi$.
Then, the modes with $|X_\alpha|\sim\xi$ are important,
so the magnitude of the dropped term is $\sim{w}\xi$
(determined by the $\delta$-function width), while
the magnitude of the remaining terms is $\sim\Omega\xi$.
 
Thus, for the diagonal matrix elements of $\hat{\mathcal{L}}$
we have
\begin{equation}
\mathcal{L}_{11}=D_0\rho^4,\quad\mathcal{L}_{22}=K_0\rho^4,
\end{equation}
with $D_0$ given by Eq.~(\ref{D0=}), and~$K_0$ by the same
expression but with the substitution
${d}_{\alpha\beta\gamma\delta}\to
{v}_{\alpha\beta\gamma\delta}$,
\begin{equation}\label{K0=}
K_0=\frac\pi{L}\sump_{\alpha,\beta,\gamma,\delta}
{v}_{\alpha\beta\gamma\delta}^2
V_{\alpha\beta\gamma\delta}^2\,\delta\!
\left(\omega_\alpha+\omega_\beta-\omega_\gamma-\omega_\delta\right),
\end{equation}
where ${v}_{\alpha\beta\gamma\delta}$ is given by
Eq.~(\ref{tildevX=}).

The off-diagonal elements
$\mathcal{L}_{12}=\mathcal{L}_{21}$ involve the product
${d}_{\alpha\beta\gamma\delta}{v}_{\alpha\beta\gamma\delta}$,
and average to zero.
This happens because the statistics is symmetric with respect
to $\omega_\alpha\to-\omega_\alpha$, and the off-diagonal
matrix elements in Eq.~(\ref{Lnoshifts=}) are odd functions of
the frequencies. To obtain a non-zero value, one has to include
the terms subleading in $1/T$, which gives
\begin{eqnarray}
&&\mathcal{L}_{12}=\mathcal{L}_{21}=-K_1\,\frac{\rho^5}T,\\
\label{K1=}
&&K_1=\frac\pi{L}\sump_{\alpha,\beta,\gamma,\delta}
(\omega_\alpha+\omega_\beta+\omega_\gamma+\omega_\delta)\,
v_{\alpha\beta\gamma\delta}d_{\alpha\beta\gamma\delta}\times{}\nonumber\\
&&\qquad\qquad{}\times{V}_{\alpha\beta\gamma\delta}^2\,\delta\!
\left(\omega_\alpha+\omega_\beta-\omega_\gamma-\omega_\delta\right).
\end{eqnarray}
In addition to this, a contribution which remains finite
at $T\to\infty$ is obtained if nonlinear frequency shifts
are included.

\subsection{Including the nonlinear frequency shifts}

If the nonlinear frequency shifts, Eq.~(\ref{omegaHF=}), are
included in the frequency $\delta$~function in
$R_{\alpha\beta\gamma\delta}$, Eq.~(\ref{Rabcd=}), then the
resulting Boltzmann equation, Eq.~(\ref{kinetic=}), conserves
the energy
\begin{equation}
\mathcal{E}=\sum_\alpha\omega_\alpha\bar{I}_\alpha
+\sum_{\alpha,\beta}V_{\alpha\beta\beta\alpha}
\bar{I}_\alpha\bar{I}_\beta.
\end{equation}
The equilibrium actions are given by Eq.~(\ref{barIeq=}),
however, with shifted frequencies:
\begin{equation}\label{IeqNLself=}
\bar{I}_\alpha^\mathrm{eq}=\frac{T}{\omega_\alpha
+2\sum_\beta{V}_{\alpha\beta\beta\alpha}
\bar{I}^\mathrm{eq}_\beta-\mu}.
\end{equation}
Unless $T\to\infty$, this is no longer an explicit expression,
but a self-consistent equation. To the first order in $1/T$,
\begin{equation}\label{IeqNLexp=}
\bar{I}_\alpha^\mathrm{eq}=-\frac{T}\mu
-\frac{1}T\left(\frac{T}\mu\right)^2
\left(\omega_\alpha-2g\,\frac{T}\mu\right)+O(1/T^2).
\end{equation}
The energy density, in addition to the expression in
Eq.~(\ref{vepxt=}), should include the nonlinear
contribution:
\begin{subequations}\begin{eqnarray}
&&\vep_\mathrm{nl}(x,t)=\int{d}t'\,\mathcal{T}(t-t')
\sum_{\alpha\beta}\mathcal{S}(x-X_{\alpha\beta})\times\nonumber\\
&&\qquad\qquad{}\times V_{\alpha\beta\beta\alpha}
\bar{I}_\alpha(t')\,\bar{I}_\beta(t'),\\
&&X_{\alpha\beta}\equiv
\left(\sum_n\phi_{\alpha{n}}^2\phi_{\beta{n}}^2\right)^{-1}
\sum_nn\phi_{\alpha{n}}^2\phi_{\beta{n}}^2.
\end{eqnarray}\end{subequations}
The current is given by the same Eq.~(\ref{Qxt=}), but
instead of ${v}_{\alpha\beta\gamma\delta}$ from
Eq.~(\ref{tildevX=}), one should use the nonlinear version,
\begin{equation}\begin{split}
\tilde{v}_{\alpha\beta\gamma\delta}=
{}&{}\omega_\alpha(X_\alpha-X_{\alpha\beta\gamma\delta})+{}\\
&{}+2\sum_\eta(X_{\alpha\eta}-X_{\alpha\beta\gamma\delta})
V_{\alpha\eta\eta\alpha}\bar{I}_\eta+{}\\
&{}+(\alpha\leftrightarrow\beta)
-(\alpha\beta\leftrightarrow\gamma\delta).
\end{split}\end{equation}
Now, to find $\mathcal{L}_{21}$, we seek for the solution in
the form $\bar{I}_\alpha=\rho-\rho^2\chi_1X_\alpha$, and calculate
the energy current. Noting that 
\begin{equation}
\sum_\eta{X}_{\alpha\eta}V_{\alpha\eta\eta\alpha}=gX_\alpha,
\end{equation}
we arrive at the final result,
\begin{equation}\label{Lall=}
\hat{\mathcal{L}}=\rho^4\left(\begin{array}{cc}
D_0 & -\rho{K}_1/T+2g\rho{D}_0 
\\ -\rho{K}_1/T+2g\rho{D}_0 & K_0 \end{array}\right).
\end{equation}

\subsection{Analysis of the coupled transport equations}
\label{ssec:coupled}

\begin{figure}
\includegraphics[width=8cm]{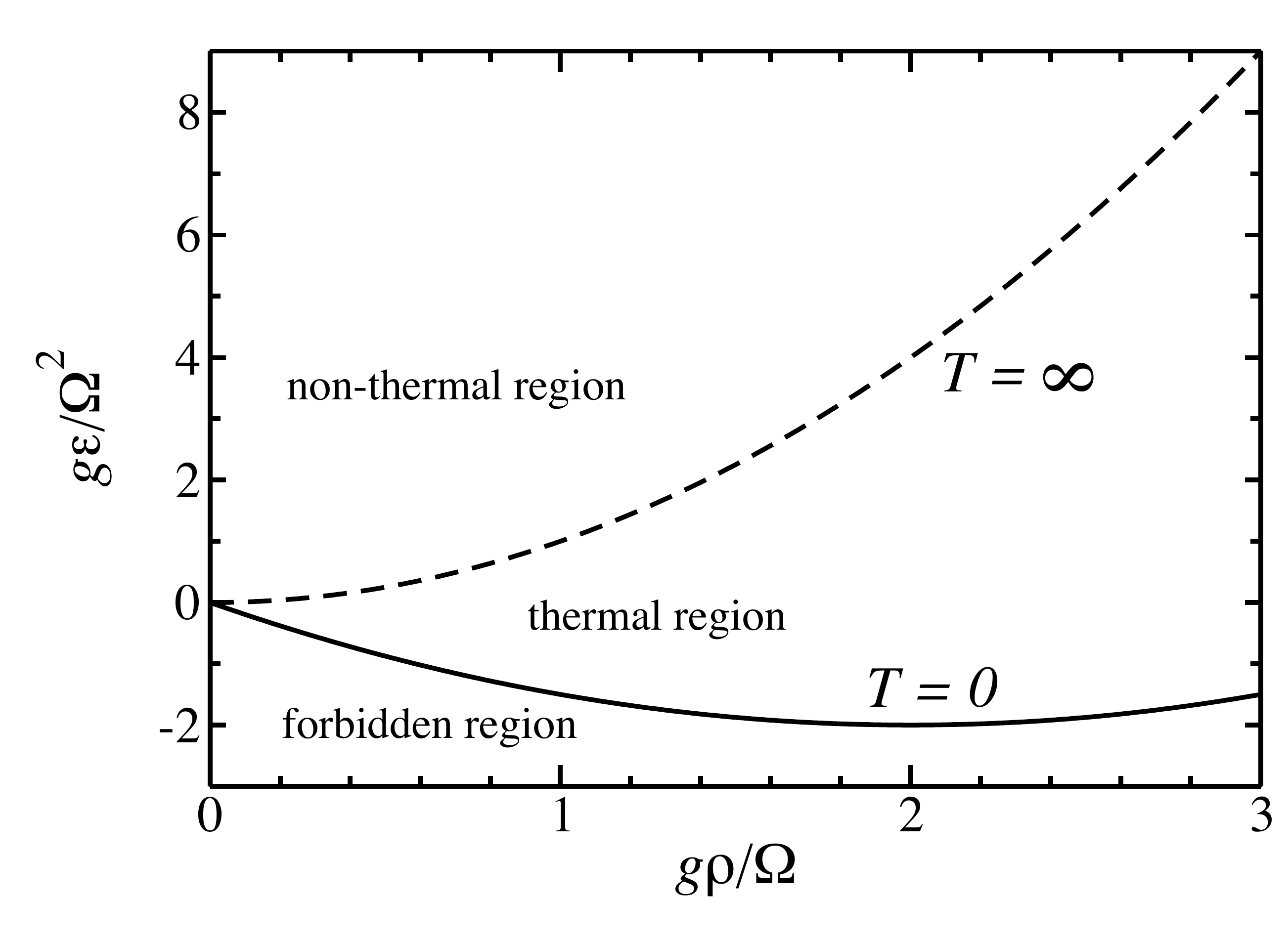}
\caption{\label{fig:phasediag}
Three regions in the $(\rho,\vep)$ plane (see text for
details).
The solid and dashed line correspond to zero and infinite
temperature, respectively.
}
\end{figure}

Let us now investigate what the coupled macroscopic equations
for the action and energy density can give at high temperatures.
The behavior of the system can be viewed in the $(\rho,\vep)$
plane (more precisely, half-plane, as $\rho\geq{0}$ by
construction), shown in Fig.~\ref{fig:phasediag}. It is
convenient to measure $\rho$ and $\vep$ in the natural units
$\Omega/g$ and $\Omega^2/g$. The properties of the grand
canonical equilibrium mapping $(\mu,T)\to(\rho,\vep)$ define
three regions in the $(\rho,\vep)$ plane~\cite{Rasmussen2000}.
(i)~The region below the $T=0$ line is forbidden: for each
fixed value of total norm $\mathcal{N}=\sum_n|\psi_n|^2=L\rho$,
the Hamiltonian~(\ref{Hpsi=}) has an absolute minimum
$H_\mathrm{min}\equiv{L}\vep_\mathrm{min}$,
so lower energies are not allowed. This is precisely the $T=0$
line. For weak disorder, $W\ll\Omega$, the $T=0$ line is close
to that for the clean case $W=0$,
$\vep_\mathrm{min}=-2\Omega\rho+g\rho^2/2$,
and it is the latter one that is shown in
Fig.~\ref{fig:phasediag}. (ii)~The region between the
$T=0$ line and the $T=\infty$ line (the latter is determined
by $\vep=g\rho^2$ regardless of the disorder) corresponds to
the usual thermal states: for
any $\rho,\vep$ in this region the corresponding values of
$\mu$ and $T$ can be found. (iii)~The region above the
$T=\infty$ line corresponds to the so-called non-thermal
states of the system, which cannot be described by the
grand-canonical ensemble with a non-negative temperature
(negative temperatures are not allowed in the thermodynamic
limit $L\to\infty$, since the upper bound for the Hamiltonian
for a fixed total norm is $\propto{L}^2$).
For the disorder-free chain, it has been shown that in the
non-thermal region the ``excess energy'' $\vep-g\rho^2$
tends to condense into localized discrete breathers
\cite{Rasmussen2000,Rumpf2004,Johansson2004,Rumpf2007,%
Rumpf2008,Rumpf2009,Iubini2013}.

The present work is concerned with the regime
\begin{equation}\label{rho<}
g\rho\ll\Omega,\quad |\vep|\ll\Omega\rho,\quad
\vep\leq{g}\rho^2.
\end{equation}
In fact, the conditions of validity discussed in
Sec.~\ref{sec:validity}, imply even stronger restrictions,
but for the discussion of this subsection
inequalities~(\ref{rho<}) suffice. Namely, they ensure
that the thermodynamic relations can be approximated by
(see Appendix~\ref{app:thermodynamics} for details)
\begin{equation}\label{thermodyn=}
\rho=\frac{1}\lambda-\frac{2g}{\lambda^3}\,\frac{1}T
+O(T^{-2}),
\quad
\vep=\frac{g}{\lambda^2}-\frac{2\Omega^2}{\lambda^2}\,\frac{1}T
+O(T^{-2}),
\end{equation}
where $\lambda\equiv-\mu/T$. Inequalities~(\ref{rho<})
imply $T\gg\Omega\rho$, which justifies the expansion in
the powers of $1/T$.

The region of the
$(\rho,\vep)$ plane defined by inequalities~(\ref{rho<})
is the lower vicinity of the $T=\infty$ line in the left
part of Fig.~\ref{fig:phasediag}. Neglecting the nonlinear
frequency shifts, as in Sec.~\ref{ssec:noshifts}, would
correspond to approximating the two parabolas $T=0,\infty$
by the straight tangents at $\rho=0$, $\vep=0$, and studying
the dynamics of $\rho$ on the horizontal tangent to the
$T=\infty$ line, $\vep=0$. However, to check how important
is the coupling between the action and the energy transport,
one should keep the $g\rho^2$ in the energy and the
off-diagonal matrix elements of $\hat{\mathcal{L}}$.
The inequality $T\gg\Omega\rho$ is not sufficient to
establish which of the two contributions to
$\mathcal{L}_{12},\mathcal{L}_{21}$ in Eq.~(\ref{Lall=})
is more important: the first one is proportionl to the
small factor $\Omega\rho/T$, the second one to
$g\rho/\Omega$, and becomes more important than the first
only at $T\gg\Omega^2/g$. Thus, both contrubutions are
kept in the equations below.

Instead of the temperature, it is convenient to introduce
the excess energy
\begin{equation}
u\equiv\vep-g\rho^2\approx\frac{2\Omega^2\rho^2}{T}.
\end{equation}
Then, using Eq.~(\ref{dt=Ldx}), Eq.~(\ref{Lall=}) (which is
valid to the order $1/T$, other corrections being of the
order $1/T^2$ due to the even-odd symmetry in the frequencies),
Eq.~(\ref{thermodyn=}) [which should be inverted keeping the
terms $O(1/T)$], the estimate $K_0\sim{K}_1\sim\Omega^2D_0$
[a natural guess from Eqs.~(\ref{D0=}), (\ref{K0=}),
(\ref{K1=}), which will be verified numerically in
Sec.~\ref{sec:numres}],
and neglecting terms which have relative smallness
$gu/\Omega^2\ll{1}$, one arrives at the following closed
system of equations:
\begin{subequations}\begin{eqnarray}
\label{drhodt=}
\frac{\partial\rho}{\partial{t}}&=&\frac\partial{\partial{x}}
\left(D_0\rho^2\,\frac{\partial\rho}{\partial{x}}+
\frac{K_1}{4\Omega^2}\,\frac{\rho{u}}{\Omega^2}\,
\frac{\partial{u}}{\partial{x}}\right),\\
\label{dudt=}
\frac{\partial{u}}{\partial{t}}&=&2gD_0
\left(\rho\,\frac{\partial\rho}{\partial{x}}\right)^2-\nonumber\\
&&{}-\frac\partial{\partial{x}}\left(\frac{2K_0-K_1}{2\Omega^2}\,
\rho{u}\,\frac{\partial\rho}{\partial{x}}
-\frac{K_0}{2\Omega^2}\,\rho^2\,
\frac{\partial{u}}{\partial{x}}\right).\quad
\end{eqnarray}\end{subequations}
Two conclusions can be drawn from these equations.
(i)~The second term in $\partial\rho/\partial{t}$ has
a relative smallness $u^2/(\Omega\rho)^2$ compared to
the first. Thus, in the regime defined by
inequalities~(\ref{rho<}) the use of the nonlinear
diffusion equation~(\ref{NLdif=}) for the density
alone is justified.
(ii)~If one prepares an initial condition with some
profile $\rho(x)$, while $u(x)=0$ everywhere, then
$\partial{u}/\partial{t}>0$, so the system is pulled
into the non-thermal region. This fact can be easily
understood: if the system tends to an
equilibrium state with $\rho(x)=\rho_\mathrm{eq}$,
$\vep(x)=\vep_\mathrm{eq}$, the equilibrium values
must be given by the spatial averages of the initial
$\rho(x)$ and $\vep(x)=g\rho^2(x)$; the convexity of
the $T=\infty$ line ensures
$\vep_\mathrm{eq}>g\rho_\mathrm{eq}^2$.

Interestingly, Eqs. (\ref{drhodt=}) and (\ref{dudt=})
remain mathematically correct even for $u>0$, i.~e.,
in the non-thermal region where the reasoning of this
section, based on the assumption of local thermal
equilibrium, is not supposed to be valid. Formally,
the origin of such behavior can be traced to the fact
that Eq.~(\ref{IeqNLself=}) still determines a
stationary solution of the kinetic equation even for
$T<0$, $-\mu/T=\lambda>0$. This stationary solution,
however, no longer corresponds to the maximum of the
entropy; the latter is reached in a state with one
discrete breather on top of the $T=\infty$
background~\cite{Rumpf2004}. Nevertheless, this true
equilibrium state will be reached only at very long
times~\cite{Johansson2004}. This should be especially
true in the regime considered here, when the excess
energy $u$ is small. At not too long times, the system
may equilibrate near a transient metastable state with
negative temperature. A straightforward analysis of
Eqs.~(\ref{drhodt=}), (\ref{dudt=}), linearized around
a homogeneous solution $\rho=0$, $u=u_0$, shows that
this homogeneous solution is stable regardless of the
sign of~$u_0$. Such equilibration has been observed 
numerically~\cite{Mulansky2009}. Thus, in the
non-thermal region Eqs.~(\ref{drhodt=}), (\ref{dudt=})
describe the system dynamics near such a
metastable state. 

\section{Numerical results for the transport coefficients}
\label{sec:numres}

\begin{figure}
\includegraphics[width=8cm]{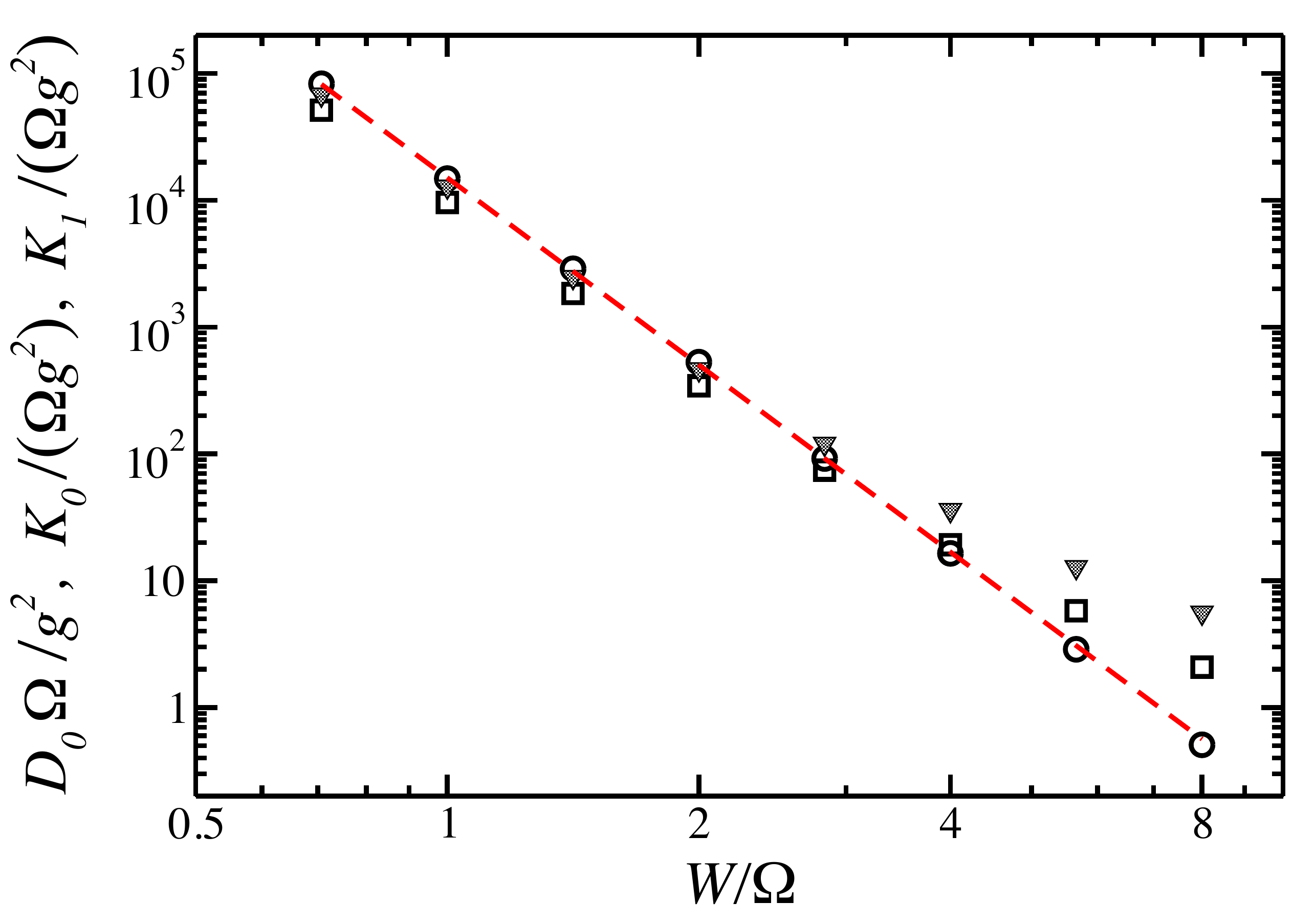}
\caption{\label{fig:transport}
Dependence of the dimensionless transport coefficients
$D_0\Omega/g^2$, $K_0/(\Omega{g}^2)$, 
$K_1/(\Omega{g}^2)$ (circles, squares, and triangles,
respectively) on the dimensionless disorder strength
$W/\Omega$. The straight line shows the dependence
$1.5\times{10}^4\,(\Omega/W)^{4.9}$.}
\end{figure}

The transport coefficients $D_0$, $K_0$, $K_1$ have been
evaluated numerically from Eqs.~(\ref{D0=}), (\ref{K0=}),
and~(\ref{K1=}). The frequency $\delta$~functions were
approximated by boxes of finite width, as discussed in
Sec.~\ref{ssec:numGamma}), and the absence of dependence
of the results on~$w$ has been checked.
The dimensionless combinations
$D_0\Omega/g^2$, $K_0/(\Omega{g}^2)$, $K_1/(\Omega{g}^2)$
depend only on the dimensionless disorder strength, and
the results of their numerical evaluation are shown in
Fig.~\ref{fig:transport}. At weak disorder, they can be
fitted by
\begin{equation}\label{D0K0K1=}
\left\{\begin{array}{c} D_0 \\ K_0/\Omega^2 \\
K_1/\Omega^2 \end{array}\right\}=\frac{g^2}\Omega
\left(\frac\Omega{W}\right)^{4.9\pm{0}.2}
\left\{\begin{array}{c} 1.5 \\ 1.0 \\ 1.3 \end{array}\right\}
\times{10}^4.
\end{equation}
One feature of these results is that
${K}_0\sim{K}_1\sim{D}_0\Omega^2$. This follows
naturally from Eqs.~(\ref{D0=}), (\ref{K0=}), (\ref{K1=}),
if the main contribution comes from modes with frequencies 
$|\omega|\sim\Omega$.
Also, comparing Eqs.~(\ref{D0K0K1=}) and (\ref{M1num=}),
one can see that $D_0\sim\Gamma\xi^2/\rho^2$, which follows
from Eqs.~(\ref{D0=}) and (\ref{Gamma=}),
if the main contribution to $D_0$ and $\Gamma$ comes from
modes with $|d_{\alpha\beta\gamma\delta}|\sim\xi$.

Does the observed dependence of $\Gamma$ or $D_0$ on the
disorder strength have a simple explanation? If one assumes
that the typical value of $V_{\alpha\beta\gamma\delta}$ for
modes located on the same localization segment scales as
$V_{\alpha\beta\gamma\delta}\sim{g}(W/\Omega)^a$ with some
exponent~$a$, and the summation over $\beta,\gamma,\delta$
in Eq.~(\ref{Gamma=}) gives a factor $\sim\xi^3$, then
$\Gamma\propto{W}^{2a-6}$. Several values for the exponent
$a$, ranging from 2 to 4, have been suggested in the
literature~\cite{Shepelyansky1994,Imry1995,Frahm1995,%
Ponomarev1997,Krimer2010,Michaely2012}. In particular,
numerical evaluation of the averages
$\overline{|V_{\alpha\beta\gamma\delta}|}$,
$\overline{V_{\alpha\beta\gamma\delta}^2}$,
gave $a=3.3$ \cite{Frahm1995} or $a=3.4$~\cite{Krimer2010}.
The result of the present work, $\Gamma\propto{W}^{-0.8}$
from Eq.~(\ref{M1num=}), is reproduced if one assumes
$a=2.6$.
The discrepancy is probably due to the fact that in
Refs.~\cite{Frahm1995,Krimer2010}
$V_{\alpha\beta\gamma\delta}$ for all eigenmodes modes were
considered, regardless of their frequencies. At the same
time, the sums in Eqs.~(\ref{Gamma=}), (\ref{D0=}) are
contributed only by those modes for which the frequency
mismatch $\varpi_{\alpha\beta\gamma\delta}$ is small, and
even within this subset there are correlations
between the overlaps and the mode frequencies, as
illustrated in Appendix~\ref{app:correlations}.
Moreover, even though the points on Fig.~\ref{fig:transport}
seem to fall well on a straight line, one cannot exclude
that Eq.~(\ref{D0K0K1=}) still does not represent the true
asymptotic behavior at weak disorder (see the discussion in
Sec.~\ref{ssec:high}).

The value of~$D_0$ determined from Eq.~(\ref{D0=}) can be
compared to the one extracted from the rate of wave packet
spreading obtained by direct numerical integration of
Eq.~(\ref{DNLS=}) in Ref.~\cite{Laptyeva2010}.
Eq.~(\ref{NLdif=}) has a well-known self-similar solution
\cite{Zeldovich1950,Barenblatt1952} (see also Ref.~\cite{Vazquez}
for a comprehensive review):
\begin{equation}\label{ZKB=}
\rho_\mathcal{N}(x,t)=\sqrt{\frac{\mathcal{N}}{\pi\sqrt{D_0t}}
-\frac{\pi\,x^2}{4D_0t}},\;\;\;
|x|<x_t\equiv\sqrt{\frac{4\mathcal{N}}\pi\,\sqrt{D_0t}},
\end{equation}
and $\rho_\mathcal{N}(x,t)=0$ for $|x|>x_t$. This solution is
parametrized by the total norm $\mathcal{N}=\int\rho(x,t)\,dx$,
which is determined by the initial conditions and remains
conserved in time. Eq.~(\ref{ZKB=}) also represents the long-time
asymptotics of the solution of Eq.~(\ref{NLdif=}) for any positive
compact initial condition. The second moment for the wave packet
described by Eq.~(\ref{ZKB=}),
\begin{equation}
m_2(\mathcal{N},t)=\frac{1}{\mathcal{N}}\int\limits_{-\infty}^\infty
x^2\,\rho_\mathcal{N}(x,t)\,dx=\frac{\mathcal{N}}\pi\,\sqrt{D_0t},
\end{equation}
can be directly compared with the numerical result of
Ref.~\cite{Laptyeva2010} for $\log_{10}m_2$, averaged
over the disorder realizations,
\[
\overline{\log_{10}m_2}=0.98+0.5\,\log_{10}(\Omega{t}),
\]
for $W/\Omega=4$ and
$g\mathcal{N}/\Omega=0.74\cdot{21}\approx{15}$. This gives
$D_0\Omega/g^2=4.0$. It should be noted, however, that
$\overline{\log_{10}m_2}\leq\log_{10}\overline{m_2}$ as the
logarithm is a concave function, so this value is likely to
underestimate~$D_0$. At the same time, evaluation of
Eq.~(\ref{D0=}) for $W/\Omega=4$ gives
$D_0\Omega/g^2=16\pm{1}$. This can be considered a
reasonable agreement, given the fact that $W/\Omega=4$
is on the borderline of the weak-disorder regime.
In addition, from the data of
Ref.~\cite{Laptyeva2010} it is seen that the wave packet
expansion with $m_2\propto{t}^{1/2}$ starts to slow down
at $m_2\sim{10}^4$, that is, at the average density in the
packet $g\rho/\Omega\sim{g}\mathcal{N}/(4\Omega\sqrt{m_2})
\approx{0}.04$, which also agrees with the low-density
boundary in Fig.~\ref{fig:rhoRange}.

\section{Conclusions}

In this paper, we have studied the discrete nonlinear Schr\"odinger
equation in the presence of weak on-site disorder. The nonlinearity
was assumed, on the one hand, to be sufficiently weak so that the
eigenmodes of the linear problem remain well-resolved,
but on the other hand, to be sufficiently strong, so that the
dynamics of the eigenmode amplitudes is chaotic for almost all modes.
It was shown that in this regime, the slow dynamics of the eigenmode
intensities can be described by a master equation of the
Fokker-Planck type. Limits of applicability of the master equation
approach have been investigated in detail.

Focusing on the transport of conserved quantities (action/norm and
energy) on macroscopic length and time scales at high temperature,
from the master equation we have derived explicit expressions for
the macroscopic transport coefficients in terms of the wave functions
and frequencies of the eigenmodes of the linear problem. Evaluation
of these expressions was performed numerically for different disorder
strengths. Analysis of the coupled macroscopic equations for the norm
and energy densities have shown that in the considered regime
(weak disorder, moderately weak nonlinearity, and high temperature)
the effect of the energy transport on the transport of action (norm) 
can be neglected, so that the norm density $\rho$ satisfies a closed
macroscopic equation, which is the nonlinear diffusion equation with
the density-dependent diffusion coefficient $D(\rho)=D_0\rho^2$.
The numerical value of~$D_0$, obtained from the present theory, is in
reasonable agreement with the result of the direct numerical
integration of the original nonlinear Schr\"odinger
equation~\cite{Laptyeva2010}.

The density dependence of the diffusion coefficient
$D(\rho)=D_0\rho^2$, obtained in the present work, translates into
the subdiffusive spreading of an initially localized wave packet
with the second moment $m_2$ growing as $m_2\propto{t}^{1/2}$.
It is known from numerical simulations, that at lower densities
or longer times this asymptotics should slow down to
$m_2\propto{t}^{1/3}$, which would mean $D(\rho)=D_0\rho^4$.
Construction of a quantitative theory for this latter regime
remains a challenge.

\section{Acknowledgements}

The author is grateful to B. Altshuler, S. Flach,
V.~Kravtsov, and A. Pikovsky for stimulating discussions.

\appendix

\section{Statistics for a linear one-dimensional chain}
\label{app:chain}

In this Appendix, several quantities are calculated for
the linear eigenvalue problem~(\ref{anderson1D=})
with the flat box distribution for $\ep_n\in[-W/2,W/2]$.

\subsection{Mode spectral density and localization length}
\label{app:DOS}

The average $N$-mode spectral density per unit length is
defined as
\begin{equation}
\nu_N(\omega)=\lim\limits_{L\to\infty}
\frac{1}{L^N}
\sum_{\alpha_1,\ldots,\alpha_N=1}^L
\delta\!\left(\sum_{i=1}^N\omega_{\alpha_i}-\omega\right),
\end{equation}
where the $\delta$~function should be approximated by a peak
of finite width, which should be set to zero \emph{after}
the limit $L\to\infty$ is taken. The $N$-mode frequency
spacing within one localization length $\xi(\omega)$ is then
$\Delta_N=1/(\nu_N\xi^N)$.

\begin{figure}
\includegraphics[width=8cm]{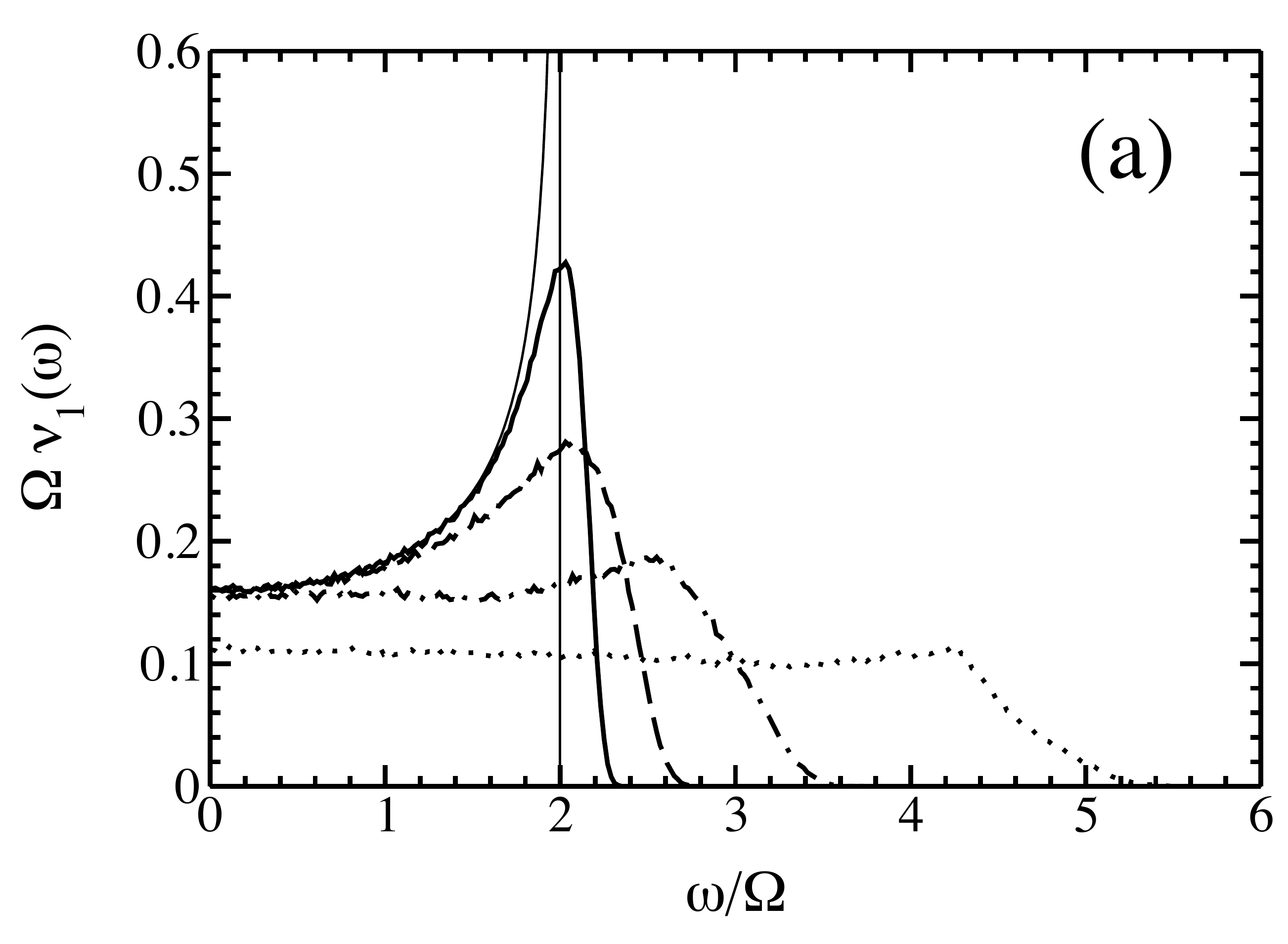}
\includegraphics[width=8cm]{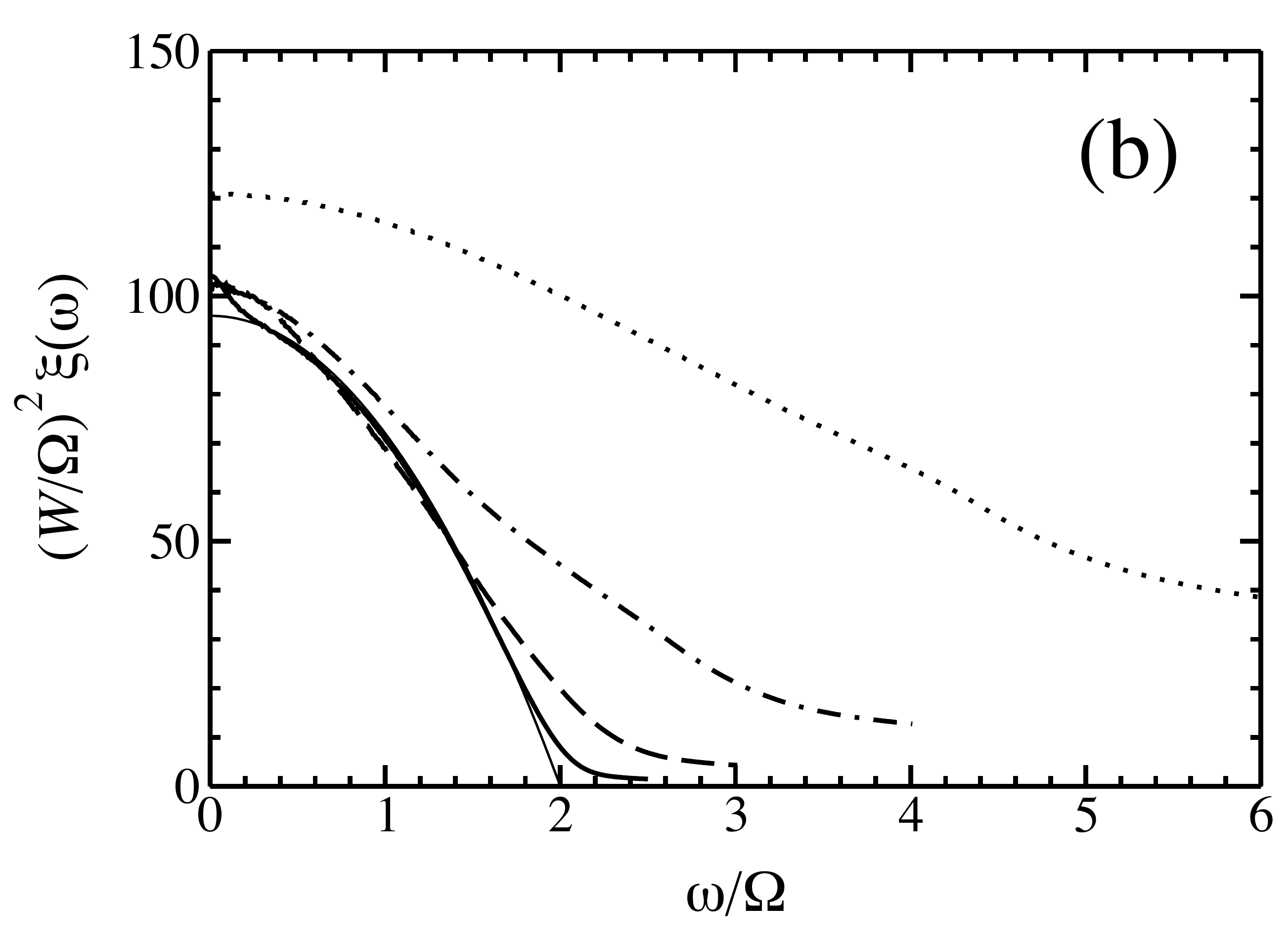}
\caption{\label{fig:DOS}
Single-mode density of states $\nu_1(\omega)$~(a) and
the localization length $\xi(\omega)$~(b) for the eigenvalue
problem~(\ref{anderson1D=}), evaluated numerically for
$W/\Omega=1$ (thick solid line), $W/\Omega=2$ (dashed line),
$W/\Omega=4$ (dash-dotted line), and $W/\Omega=8$ (dotted
line). 
The thin solid lines correspond to the analytical expressions,
Eq.~(\ref{nu1=})~(a), and Eq.~(\ref{xiw=})~(b).
}
\end{figure}

In the weak-disorder limit, for most frequencies $\omega$,
the density of modes can be well approximated by that of
the clean chain. $\nu_1(\omega)$ is straightforwardly
calculated,
\begin{equation}\label{nu1=}
\nu_1(\omega)=\int\limits_{-\pi}^\pi\frac{dk}{2\pi}\,
\delta(\omega-2\Omega\cos{k})
=\frac{1}{2\pi}\,\frac{1}{\sqrt{\Omega^2-(\omega/2)^2}},
\end{equation}
and has square-root type singularities at the band edges,
$\omega=\pm{2}\Omega$. The main effect of weak disorder
is to smear these singularities, as seen from the numerical
results shown in Fig.~\ref{fig:DOS}(a). The smearing occurs
on the frequency scale
$||\omega|-2\Omega|\sim(W/10)^{4/3}\Omega^{-1/3}$.

Given the analytical expression, Eq.~(\ref{nu1=}), the
two-mode spectral density can also be calculated analytically:
\begin{eqnarray}
\nu_2(\omega)&=&\int\limits_{-\pi}^\pi
\frac{dk_1}{2\pi}\,\frac{dk_2}{2\pi}\,
\delta(\omega-2\Omega\cos{k}_1-2\Omega\cos{k}_2)=\nonumber\\
&=&\frac{(2/\pi)^2}{4\Omega+|\omega|}\,\mathbf{K}\!
\left(\frac{(4\Omega-|\omega|)^2}{(4\Omega+|\omega|)^2}\right),
\end{eqnarray}
expressed in terms of the complete elliptic integral
$\mathbf{K}(m)=\int\limits_0^{\pi/2}(1-m\sin^2\phi)^{-1/2}d\phi$.
The singularities at $\omega=\pm{2}\Omega$ are logarithmic,
i.~e., weaker than for $\nu_1(\omega)$, as they are
smeared by the convolution.

The three-mode spectral density,
\begin{equation}
\nu_3(\omega)=\int\limits_{\max\{-2\Omega,-4\Omega+|\omega|\}}^2
d\omega'\,\nu_1(\omega')\,\nu_2(\omega-\omega'),
\end{equation}
is regular at $\omega\to{0}$, $\Omega\nu_3(0)=0.1426\ldots$,
while at $\omega=\pm{2}\Omega$ it has a cusp,
$\Omega\nu_3(2\Omega)=0.1447\ldots$. Thus, in the most
relevant interval $|\omega|<2\Omega$, $\nu_3(\omega)$ is
almost a constant.
For $2\Omega<|\omega|<6\Omega$ it monotonously decreases
to zero. In general, $\nu_N(\omega)$ have weaker singularities
for larger $N$, and in the limit $n\gg{1}$ one obains the
Gaussian shape of $\nu_N(\omega)$ due to the central limit
theorem. 


For the localization length, the following analytical
expression is available in the weak disorder
limit~\cite{Thouless1979}:
\begin{equation}\label{xiw=}
\xi(\omega)=
96\,\frac{\Omega^2}{W^2}
\left(1-\frac{\omega^2}{4\Omega^2}\right).
\end{equation}
As seen from Fig.~\ref{fig:DOS}(b), for $W/\Omega<2$ this
expression works well for most values of~$\omega$, except
for the band edges and the band center. The latter behavior
is a consequence of the well-known anomaly \cite{Gorkov1976,
Czycholl1981,Kappus1981,Schomerus2003,Deych2003,Kravtsov2010}.
Other anomalies at frequencies corresponding to wave vectors
being rational multiples of~$\pi$
\cite{Derrida1984,Kravtsov2011} are beyond the numerical
precision of the present calculation.

\subsection{Fluctuations of nonlinear frequency shifts}
\label{app:shifts}

\begin{subequations}
In this subsection, we present the results for the average
relative dispersion of the nonlinear frequency shifts, which
was defined in Eq.~(\ref{omegaHFfluc=}) as the average over
all eigenmodes~$\alpha$,
\begin{equation}\label{sigma2=}
\sigma^2=\frac{1}L\sum_{\alpha,\beta}
\sum_{n,n'}\phi_{\alpha{n}}^2\phi_{\beta{n}}^2
\phi_{\alpha{n}'}^2\phi_{\beta{n}'}^2,
\end{equation}
as well as for
\begin{equation}\label{sigma2omega=}
\sigma^2_\omega=\frac{1}{L\nu_1(\omega)}\,
\sum_{\alpha,\beta}\delta(\omega-\omega_\alpha)
\sum_{n,n'}\phi_{\alpha{n}}^2\phi_{\beta{n}}^2
\phi_{\alpha{n}'}^2\phi_{\beta{n}'}^2,
\end{equation}
which restricts the average to eigenmodes~$\alpha$ at a given
frequency~$\omega$.
\end{subequations}

\begin{figure}
\begin{center}
\includegraphics[width=8cm]{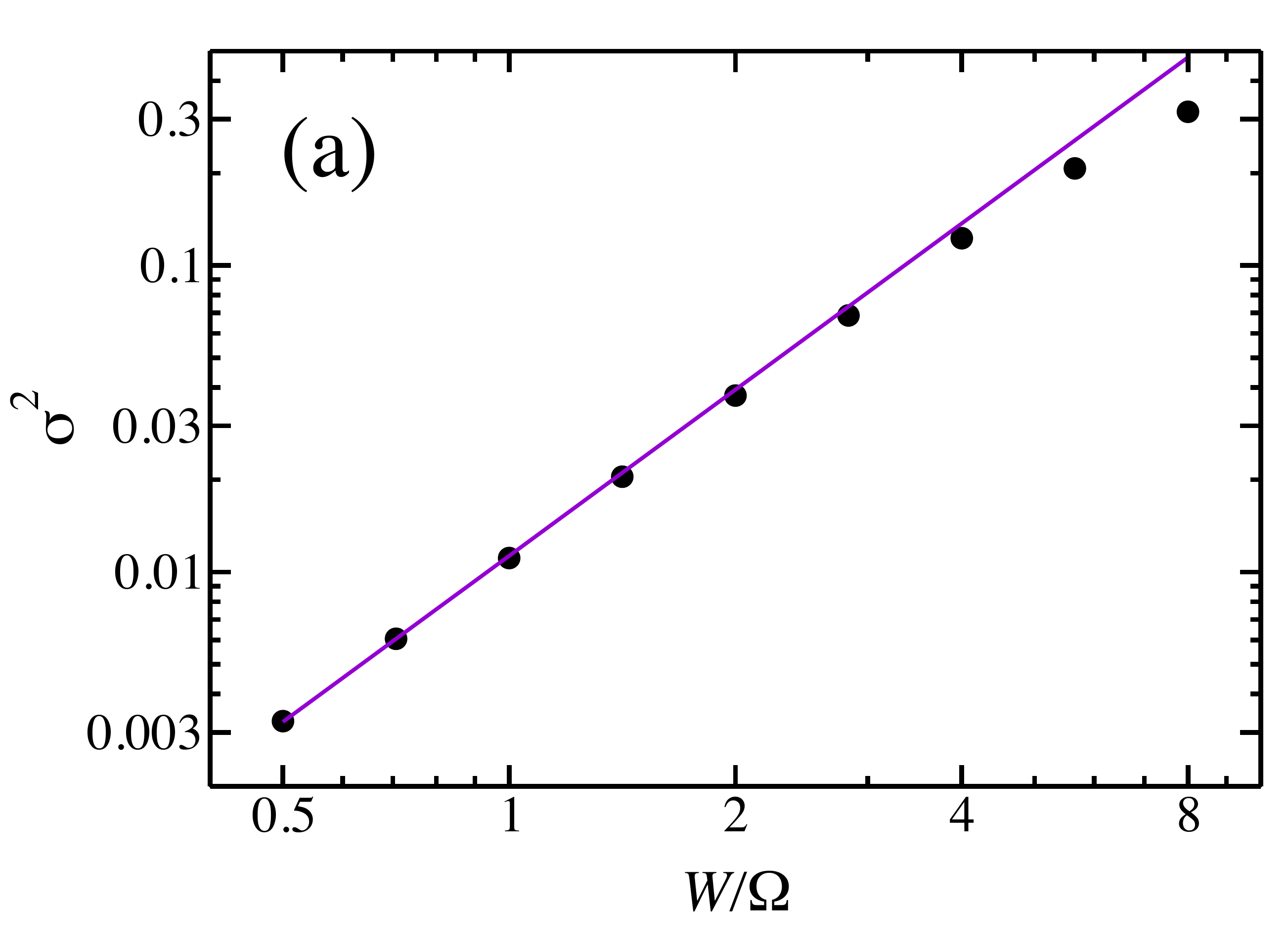}
\includegraphics[width=8cm]{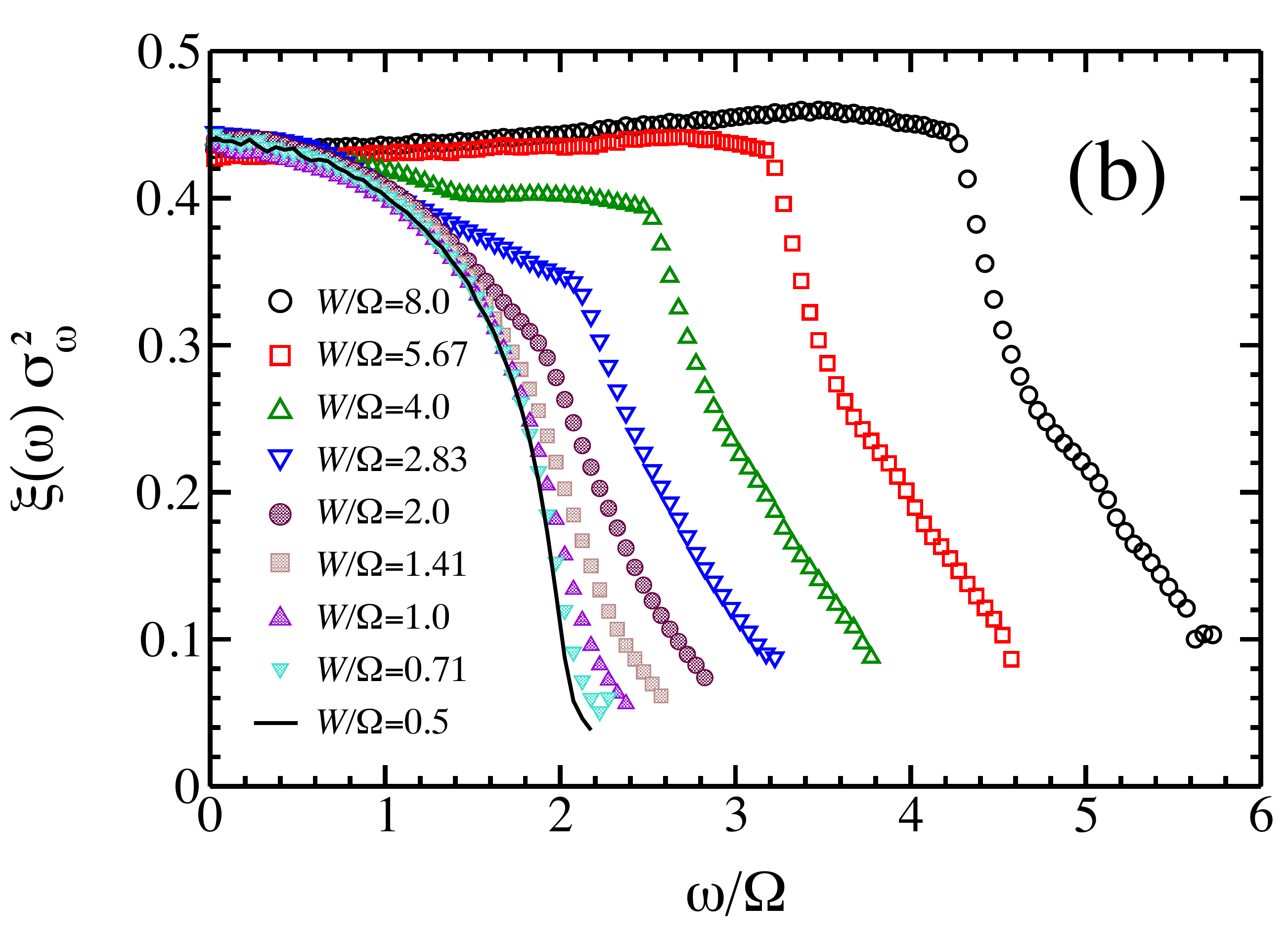}
\end{center}
\caption{\label{fig:shifts}(color online)
(a)~The average relative dispersion $\sigma^2$ of the nonlinear
frequency shift, Eq.~(\ref{sigma2=}), as a function of disorder
strength (circles), fitted by the dependence in Eq.~(\ref{sigma2fit=})
(straight line).
(b)~The rescaled frequency-resolved average relative dispersion,
$\sigma^2_\omega\xi(\omega)$, for different disorder strengths.
}
\end{figure}

The numerical results for $\sigma^2$ are shown
in Fig.~\ref{fig:shifts}(a).
At weak disorder, the dependence on the disorder strength
can be fitted by the expression
\begin{equation}\label{sigma2fit=}
\sigma^2=0.011\,\left(\frac{W}\Omega\right)^{1.8},
\end{equation}
that is, $\sigma^2\propto\xi^{-0.9}$. This dependence is
slightly weaker than $\sigma^2\propto{1}/\xi$ expected from
the simple reasoning of Sec.~\ref{ssec:shifts}. The reason
for this can be understood by looking at the frequency-resolved
fluctuation, Eq.~(\ref{sigma2omega=}), scaled by the
localization length at the same frequency,
$\sigma^2_\omega\xi(\omega)$, plotted in Fig.~\ref{fig:shifts}(b).
At $W/\Omega\gtrsim{3}$, the curves clearly show different
behavior in two distinct frequency intervals, as was also observed
in Ref.~\cite{JohriBhatt}.
At $W/\Omega<2$ the curves show a clear tendency to collapse
on a single universal curve, but the limit is reached quite slowly.
Thus, the asymptotics $\sigma^2\propto{1}/\xi$ is expected to set
in at small $W/\Omega$, corresponding to extremely large
$\xi\gtrsim{1000}$. Similar behavior in the statistics of values
of a single eigenfunction was observed in Ref.~\cite{Kravtsov2011}.
This is also in agreement with discussion of the relaxation rates
in Sec.~\ref{ssec:high}.

\subsection{Correlations between frequencies and overlaps}
\label{app:correlations}

\begin{figure}
\begin{center}
\includegraphics[width=8cm]{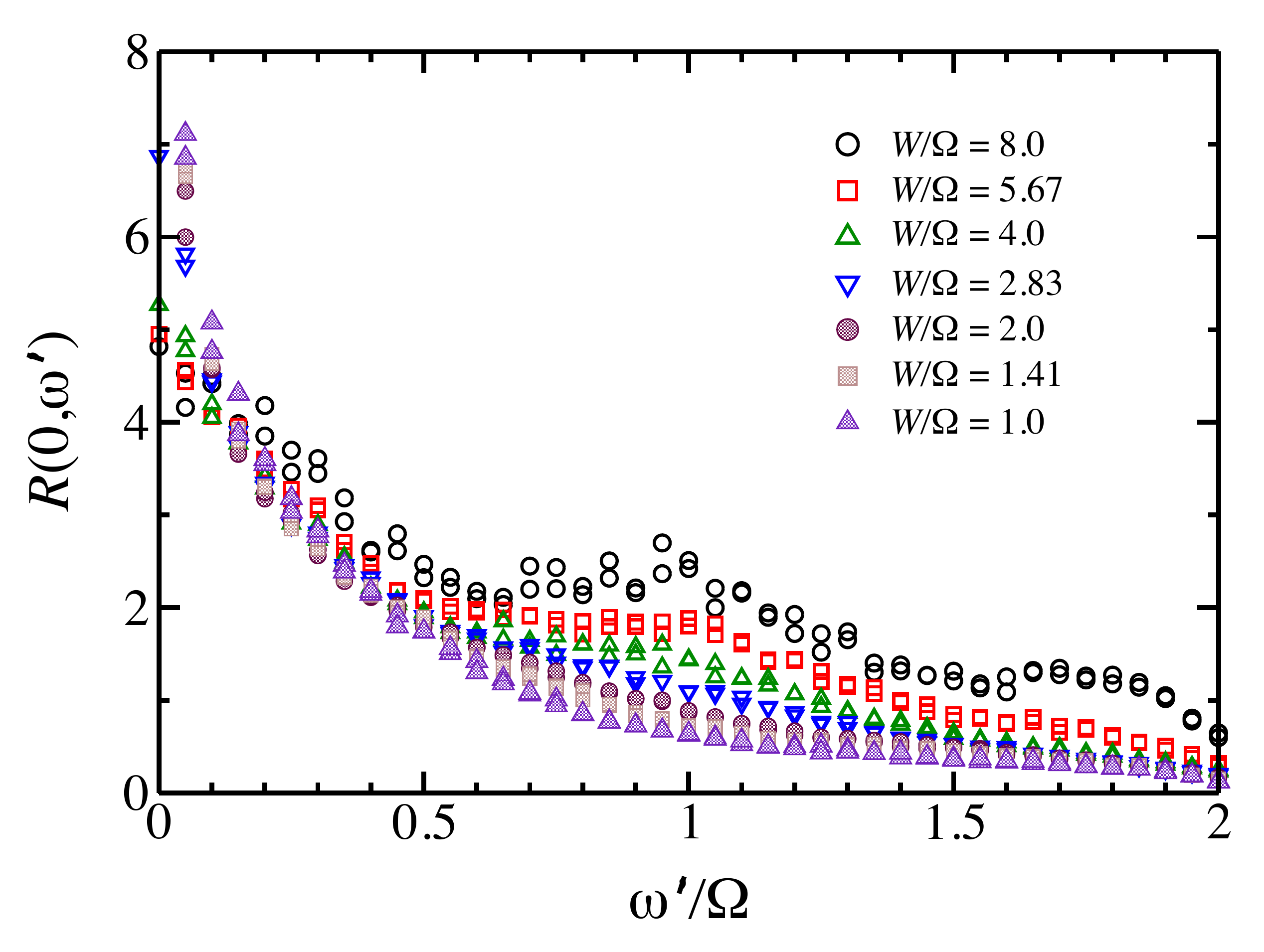}
\end{center}
\caption{\label{fig:Vomega}(color online)
The reltive weight of modes with frequency $\omega'$ in the
decay rate $\overline{\Gamma}_{\omega=0}$ of the modes at
$\omega=$ for different disorder strengths.
}
\end{figure}

To see how the overlaps $V_{\alpha\beta\gamma\delta}$ which
contribute to $\Gamma_\alpha$ are correlated with the frequencies
of the corresponding modes, we calculate the following quantity:
\begin{equation}\begin{split}
{R}(\omega,\omega')={}&{}\frac{4\pi\rho^2}{L\overline{\Gamma}_\omega}
\sump_{\alpha,\beta,\gamma,\delta}
V^2_{\alpha\beta\gamma\delta}
\delta(\omega_\alpha+\omega_\beta-\omega_\gamma-\omega_\delta)\times{}\\
&{}\times{}\frac{\delta(\omega_\alpha-\omega)}{\nu_1(\omega)}\,
\frac{\delta(\omega_f-\omega')}{\nu_1(\omega')},
\end{split}\end{equation}
where $\omega_f=\omega_\gamma$ if
$|\omega_\gamma-\omega_\alpha|<|\omega_\delta-\omega_\alpha|$,
and $\omega_f=\omega_\delta$ in the opposite case ($\omega_f$
is introduced in order to take care of the symmetry
$\omega_\gamma\leftrightarrow\omega_\delta$ by choosing the
one which is closer to $\omega_\alpha$).
Thus defined ${R}(\omega,\omega')$ represents the relative
weight with which partner modes at some frequency~$\omega'$
contribute to  $\overline{\Gamma}_\omega$, with the
normalization
\begin{equation}
\int{R}(\omega,\omega')\,\nu_1(\omega')\,d\omega'=1.
\end{equation}
$R(\omega=0,\omega')$ for different disorder strengths is
shown in Fig.~\ref{fig:Vomega}. At weak disorder, the symbols
have a tendency to collapse on a single curve. At
$W/\Omega\gtrsim{3}$, $R(0,\omega')$ has a singularity
at $|\omega-\omega'|\to{0}$. The present data are not sufficient
to establish its precise character (power-law or logarithmic);
one can only conclude that the singularity is not stronger than
$|\omega-\omega'|^{-0.5}$, and is thus integrable.
At the same time, in the disorder-free system, the singularity
is $|\omega-\omega'|^{-1}$, as was seen in Sec.~\ref{ssec:high}.
Again, as in Sec.~\ref{ssec:high} and Appendix~\ref{app:shifts},
it appears that the values of disorder corresponding for which
the results are presented in Fig.~\ref{fig:Vomega} are still too
large to be in the true weak-disorder limit.

\section{Damped oscillator subject to noise}
\label{app:oscnoise}

Consider a single oscillator, described by the complex
amplitude~$c$ which satisfies the equation of motion,
\begin{equation}\label{oscEOM=}
i\,\frac{dc}{dt}=\left(\omega+g|c|^2\right)c,
\end{equation}
where $\omega$ is the frequency of linear oscillations,
and $g$~is the anharmonicity. The solution of
Eq.~(\ref{oscEOM=}) is
\begin{equation}
c(t)=\sqrt{I}\,e^{-i(\omega+gI)t-i\theta^0}.
\end{equation}
Let now the oscillator be subject to a white noise and
friction:
\begin{equation}\label{Langevin=}
\frac{dc}{dt}=-i\left(\omega+g|c|^2\right)c-\frac\Gamma{2}\,c
+\eta_x(t)+i\eta_y(t),
\end{equation}
where the noise amplitudes satisfy
\begin{equation}
\langle\eta_i(t)\,\eta_j(t')\rangle=\nu\delta_{ij}\delta(t-t'),
\quad i,j=x,y,
\end{equation}
$\nu$~measures the strength of the noise, and $\Gamma$ is
the friction coefficient.
Writing $c=x+iy$, one can introduce the probability
distribution function $P(x,y)$ in the complex plane of~$c$,
and write down the Fokker-Planck equation, corresponding
to the Langevin equation~(\ref{Langevin=}):
\begin{equation}\begin{split}\label{FPxy=}
\frac{\partial{P}}{\partial{t}}={}&{}
-\frac{\partial}{\partial{x}}\left[\omega+g(x^2+y^2)\right]yP-\\
&{}+{}\frac{\partial}{\partial{y}}\left[\omega+g(x^2+y^2)\right]xP+\\
&{}+{}\frac\Gamma{2}\left[\frac{\partial}{\partial{x}}\,xP
+\frac{\partial}{\partial{y}}\,yP\right]+\\
&{}+{}\frac{\nu}2\left(\frac{\partial^2}{\partial{x}^2}
+\frac{\partial^2}{\partial{y}^2}\right)P.
\end{split}\end{equation}
If $P(x,y)$ is interpreted as the density in a cloud of
particles, the first two lines of Eq.~(\ref{FPxy=})
correspond to the clockwise rotation of the cloud around
the origin, the third line to the uniform squeezing
towards the origin, and the last line to the uniform
spread of the cloud. In the action-angle variables,
$x+iy=\sqrt{I}\,e^{-i\theta}$, the same equation
becomes
\begin{equation}\begin{split}\label{FPItheta=}
\frac{\partial{P}}{\partial{t}}={}&{}
-(\omega+gI)\,\frac{\partial{P}}{\partial\theta}+\\
&{}+{}\frac{\Gamma\rho}{4I}\,\frac{\partial^2P}{\partial\theta^2}
+\Gamma\,\frac{\partial}{\partial{I}}\,I
\left(\rho\,\frac{\partial}{\partial{I}}+1\right)P,
\end{split}\end{equation}
where we introduced $\rho=2\nu/\Gamma$ [which is the
equilibrium average value of~$I$, since the stationary
solution of Eq.~(\ref{FPItheta=}) is $e^{-I/\rho}$],
and $\Gamma\rho{I}$ plays the role of the action-dependent
diffusion coefficient. Upon averaging over the phase~$\theta$,
Eq.~(\ref{FPItheta=}) becomes identical to Eq.~(\ref{diff1=}).

\section{Nonlinear oscillator under a quasi-periodic force}
\label{app:fraction}

Consider a single oscillator, described by the complex
amplitude~$c$ which satisfies the equation of motion,
\begin{equation}\label{ipsidot=f}
i\,\frac{dc}{dt}=\left(\omega+g|c|^2\right)c
+\sum_kf_ke^{-i\omega_kt-i\theta_k},
\end{equation}
where the real $f_k,\theta_k$ are the amplitude and the phase
of the $k$th external force oscillating at frequency~$\omega_k$.
Without loss of generality, we assume $g>0$, $f_k>0$.
When will the motion of this oscillator be chaotic?

\begin{figure}
\begin{center}
\includegraphics[width=7cm]{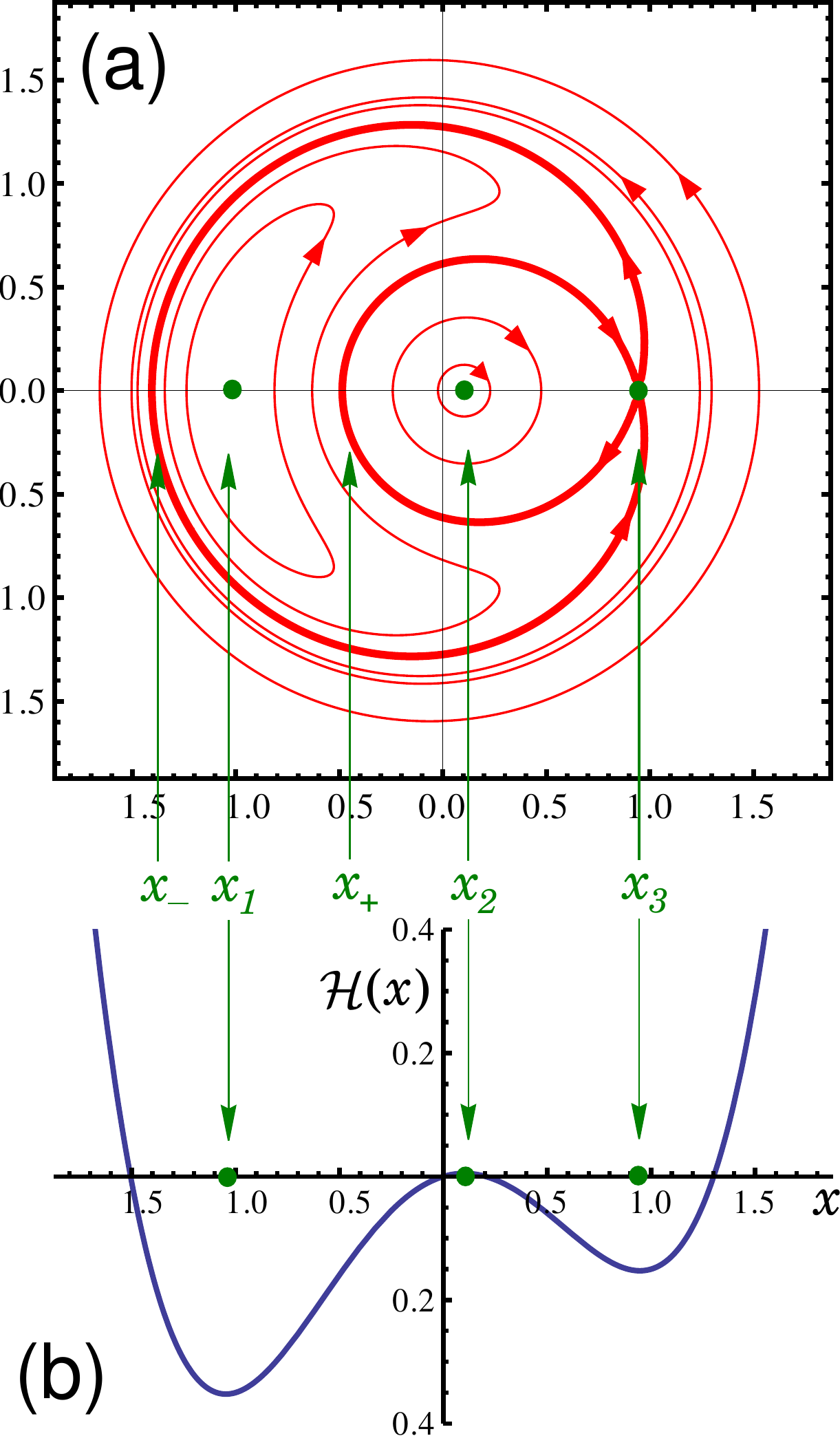}
\end{center}
\caption{\label{fig:oscillator}
(a) The phase portrait corresponding to the Hamiltonian 
(\ref{Hosc=}) (the thick line shows the separatrix, the
the full circles show the stationary points), and
(b)~the function $\mathcal{H}(x)$ for $\mathcal{F}=0.1$.
}
\end{figure}

Let us first analyze the case when only one term is present.
Substituting
\begin{equation}
c(t)=\sqrt{I(t)}\,e^{-i\omega_kt-i\theta_k-i\theta(t)},
\end{equation}
we arrive at equations of motion which have a Hamiltonian
form:
\begin{subequations}\begin{eqnarray}
&&\frac{d\theta}{dt}
=\omega+gI-\omega_k+\frac{1}{\sqrt{I}}\,f_k\cos\theta
=\frac{\partial\mathcal{H}}{\partial{I}},\\
&&\frac{dI}{dt}=2\sqrt{I}\,f_k\sin\theta=
-\frac{\partial\mathcal{H}}{\partial\theta},\\
&&\mathcal{H}(I,\theta)=-\varpi_k{I}+\frac{gI^2}2
+2\sqrt{I}\,f_k\cos\theta,\label{Hosc=}
\end{eqnarray}\end{subequations}
where we denoted $\varpi_k\equiv\omega_k-\omega$.
The stationary points are located at $\theta=0$ or
$\pi$, and $I$ can be found from the cubic equation,
\begin{equation}
\frac{(gI-\varpi_k)^2}{f_k^2}=\frac{1}{I}.
\end{equation}
At $\varpi_k^3<(27/4)gf_k^2$, there is only one elliptic
stationary point at $\theta=\pi$, and the phase portrait
of the oscillator motion in the $(\Re{c},\Im{c})$ plane
has the same topology as in the absence of the external
force.
At $\varpi_k^3>(27/4)gf_k^2$, two more stationary points
appear at $\theta=0$, one elliptic and one hyperbolic.
The phase portrait of the oscillator in this case is
shown in Fig.~\ref{fig:oscillator}(a). It has a separatrix,
which corresponds to the standard case of the nonlinear
resonance~\cite{Chirikov1979,Zaslavsky}.

Let us study this separatrix in more detail. 
As the stationary points are at $\theta=0,\pi$ (corresponding
to $\Im{c}=0$) it is convenient to study the Hamiltonian
$\mathcal{H}$ as a function of the dimensionless variable
$x=\sqrt{g/\varpi_k}\,\Re{c}$:
\begin{equation}
\mathcal{H}(x)=\frac{2\varpi_k^2}g
\left(\frac{x^4}4-\frac{x^2}2+{F}x\right),\quad
{F}\equiv\sqrt{\frac{gf^2}{\varpi_k^3}},
\end{equation}
whose behavior is determined by a single dimensionless
parameter~${F}$.
Let $x_1<x_2<x_3$ be the stationary points, i.~e., the
roots of the cubic equation $\mathcal{H}'(x)=0$ (the
prime indicates the derivative). Noting that
$\partial^2\mathcal{H}(I,\theta)/\partial\theta^2
\propto{-}\cos\theta$
changes sign at $x=0$, we obtain that the points $x_{1,2}$
are elliptic, while the point $x_3$ is hyperbolic.

The points $x_\pm$ at which the separatrix crosses the
negative real semiaxis of~$c$ (that is, $\theta=\pi$) should
be found from the equation $\mathcal{H}(x)=\mathcal{H}(x_3)$.
Since
\begin{equation}
\mathcal{H}(x)-\mathcal{H}(x_3)=\frac{2\varpi_k^2}g\,
(x-x_3)^2\left[\frac{(x+x_3)^2}4-\frac{{F}}{2x_3}\right],
\end{equation}
they are given by $x_\pm=-x_3\pm\sqrt{2{F}/x_3}$.
Both solutions are negative when $\mathcal{H}(x_2)<0$, which is
the case when ${F}<\sqrt{2/27}$.

An exact analytic expression for $x_3$ (which depends on the
dimensionless parameter~${F}$) is not available.
Analyzing the cubic parabola $\mathcal{H}'(x)$, one can
see that $x_1,x_2,x_3$ satisfy
\begin{equation}
-\sqrt{4/3}<x_1<-1,\quad 0<x_2<\sqrt{1/3}<x_3<1,
\end{equation}
where $0,\pm{1}$ are the roots at ${F}=0$,
the points $\pm{1}/\sqrt{3}$ are the extrema of
$\mathcal{H}'(x)$, and $-\sqrt{4/3}$ is the solution of 
$\mathcal{H}'(x)=\mathcal{H}'(1/\sqrt{3})$. Numerically,
$x_1,x_2,x_3$ can be efficiently found by the Newton's
method starting from $-\sqrt{4/3},0,1$, respectively.
It turns out that $x_3$ is well approximated by
\begin{equation}
x_3\approx
1/\sqrt{3}+\sqrt{2\sqrt{3}-3}\,\sqrt{\sqrt{4/27}-{F}},
\end{equation}
whose relative error does not exceed 1.4\% in the whole
available range $0<{F}<\sqrt{4/27}$. It is this
approximate expression that is used in the numerical
calculation.

Let us now consider several forces acting on the oscillator.
Treating each of them separately,
for those of them which have $\varpi_k^3>(27/4)gf_k^2$ and
thus produce a separatrix, we can define the values
$I_{3k}$ (the action corresponding to the hyperbolic point)
and $I_{\pm,k}$ (the actions corresponding to the points
where the separatrix passes through~$\theta=\pi$). If the
separatrices are well separated, that is, the width
$I_{-,k}-I_{+,k}$ of the separatrix is much smaller than
the typical distance $|I_{3k}-I_{3k'}|$ for different $k,k'$,
the different resonances do not interfere. If the opposite
happens, that is, for some $k,k'$ the intervals
$I_{+,k}<I<I_{-,k}$ and $I_{+,k'}<I<I_{-,k'}$ overlap, the
motion becomes chaotic, and the chaotic region roughly
corresponds to the interval
$\min\{I_{+,k},I_{+,k'}\}<I<\max\{I_{-,k},I_{-,k'}\}$.
This is essentially the Chirikov's criterion
\cite{Chirikov1959,Chirikov1979,Zaslavsky}.

Thus, for each normal mode~$\alpha$, whose action $I_\alpha$
is determined by the initial condition and the anharmonicity
is given by
$V_{\alpha\alpha\alpha\alpha}=g\sum_n\phi_{\alpha{n}}^4$,
we determine the forces
$f_{\alpha\beta\gamma\delta}=V_{\alpha\beta\gamma\delta}
\sqrt{I_\beta{I}_\gamma{I}_\delta}$ and the corresponding
separatrix intervals
$I_{+,\beta\gamma\delta}<I<I_{-,\beta\gamma\delta}$.
If among various terms at least two distinct triples
$\beta,\gamma,\delta$ can be found such that
$I_{+,\beta\gamma\delta}<I_\alpha<I_{-,\beta\gamma\delta}$,
the mode~$\alpha$ is counted as chaotic.

\section{Decoupling of higher moments}
\label{app:moments}

Let us neglect the correlations in zero approximation,
$\langle{I}_\alpha{I}_\beta\rangle\to
\langle{I}_\alpha\rangle\langle{I}_\beta\rangle$,
and check whether in the next approximation the cumulant
$\langle{I}_\alpha{I}_\beta\rangle-
\langle{I}_\alpha\rangle\langle{I}_\beta\rangle$
for $\alpha\neq\beta$ is smaller than the main average
$\langle{I}_\alpha\rangle\langle{I}_\beta\rangle$.
The master equation, Eq.~(\ref{master=}), gives
\begin{equation}\begin{split}
&\frac{d}{dt}\left(\langle{I}_\alpha{I}_\beta\rangle-
\langle{I}_\alpha\rangle\langle{I}_\beta\rangle\right)={}\\
&{}=2\sump_{\gamma,\delta}
\left(R_{\alpha\beta\gamma\delta}-2R_{\alpha\gamma\delta\beta}\right)
\bar{I}_\alpha\bar{I}_\beta\bar{I}_\gamma\bar{I}_\delta,
\end{split}\end{equation}
containing a double sum. At the same time, for the main
average, $d(\bar{I}_\alpha\bar{I}_\beta)/dt$, from
Eq.~(\ref{kinetic=}) one straightforardly obtains an
expression involving a triple sum. Since each summation
involves a large number of terms, the cumulant is smaller
then the main average.

\section{Thermodynamics at high temperatures}
\label{app:thermodynamics}

Let us denote $\mu/T\equiv-\lambda$ for brevity. Then the
partition function,
\begin{equation}
Z(\lambda,T)=\int{e}^{-H_\lambda/T}
\prod_n\frac{d\Re\psi_n\,d\Im\psi_n}\pi,
\end{equation}
for the Hamiltonian
\begin{equation}\begin{split}\label{Hlambda=}
H_\lambda={}&{}
-\Omega\sum_n\left(\psi_n^*\psi_{n+1}+\psi_{n+1}^*\psi_n\right)
+\\ &{}+{}\sum_n(\ep_n+\lambda{T})|\psi_n|^2
+\frac{g}2\sum_n|\psi_n|^4,
\end{split}\end{equation}
can be straightforwardly calculated in the high-temperature
limit:
\begin{equation}
\frac{\ln{Z}}{L}=\ln\frac{1}\lambda-\frac{g}{\lambda^2}\,\frac{1}T
+\left(\frac{\Omega^2+\overline{\ep_n^2}}{\lambda^2}
+\frac{5}2\,\frac{g^2}{\lambda^4}\right)\frac{1}{T^2}+O(T^{-3}),
\end{equation}
where $\overline{\ep_n^2}=W^2/12$ is the second moment
of the disorder potential, $L\to\infty$ is the chain length,
and the limit $T\to\infty$ is taken at constant $\lambda$.
Differentiating this expression with respect to $\lambda$
and $1/T$, we obtain the norm and energy densities,
\begin{subequations}\begin{eqnarray}
&&\rho(\lambda,T)=\frac{1}\lambda-\frac{2g}{\lambda^3}\,\frac{1}T
+O(T^{-2}),\label{rholT=}\\
&&\vep(\lambda,T)=\frac{g}{\lambda^2}-
\left(2\,\frac{\Omega^2+\overline{\ep_n^2}}{\lambda^2}
+5\,\frac{g^2}{\lambda^4}\right)\frac{1}{T}+O(T^{-2}).
\nonumber\\ \label{veplT=}
\end{eqnarray}\end{subequations}
The $T\to\infty$ limit is reached at the line $\vep=g\rho^2$.
Since the temperature $T>0$ (otherwise the partition function
diverges, as the Hamiltonian is not bounded from above), the
states of the system for which $\vep>g\rho^2$ are
non-Gibbsian, that is, it is impossible to find
$\lambda$~and~$T$ which would produce such $\rho$~and~$\vep$
in the grand canonical ensemble~\cite{Rasmussen2000}.
It is convenient to introduce the amount of ``non-thermal''
energy in the system,
\begin{equation}
u\equiv{g}\rho^2-\vep=
\left(2\,\frac{\Omega^2+\overline{\ep_n^2}}{\lambda^2}
+\frac{g^2}{\lambda^4}\right)\frac{1}{T}+O(T^{-2}).
\end{equation}

Let us now see what thermodynamic relations are obtained from
Eq.~(\ref{barIeq=}), the equilibrium solution of the kinetic
equation~(\ref{kinetic=}) which neglects the nonlinear shifts.
Expanding in $1/T$, we obtain
\begin{subequations}\begin{eqnarray}
&&\rho(\lambda,T)=\frac{1}L\sum_{\alpha=1}^L\bar{I}_\alpha
=\frac{1}\lambda+O(T^{-2}),
\label{rholTap=}\\
&&\vep(\lambda,T)
=\frac{1}L\sum_{\alpha=1}^L\omega_\alpha\bar{I}_\alpha
=\frac{1}L\sum_{\alpha=1}^L
\frac{\omega_\alpha^2}{\lambda^2}\,\frac{1}{T}
+O(T^{-2})=\nonumber\\
&&\qquad\quad=
\frac{2\Omega^2+\overline{\ep_n^2}}{\lambda^2T}+O(T^{-2}),
\label{veplTap=}
\end{eqnarray}\end{subequations}
where the average of $\omega_\alpha^2$ is calculated by noting
that it is the trace of the square of the linear operator on
the right-hand side of Eq.~(\ref{anderson1D=}).
Eqs.~(\ref{rholTap=}), (\ref{veplTap=}) differ from
Eqs.~(\ref{rholT=}), (\ref{veplT=}) by the absence of the
nonlinear terms proportional to~$g$, but also by the
coefficients at $\overline{\ep_n^2}$. The origin of this
latter difference can be traced back to non-vanishing
correlations, $\langle{I}_\alpha{I}_\beta\rangle\neq
\langle{I}_\alpha\rangle\langle{I}_\beta\rangle$
when $\alpha=\beta$. Indeed, Eq.~(\ref{veplT=}) keeps track of
all correlations, while Eq.~(\ref{veplTap=}) follows from
Eq.~(\ref{kinetic=}) which was obtained by neglecting
correlations. Still, the relative error of Eq.~(\ref{veplTap=})
with respect to Eq.~(\ref{veplT=}) is
$\overline{\ep_n^2}/\Omega^2\sim{1}/\xi\ll{1}$, which was
precisely the justification for neglecting the correlations
in Eq.~(\ref{kinetic=}).

If we use the equilibrium solution which includes the nonlinear
frequency shifts, Eq.~(\ref{IeqNLexp=}), and calculate the
average action and energy densities, it reproduces
Eq.~(\ref{rholT=}) for $\rho$, while for $\vep$ it gives
\begin{equation}
\vep(\lambda,T)=\frac{g}{\lambda^2}-
\left(\frac{2\Omega^2+\overline{\ep_n^2}}{\lambda^2}
+4\,\frac{g^2}{\lambda^4}\right)\frac{1}{T}+O(T^{-2}).
\end{equation}
This expression differs from Eq.~(\ref{veplT=}) by the numerical
coefficients at $\overline{\ep_n^2}/\lambda^2$ (discussed above),
and at $g^2/\lambda^4$. The latter produces a relative error
$(g\rho/\Omega)^2\ll{1}$ in the coefficient at $1/T$.

\section{Spatio-temporal smoothing}
\label{app:smooth}

Let us express $|\psi_n(t)|^2$ in Eq.~(\ref{rhoxt=}) in terms of
the normal mode wave functions $\phi_{\alpha{n}}$ and amplitudes
$c_\alpha(t)=\sqrt{I_\alpha(t)}\,e^{-i\theta_\alpha(t)}$:
\begin{equation}
|\psi_n(t)|^2=\sum_{\alpha\beta}
\phi_{\alpha{n}}\phi_{\beta{n}}\sqrt{I_\alpha(t)\,I_\beta(t)}\,
e^{-i\theta_\alpha(t)+i\theta_\beta(t)}.
\end{equation}
Only terms with $\alpha=\beta$ survive the convolution with
$\mathcal{T}(t)$ when $\tau\gg{1}/\Delta_1$. Then, averaging
over $\mathcal{F}$, we obtain
\begin{equation}
\rho(x,t)=\int{d}t'\,\mathcal{T}(t-t')
\sum_{\alpha,n}\mathcal{S}(x-n)\,\phi_{\alpha{n}}^2
\bar{I}_\alpha(t').
\end{equation}
In the spatial sum, we expand $\mathcal{S}(x-n)$ around
$x-X_\alpha$ to the second order:
\begin{equation}\begin{split}
\sum_n\mathcal{S}(x-n)\,\phi_{\alpha{n}}^2
={}&{}\sum_n\mathcal{S}(x-X_\alpha)\,\phi_{\alpha{n}}^2-{}\\
&-\frac{d\mathcal{S}(x-X_\alpha)}{dx}\sum_n
(n-X_\alpha)\phi_{\alpha{n}}^2+\\
&{}+{}\frac{d^2\mathcal{S}(x-X_\alpha)}{dx^2}
\sum_n\frac{(n-X_\alpha)^2}2\,\phi_{\alpha{n}}^2.
\end{split}\end{equation}
The second term vanishes by the definition of~$X_\alpha$, and
in the last term the first factor gives $1/\ell^2$, while the
sum over $n$ is $\sim\xi^2$.

Equations (\ref{dSdx1=}), (\ref{dSdx2=}) follow from the fact
that for any smooth function $A(x)$, depending on~$x$ on the
scale~$\ell\gg\xi$,
\begin{equation}
\sum_\alpha{A}(X_\alpha)=\int{A}(x)\,dx
\left[1+O(\xi^2/\ell^2)\right].
\end{equation}
This can be seen by first noting that
\begin{equation}
\sum_\alpha{A}(X_\alpha)=\sum_nA(n)\sum_\alpha\phi_{\alpha{n}}^2
\left[1+O(\xi^2/\ell^2)\right],
\end{equation}
obtained analogously to the previous paragraph. Then,
\begin{equation}
\sum_n{A}(n)=\int{A}(x)\,dx,
\end{equation}
with exponential in $\ell$ precision, if $A(x)$ is infinitely
differentiable.

\section{Toy model of an electric $RC$-cirquit}
\label{app:circuit}

\begin{figure}
\begin{center}
\includegraphics[width=8cm]{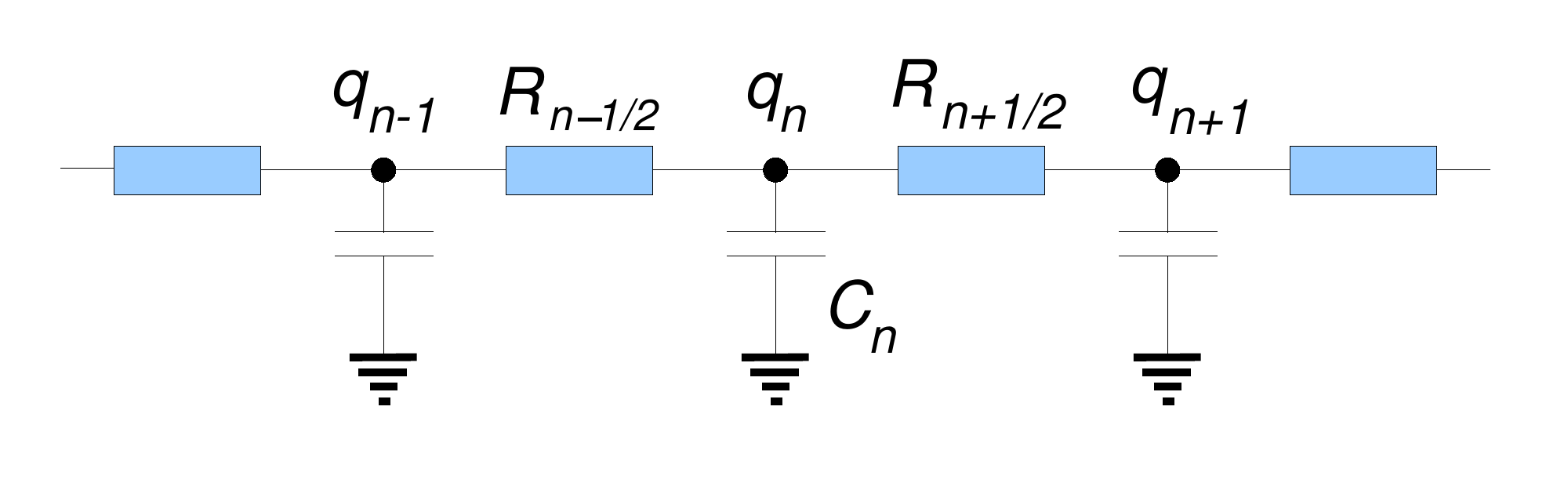}
\end{center}
\caption{\label{fig:circuit}(color online)
The electric circuit with resistors and capacitors, described by
Eq.~(\ref{RCcirquit=}). The bottom plate of each capacitor is grounded.
}
\end{figure}

It is instructive to see how the formalism of Sec.~\ref{sec:macroscopic}
works for a very simple toy model, that of an electric
$RC$-circuit, shown in Fig.~\ref{fig:circuit}. The dynamics of the
charge $q_n$ on the $n$th capacitor is governed by the equation 
\begin{equation}\begin{split}\label{RCcirquit=}
\frac{dq_n}{dt}={}&{}\frac{1}{R_{n+1/2}}
\left(\frac{q_{n+1}}{C_{n+1}}-\frac{q_n}{C_n}\right)+\\
&{}+\frac{1}{R_{n-1/2}}
\left(\frac{q_{n-1}}{C_{n-1}}-\frac{q_n}{C_n}\right).
\end{split}\end{equation}
Indeed, $\varphi_n\equiv{q}_n/C_n$ is the electrostatic
potential on the upper plate of the $n$th capacitor, and 
$(\varphi_{n-1}-\varphi_n)/R_{n-1/2}$ is the current flowing
through the resistor between the capacitors $n$ and $n-1$.
In equilibrium, the potential is constant along the chain.

For the simple model of Eq.~(\ref{RCcirquit=}), the exact
relation between the current and the macroscopic charge
density or potential gradient can be derived straightforwardly.
Indeed, noticing that in a stationary situation the currents
through all resistors should be the same, we the potential
drop is determined by
\begin{equation}
\frac{\varphi_{n+1}-\varphi_n}{R_{n+1/2}}=-J=\mathrm{const},
\end{equation}
so the potential differents between any two capacitors
$n$ and $n+\ell$ is given by
\begin{equation}
\varphi_{n+\ell}-\varphi_n=-J\sum_{n'=n}^{n+\ell-1}R_{n'+1/2}.
\end{equation}
Taking $\ell\gg{1}$, from this we can calculate the macroscopic
potential gradient
\begin{equation}
\frac{\partial\varphi}{\partial{x}}\approx
\frac{\varphi_{n+\ell}-\varphi_n}\ell=-J\overline{R},
\end{equation}
where $\overline{R}$ is the average resistance,
\begin{equation}
\overline{R}=\lim\limits_{\ell\to\infty}\frac{1}\ell
\sum_{n'=n}^{n+\ell-1}R_{n'+1/2}.
\end{equation}
The ``thermodynamic equation of state'', relating the charge
density~$\rho$ to the potential, is 
\begin{equation}
\rho=\frac{1}\ell\sum_{n'=n}^{n+\ell}q_{n'}=
\frac{1}\ell\sum_{n'=n}^{n+\ell}C_{n'}\varphi=\overline{C}\varphi,
\end{equation}
where $\overline{C}$ is the average capacitance, defined
analogously to $\overline{R}$. As a result, we obtain the
sought relation between the current and the gradient of the
potential or charge density:
\begin{equation}
J=-\frac{1}{\overline{R}}\,\frac{\partial\varphi}{\partial{x}}
=-\frac{1}{{\overline{R}}\,{\overline{C}}}\,
\frac{\partial\rho}{\partial{x}}.
\end{equation}

Let us now see how the approach of Sec.~\ref{sec:macroscopic}
works for Eq.~(\ref{RCcirquit=}). The
macroscopic density and current are defined as
\begin{widetext}\begin{subequations}\begin{eqnarray}
&&\rho(x,t)=\int{d}t'\,\mathcal{T}(t-t')
\sum_n\mathcal{S}(x-n)\,q_n(t'),\\
&&J(x,t)=\int{d}t'\,\mathcal{T}(t-t')
\sum_n\tilde{\mathcal{S}}(x-n)\,\frac{dq_n(t')}{dt'}=\nonumber\\
&&\qquad\quad{}=\int{d}t'\,\mathcal{T}(t-t')
\sum_n\tilde{\mathcal{S}}(x-n)\left[\frac{1}{R_{n+1/2}}
\left(\frac{q_{n+1}}{C_{n+1}}-\frac{q_n}{C_n}\right)
+\frac{1}{R_{n-1/2}}\left(\frac{q_{n-1}}{C_{n-1}}
-\frac{q_n}{C_n}\right)\right]=\nonumber\\
&&\qquad\quad{}=\int{d}t'\,\mathcal{T}(t-t')
\sum_n\frac{\tilde{\mathcal{S}}(x-n)-\tilde{\mathcal{S}}(x-n-1)}%
{R_{n+1/2}}\left(\frac{q_{n+1}}{C_{n+1}}-\frac{q_n}{C_n}\right)
\approx\nonumber\\
&&\qquad\quad{}\approx\int{d}t'\,\mathcal{T}(t-t')
\sum_n\frac{\mathcal{S}(x-n-1/2)}{R_{n+1/2}}
\left[\frac{q_n(t')}{C_n}-\frac{q_{n+1}(t')}{C_{n+1}}\right].
\label{Jcirquit=}
\end{eqnarray}\end{subequations}\end{widetext}
To find the current response to a small gradient of the potential,
let us look for a stationary solution of Eq.~(\ref{RCcirquit=})
in the form
\begin{equation}
q_n=C_n(\varphi_\mathrm{eq}-\chi{n}+r_n),
\end{equation}
with the requirement $\overline{r_n}=0$. The corrections~$r_n$
can be found from the equations
\begin{equation}
\frac{r_{n+1}-r_n}{R_{n+1/2}}-\frac{r_n-r_{n-1}}{R_{n-1/2}}
=\frac\chi{R_{n+1/2}}-\frac\chi{R_{n-1/2}},
\end{equation}
which are satisfied when 
\begin{equation}
\frac{r_{n+1}-r_n}{R_{n+1/2}}=\frac\chi{R_{n+1/2}}+C.
\end{equation}
The constant $C$ should be chosen to ensure $\overline{r_n}=0$.
This condition yields $C=-\chi/\overline{R}$ and the solution
\begin{equation}
r_n=r_0+\sum_{n'=0}^{n-1}\chi\left(1-\frac{R_{n'+1/2}}{\overline{R}}\right).
\end{equation}
With this solution, the current is obtained from
Eq.~(\ref{Jcirquit=}), which gives $J=\chi/\overline{R}$.

\end{document}